\documentclass[a4paper,fleqn]{cas-sc}
\usepackage[numbers]{natbib}
\def\tsc#1{\csdef{#1}{\textsc{\lowercase{#1}}\xspace}}
\tsc{WGM}
\tsc{QE}
\tsc{EP}
\tsc{PMS}
\tsc{BEC}
\tsc{DE}
\usepackage{supertabular}
\usepackage{longtable}
\usepackage{lscape} 
\usepackage{comment}
\usepackage{url}
\usepackage{graphicx}
\usepackage{amsfonts}
\usepackage{color,soul}
\usepackage{multicol}
\usepackage{multirow}
\usepackage{color, colortbl}
\usepackage{blindtext}
\usepackage{rotating}
\usepackage{array}
\usepackage{tikz}
\usepackage{longtable}
\usepackage{multicol}
\usepackage{gensymb}
\usepackage{longtable}
\usepackage{placeins}
\usepackage{varioref}

\begin{document}
\let\WriteBookmarks\relax
\def\floatpagepagefraction{1}
\def\textpagefraction{.001}

\shorttitle{A Survey on Privacy for B5G/6G: New Privacy Challenges, and Research Directions}

\title [mode = title]{A Survey on Privacy for B5G/6G: New Privacy Challenges, and Research Directions}        

\shortauthors{C Sandeepa et~al.}
\author[1]{Chamara Sandeepa}
[orcid=0000-0002-3101-7097]
\cormark[1]

\ead{abeysinghe.sandeepa@ucdconnect.ie}

\affiliation[1]{organization={School of Computer Science, University College Dublin},
    city={Dublin},
    country={Ireland}}

\author[1]{Bartlomiej Siniarski}

\author[2]{Nicolas Kourtellis}

\author[1]{Shen Wang}

\author[1,3]{Madhusanka Liyanage}

\affiliation[2]{organization={Telefonica Research},
    city={Barcelona},
    country={Spain}}

\affiliation[3]{organization={Centre for Wireless Communications, University of Oulu},
    city={Oulu},
    country={Finland}}
    
\cortext[cor1]{Corresponding author}

\begin{abstract}
Massive developments in mobile wireless telecommunication networks have been made during the last few decades. At present, mobile users are getting familiar with the latest 5G networks, and the discussion for the next generation of Beyond 5G (B5G)/6G networks has already been initiated.
It is expected that B5G/6G will push the existing network capabilities to the next level, with higher speeds, enhanced reliability and seamless connectivity. 
To make these expectations a reality, research is progressing on new technologies, architectures, and intelligence-based decision-making processes related to B5G/6G.
Privacy considerations are a crucial aspect that requires further attention in such developments, as billions of people and devices will be transmitting data through the upcoming network. However, the main recognition remains biased towards the network security. A discussion focused on privacy of B5G/6G is lacking at the moment.
To address the gap, this paper provides a comprehensive survey on privacy-related aspects of B5G/6G networks. 
First, it discusses a taxonomy of different privacy perspectives. 
Based on the taxonomy, the paper then conceptualizes a set of challenges that appear as barriers to reach privacy preservation.
Next, this work provides a set of solutions applicable to the proposed architecture of B5G/6G networks to mitigate the challenges.
It also provides an overview of standardization initiatives for privacy preservation.
Finally, the paper concludes with a roadmap of future directions, which will be an arena for new research towards privacy-enhanced B5G/6G networks. This work provides a basis for privacy aspects that will significantly impact peoples' daily lives when using these future networks.
\end{abstract}


\maketitle

\section{Introduction}
Wireless mobile communication networks perform the transmission of information through radio waves. With the proliferation of wireless network connectivity, we observe tremendous improvements in the techniques that derive a seamless experience of smooth connectivity to daily activities. We can see that the first ``generation'' of wireless networks named ``1G'' has evolved to the present ``4G'' within four decades. It gradually upgraded with increasing usability and applications. The roll-out of ``5G'' worldwide in this decade has already started, eventually becoming a part of our daily life.  It can be observed that the time difference between two generations is also declining~\cite{tataria20216g}. Therefore, we can expect this trend will continue to B5G/6G and deliver even more speed opening new doors to a wide range of associated technologies. 

The importance of B5G/6G lies in its impact on the lifestyle of the general public. According to the white paper from ITU~\cite{han2018network}, citizens and humans will be placed at the very centre of the next-generation internet by increasing the capabilities of network technology for all aspects of human interactions. It is expected that the 6G networks will be highly scalable~\cite{ray2021perspective}. Therefore, users will get a seamless range of high-speed connectivity of B5G/6G wherever they are located. According to~\cite{han2018network}, the critical drivers for lifestyle and societal changes with futuristic networks in the 2030s will be:
\begin{itemize}
    \item New types of communication modes: Technologies to reduce the overhead of being physically present, provide more interactive telepresence and collaborations, create a range of applications such as industry-based digital twin and medical imaging volumetric data.
    \item Multi-sense networks: Create a fully immersive experience involving multiple physical senses including smell and taste.
    \item Time-engineered applications: Experience ultra-smooth connectivity with better Quality of Service (QoS).
    \item Critical infrastructure: Providing users with secure, reliable connectivity despite the location barriers.
\end{itemize}

The limitations in current 5G networks are also a key driver for the progression to B5G/6G networks. For instance, the existing 5G networks have a latency of 1 ms, but this is too long for Industrial Internet-of-Things (IIoT) applications~\cite{yang20196g}. Therefore, B5G/6G networks will fill this gap by introducing high-speed, reliable connections for IIoT. 

We can see that the upcoming changes focus on providing new mobility experiences with enhanced QoS. The network providers should develop the underlying technologies supporting B5G/6G networks to achieve these needs. Furthermore, it is better to know ahead the potential applications that may benefit from these technologies, features, and functionalities for many stakeholders, including the end-users. The limitations of these applications need to be defined by B5G/6G requirements to provide requirements for these applications. 

\subsection{Why Privacy is Critical in B5G/6G}

The possibility of achieving harmony among hundreds of devices connected around a person is underway with B5G/6G and will be a common ground or truth in future. Since these devices are connected, they may communicate with nearby devices, and stream data continuously. It will digitally create a very sophisticated environment that makes it virtually possible to track every user action. Therefore, B5G/6G networks will eventually lead the way to the creation of a fully automated smart environment that may span over a vast population.

Since this seems inevitable, it is obvious that privacy is crucial and has to be considered. The 6G applications drastically increase the possibility of identifying individuals, their health status, current actions, prediction of decisions, motion, interests, habits, personal beliefs, ideologies, etc. Applications may do this by analysing the data output through sensors, smartphones, and other personal electronic devices with network connectivity. As B5G/6G will facilitate their communication, these devices will not be isolated anymore. Therefore, third parties may accumulate a wide range of signals and extract this information on the data subjects/users. For example, a smart light connected wirelessly with a remote server or smart device can increase the house's energy efficiency.  However, at the same time, it can collect data such as the times that a user is at home, which rooms are used often, and whether there are people in the house. Then, this can provide insights into user habits, preferences, and daily routine~\cite{ylianttila20206g}. 

With the facilitation of capabilities by B5G/6G networks, new technology applications such as extended reality, smart medical systems and autonomous driving, can be expected to improve soon. Then, there will be many protocols, algorithms, and different device brands that will increase the complexity of the network. The work in~\cite{nguyen2021security} shows that since B5G/6G networks will be having this increased network heterogeneity, more concerns will be raised on privacy compared to previous generations. For example, the authors claim the involvement of connected devices in every aspect of humans, such as medical implants, that could pose serious concerns of potential leaks of personal information such as health records. Specifically, this could happen if an adversary detects a privacy vulnerability in algorithms of such systems or transfer data through protocols with lesser data privacy. However, we noted the discussions of privacy in B5G/6G and verticals is still at its early stage, though there is a significant attention on these networks. 

\subsection{Motivation}

As the advent of the study of the next generation of mobile networks has already begun in the research community, we see a lot of related work in the B5G/6G. All of the surveys in~\cite{
nguyen2021security,
sun2020machine,
wang2020security,
porambage2021roadmap,
zhao2020comprehensive,
shahraki2021comprehensive,
dogra2020survey,
huang2019survey,
lu20206g,
de2021survey} are based on the B5G/6G networks, in the areas such as possible architecture or use cases.
The work in~\cite{nguyen2021security} gives an emphasis on security and a discussion on selected privacy issues related to 6G networks.
In~\cite{sun2020machine}, the authors have discussed Machine Learning (ML) privacy aspects, such as issues and protection related to ML in 6G networks. However, ML is only a part of AI, and their work does not cover related aspects other than ML in 6G. Studies on the progression of mobile networks up to the present 5G networks and beyond, their technologies, and challenges are discussed in multiple works including~\cite{
wang2020security, 
zhao2020comprehensive,
shahraki2021comprehensive,
dogra2020survey,
huang2019survey,
lu20206g,
de2021survey}. However, these are not mainly focused on the privacy, but rather on the B5G/6G networks in general.
The survey in~\cite{porambage2021roadmap} provides a vision and a set of Key Performance Indicators (KPI) for 6G networks. This work consists of only a limited discussion on privacy. The work in~\cite{zhao2020comprehensive} discusses several potential privacy issues related to technologies associated with 6G networks. However, their work uses privacy and security together, making it challenging to differentiate. The survey in~\cite{shahraki2021comprehensive} focuses on the 6G network requirements and trends, associated technologies, and challenges. Similarly, the works in~\cite{dogra2020survey,huang2019survey,lu20206g, de2021survey} discuss the technologies associated with 6G networks and potential challenges. The authors in~\cite{de2021survey} also discuss some standardization approaches for 6G networks, yet not specifically targeting privacy.
 
The key highlight in all of these related surveys is the lack of in-depth investigation and discussion on 6G privacy-related aspects. To summarize, these works discuss security issues and only partially consider privacy; it is clear that the research on privacy for B5G/6G is still at an early stage, as already mentioned in~\cite{nguyen2021security}. None of the surveyed literature stands out with a complete focus on privacy, yet discusses only limited aspects of it.

\subsection{Selection Method}

The main objective of this paper is to bring out the discussion of privacy aspects of 6G to highlight its importance for future networks. The approach to reaching the objective is made by investigating answers to a set of research questions that are relevant to the privacy of B5G/6G. Table \ref{tab:method} provides section-wise research questions along with the keywords that are used in the process of finding answers. For this, we used numerous databases such as Elsevier, IEEE xplore, and ACM that contain related research works in privacy and B5G/6G. It is observed that the topics of B5G/6G and privacy are emerging as interesting areas in recent research. We observed an increasing trend of publications in the combinations of keywords privacy, B5G, and 6G. The availability of those topics in publications from either all relevant works or up to the top most relevant works in years from 2018 to 2022 (August) is shown in Figure \ref{trendsAnalysis}. There was a relatively lesser discussion on B5G/6G before 2020, as shown from the graph. The majority of these references (112 out of 141) are from the year 2020 onward. The frequency of using keyword privacy with B5G/6G has increased over time. Especially considering the publications as of August 2022, the discussion on privacy in B5G/6G has already surpassed 2021. These factors indicate increasing attention to B5G/6G privacy is shown by the research community, especially more recently since 2020. Therefore, this paper discusses a comprehensive survey on the privacy aspects of B5G/6G, which we identify as a timely concern.

\begin{center}

\scriptsize
\begin{longtable}[hb!]{|m{3cm}|m{7.5cm}|m{4.5cm}|} 
  \caption{Section-wise research questions and search keywords}
  \label{tab:method}
 \\
  \hline
 \rowcolor[HTML]{CBCEFB} 
\multicolumn{1}{|c|}{\textbf{Section}} &
\multicolumn{1}{|c|}{\textbf{Research Questions}}  &\multicolumn{1}{|c|}{\textbf{Search Keywords}}
  \\  
  \hline\hline
   Section \ref{sec:6GNetwork} - 6G Network & 
   \begin{itemize}
   \item What are different proposed architectures in 6G that builds up the network $?$
   \item What are the key requirements in 6G that is expected to provide better service than the existing networks$?$
   \item What are the technologies that build up and support establishing 6G services$?$ 
   \item What applications are envisioned to emerge with 6G$?$ 
   \end{itemize} &
   6G architecture, 6G technologies, 6G requirements, 6G vision, and 6G applications. \\
   \hline
   
   Section \ref{sec:TM} - Privacy Taxonomy & 
   \begin{itemize}
   \item What is the definition of privacy$?$
   \item How privacy is relevant in the 6G context$?$
   \item What are different categories of privacy that are based on the actions performed on data$?$
   \item What are different types of privacy that is applicable in 6G networks$?$
   \end{itemize}
   & 
   privacy, actions on data, and 6G privacy types. \\
   \hline
   
   Section \ref{sec:issues} - 6G Privacy Challenges and Issues & 
   \begin{itemize} 
   \item Based on the proposed architecture of 6G, what are the possible 6G privacy issues that can arise with either existing and future technologies$?$
   \item What adverse effects can these issues cause to the privacy of individuals, organizations, and the data in the 6G network?$?$
   \end{itemize}
   & 
   6G privacy issues, data leakage, 6G orchestration limitations, privacy attacks, AI adversarial attacks, edge computing issues, XAI limitations, data ownership, confidentiality, private data access, and privacy understanding of the public.  
   \\
   \hline
   
   Section \ref{sec:solutions} - Privacy Solutions & 
   \begin{itemize}  
   \item What are the potential solutions that can eliminate or mitigate the privacy issues in 6G$?$
   \item To which privacy issues the proposed solutions can be applied$?$
   \item What are the possible costs of the proposed techniques when applying as solutions in 6G$?$
   \end{itemize}
   & 
   decentralized AI, edge AI, intelligent control, XAI, PII, blockchain, homomorphic encryption, quantum-resistant encryption, anonymization, differential privacy, privacy by design, and regulations.  
   \\
   \hline

   Section \ref{sec:projects} - 6G Privacy Projects and Standardization & 
   \begin{itemize} 
   \item What are the initiatives of research from both industry and academia focused on enhancing privacy in the 6G era$?$
   \item What scope in privacy is covered in the projects and standardization activities$?$
   \end{itemize}
   & 
   privacy standardization, research projects, privacy guidelines, and privacy verification mechanisms.
   \\
   \hline
   
   Section \ref{sec:lesson} - Lessons Learned and Future Research Directions & 
   \begin{itemize}  
   \item What are the key points to be taken into consideration when evaluating the 6G privacy$?$
   \item What are the open research problems to be addressed$?$
   \item What future directions can be taken to solve the open problems and improve privacy in 6G$?$
   \end{itemize}
   & 
  open problems 6G privacy, and 6G potential future directions.
   \\
   \hline
   
\end{longtable}

\begin{figure}[ht]
    \centering
    \includegraphics[width=0.5\linewidth]{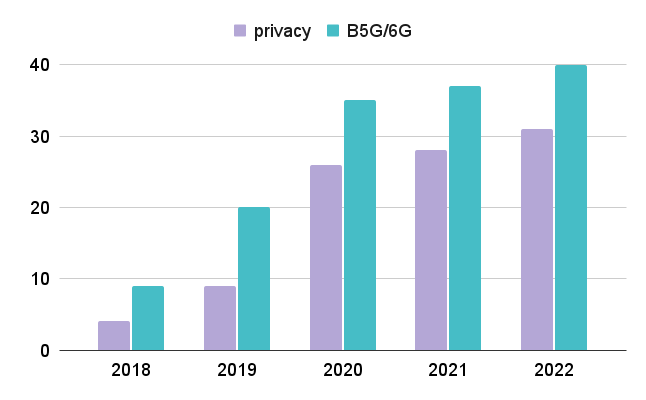} 
    \vspace{1mm}
    \caption{Keywords Frequency Trend of Recent Publications Related to B5G/6G Privacy. (Note that the frequency of the keywords for the ongoing year 2022 is from January to August.) }
    \label{trendsAnalysis}
\end{figure}
\end{center}

Therefore, we consider this work as a useful starting point to identify key privacy issues that may arise in the B5G/6G networks. This work focuses on the current issues of privacy of B5G/6G and potential solutions to these issues. Also, we see numerous future directions also available for the privacy of B5G/6G that could be used to mitigate identified problems, enhance the solutions and be used as a potential starting point.

\subsection{Our Contribution}

Our work is exclusively focused on the privacy aspects of B5G/6G networks and its verticals, within a wide range of topics. Though there are many security focused surveys of B5G/6G architecture, we found it is lacking a comprehensive, and detailed discussion on privacy as indicated by Table \ref{tab:surveys}. The table also includes the limitations of the related works in the context of privacy, that we have already addressed in this paper. Some of the surveys give main attention to security of 6G networks, which should be distinguished from the concept of privacy. Other works in this table only focus on certain aspects of privacy, which may not necessarily be related to B5G/6G. Therefore, we contribute to address this gap in our work on comprehensive discussion on privacy of B5G/6G.
The following key points highlight our contributions to this survey:

\begin{itemize}
    \item \textbf{Identify privacy taxonomy:} Investigate different taxonomies for privacy defined by consideration of various aspects of privacy requirements.
    \item \textbf{Explore privacy issues in B56/G6:} We identify possible privacy issues that could be affected when reaching them.
    \item \textbf{Discuss privacy solutions for the issues:} The potential solutions that could be used to address the issues are also identified.
    \item\textbf{Discuss 6G privacy projects and standardization:} We summarize the existing privacy projects applicable for 6G and also the standardization efforts made to achieve privacy preservation and enhancement. 
    \item \textbf{Summarize lessons learned and propose future directions for B5G/6G privacy:} The key messages from the research are discussed, and the future actions for them are presented, with the possible challenges to be addressed.
\end{itemize}

\renewcommand{\arraystretch}{1}
\begin{center}
\scriptsize
  \begin{longtable}{| m{0.2cm}|m{0.38cm}|m{0.38cm}|m{0.38cm}|m{0.38cm}|m{0.38cm}|p{4cm}|p{5.5cm}|}
    \caption{Summary of Important Surveys on 6G Privacy }
    \label{tab:surveys}
    \\

  \hline
      \rowcolor[HTML]{CBCEFB}
    	\multicolumn{1}{|c|}{\textbf{Ref.}} 
         & \multicolumn{1}{|c|}{\textbf{{\rotatebox[origin=c]{90}{6G Network}}}}
         &\multicolumn{1}{|c|}{ \textbf{{\rotatebox[origin=c]{90}{Privacy Taxonomy}}}}
         &\multicolumn{1}{|c|}{\textbf{{\rotatebox[origin=c]{90}{Privacy issues}}}}
         &\multicolumn{1}{|c|}{\textbf{{\rotatebox[origin=c]{90}{Privacy solutions}}}}
         &\multicolumn{1}{|c|}{\textbf{{\rotatebox[origin=c]{90}{Research Directions}}}}
         &\multicolumn{1}{|c|}{\textbf{Remarks}}
         &\multicolumn{1}{|c|}{\textbf{Limitations}}
         \\ [5ex]
    \hline
    \hline
        \multicolumn{1}{|c|}{\cite{nguyen2021security}} & \cellcolor{yellow!20} M & \cellcolor{red!20} L & \cellcolor{red!20} L & \cellcolor{red!20} L & \cellcolor{yellow!20} M &
        Overview of vulnarabilities in legacy networks, 
required security and privacy enhancements for 6G, 
security and privacy solutions for 6G, technical challenges. &
Mainly focuses on 6G security issues rather than privacy. Only a limited discussion on the open challenges for privacy in 6G, privacy solutions are not available for most issues, lack of distinction between privacy and security.
\\ [2ex]
    \hline
    \multicolumn{1}{|c|}{\cite{sun2020machine}} & \cellcolor{yellow!15} M & \cellcolor{red!20} L & \cellcolor{red!20} L & \cellcolor{yellow!20} M & \cellcolor{yellow!15} M & 
        Privacy violations in ML, privacy protection related to ML,
applications of ML in 6G, new trends and open issues in ML privacy. & The work is based on the role and the privacy of ML models. ML, being a subset of AI, is not necessarily be completely related with 6G. The work does not cover the aspects of privacy in 6G other than ML.\\ [2ex]
        \hline
        \multicolumn{1}{|c|}{\cite{wang2020security}} & 
        \cellcolor{green!15} H & \cellcolor{yellow!20} M & \cellcolor{yellow!20} M & \cellcolor{red!20} L & \cellcolor{yellow!20} M &
        Mobile network evolution to 6G, 
security and privacy issues,
future research challenges. & Does not differentiate privacy from security. Both terms are used together, making it difficult to identify privacy issues separately. Limited to 6G associated technologies.\\ [2ex]
    \hline
        \multicolumn{1}{|c|}{\cite{porambage2021roadmap}} & 
        \cellcolor{yellow!20} M & \cellcolor{yellow!20} M & \cellcolor{red!20} L & \cellcolor{red!20} L & \cellcolor{yellow!20} M &
        6G security and privacy requirements, 
vision and KPIs, security and privacy challenges, discussion on privacy aspects for 6G. & This work is focused on security aspects of 6G. Contains a limited discussion on privacy issues and potential solutions. Unclear on how the solutions connect with the issues.\\ [2ex]
    \hline
    \multicolumn{1}{|c|}{\cite{zhao2020comprehensive}} & 
        \cellcolor{green!15} H & \cellcolor{red!20} L & \cellcolor{red!20} L & \cellcolor{red!20} L & \cellcolor{yellow!20} M &
        6G networks technologies, 
potential security and privacy  
issues from the technologies. & Mainly discuss the 6G network and its architecture, only a small discussion of security and privacy together and several attacks, most of them are not directly related to privacy, but rather security.\\ [2ex]
    \hline
    \multicolumn{1}{|c|}{\cite{shahraki2021comprehensive}} & 
        \cellcolor{green!15} H & \cellcolor{red!15} L & \cellcolor{yellow!20} M &
        \cellcolor{red!20} L &
        \cellcolor{red!15} L & 
        6G requirements and trends, 
revolutionary technologies, 
6G challenges and future research directions. & Discusses mainly on the 6G network applications, services, and technologies. Only a small consideration of privacy issues in smart services. Does not significantly discusses privacy solutions.\\ [2ex]
    \hline
    \multicolumn{1}{|c|}{\cite{dogra2020survey}} & 
        \cellcolor{green!15} H & \cellcolor{red!15} L & 
        \cellcolor{red!15} L &
        \cellcolor{red!15} L &
        \cellcolor{red!15} L & 
        5G NR features and use-cases,
migration towards B5G/6G,
6G networks requirements, and
research challenges. & Discusses B5G network architecture and associated technologies. Lack of discussion on privacy issues and solutions \\ [2ex]
    \hline
    \hline
    \multicolumn{1}{|c|}{\cite{de2021survey}} & 
        \cellcolor{green!15} H & \cellcolor{red!15} L & 
        \cellcolor{yellow!20} M &
        \cellcolor{red!20} L &
        \cellcolor{red!15} L & 
        Current developments of 6G, 
Limitations of existing 5G mobile networks,
new technology enablers for 6G,
standardization approaches &  Focuses on 6G trends, applications, and requirements and does not have the main consideration of privacy issues. Mentions the importance of providing privacy for 6G but does not include a sufficient discussion on solutions.\\ [2ex]
    \hline
        \multicolumn{1}{|c|}{\textbf{This paper}} & \cellcolor{green!15} \textbf{H} & \cellcolor{green!15} \textbf{H} & \cellcolor{green!15} \textbf{H} & \cellcolor{green!15} \textbf{H} & \cellcolor{green!15} \textbf{H} & \cellcolor{green!15}  \textbf{A comprehensive survey on privacy of B5G/6G, taxonomy, challenges and issues, potential solutions, standardization, and future research directions.}&\\[2ex]
    \hline

  \end{longtable}
  
\begin{flushleft}
\begin{center}
\begin{tikzpicture}

\node (rect) at (1,0) [draw,thick,minimum width=1cm,minimum height=0.5cm, fill= red!15, label={[align=left]right:Low Coverage: The paper did not consider this area or only very briefly discussed it through mentioning it \\in passing}] {L};
\node (rect) at (1,0.8) [draw,thick,minimum width=1cm,minimum height=0.5cm, fill= yellow!20, label={[align=left]right:Medium Coverage: The paper partially considers this area (leaves out vital aspects or discusses it in relation to \\other areas without a specific focus on it)} ] {M};
\node (rect) at (1,1.6) [draw,thick,minimum width=1cm,minimum height=0.5cm, fill= green!15, label={[align=left]right:High Coverage: The paper considers this area in reasonable or high detail}] {H};
\end{tikzpicture}
\end{center}

\end{flushleft}

\end{center}

\subsection{Outline}

\begin{figure}[htb]
    \centering
    \includegraphics[width=0.75\textwidth]{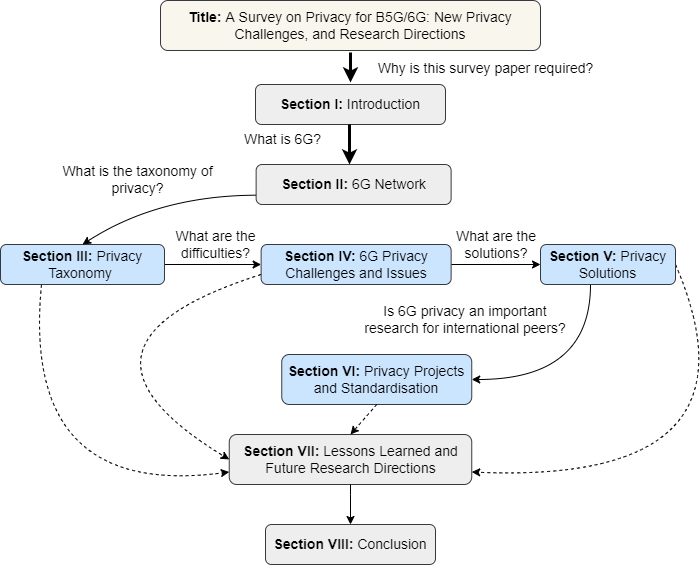} 
    \vspace{1mm}
    \caption{Outline of the present study}
    \label{Figure_outline}
\end{figure}

The rest of the paper is organized as follows. Section~\ref{sec:6GNetwork} outlines the details of the possible 6G network and its architectures available in literature. Section~\ref{sec:TM} discusses the privacy taxonomy. Privacy issues that could arise in B5G/6G are discussed in Section~\ref{sec:issues}. The proposed solutions to address these issues are described in Section~\ref{sec:solutions}. The standardization and 6G privacy projects are summarized in Section~\ref{sec:projects}, whereas Section~\ref{sec:lesson} provides insights on lessons learned and possible future directions. 
Finally, Section~\ref{conclusion} concludes the paper.
Figure~\ref{Figure_outline} provides an outline of the Sections and the arrangement of the paper.

\section{6G Network}\label{sec:6GNetwork}

\subsection{6G Architectures}

\begin{figure*}[htb]
    \centering
    \includegraphics[width=0.8\textwidth]{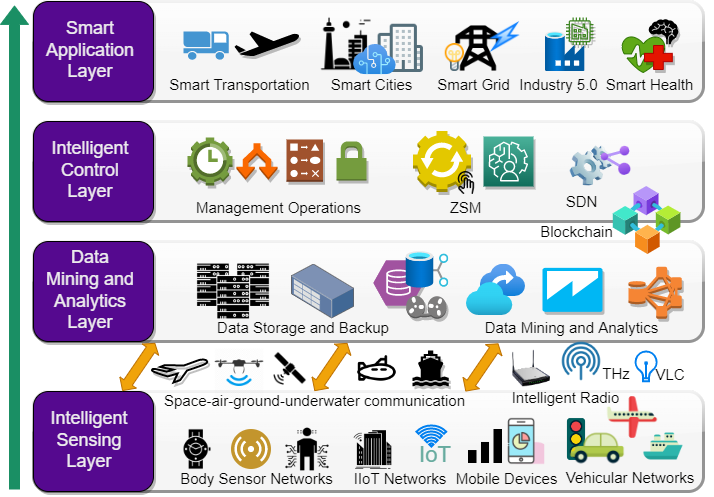} 
    \vspace{1mm}
    \caption{B5G/6G Architecture Layers with Different Layers showing Technologies, and Applications}
    \label{Figure_6G_AI_architecture}
\end{figure*}

To meet the expectations for B5G/6G, novel architectures are designed by adding AI as a key component. The work in~\cite{yang2020artificial} provides a layered architecture for B5G/6G considering AI-enabled functions. It includes four layers for 6G architecture which consists of:
\begin{itemize}
    \item \textbf{Intelligent Sensing Layer} - Using AI-enabled devices or a crowd, sense data from physical locations.
    \item \textbf{Data Mining and Analytics Layer} - Process and analyze raw data from a massive number of devices for knowledge discovery, with the aid of AI-based data mining approaches.
    \item \textbf{Intelligent Control Layer} - Learning, optimization and decision-making process for 6G network functionalities, with the prominent use of AI models that can adapt and automatically make decisions by itself.
    \item \textbf{Smart Application Layer} - Deliver application-specific functionalities with the application of high-level industry AI-based concepts such as intelligent transportation, and smart health.
\end{itemize}
This layered architecture is represented in Figure~\ref{Figure_6G_AI_architecture}. Therefore, there is a ubiquitous use of AI throughout the B5G/6G networks associated ecosystem. 

Considering other works on architecture, the authors of~\cite{ziegler20206g} discuss four building blocks for 6G architecture from the physical layer to service layer with secure automation of orchestrated functions. The building blocks are: 1) \textit{platform} that consist of the infrastructure, 2) \textit{functional} components, that cover the network operations and AI functions, 3) \textit{specialized} component that enables operations such as flexible offloading and slicing, and 4) \textit{orchestration} to provide open service functionalities and monetization and closed-loop automation. Another work in ~\cite{zhang20196g} represents a four-tier 6G model with AI-enabled space-air-ground-underwater networks. The space networks consist of densely deployed satellites in different orbits to cover the under-served areas in terrestrial networks. 

\subsection{6G Requirements}

The 6G networks are expected to offer enhanced connectivity along with substantially higher data delivery rates for regular tasks. These tasks may include quicker downloads, uploads, and higher resolution streaming. However, when a new generation of wireless networks is introduced, data rate is just one of the primary expected needs that is enhanced.  As discussed in~\cite{yang20196g}, the adoption of 6G technologies is expected to be available around the 2030s with 100 to 1000 times faster than 5G with up to terabit-per-second speeds. Also, the authors mention that B5G/6G networks should be able to handle extremely dense connections with trillion-level objects, compared with the current billion-level. B5G/6G networks also consist of very low latencies (less than 1 ms) to support latency-sensitive future applications. The following key points can be shown as the estimated possible key requirements for 6G networks~\cite{zhang20196g}:
\begin{itemize}
    \item Peak data rate of at least $1 Tb/s$
    \item User-experienced data rate of $1 Gb/s$.
    \item Over the air latency of $10-100 \mu s$ and high mobility ($\leq 1,000 km/h$).
    \item $10^7$  devices/$km^2$ connection density and area traffic capacity of up to $1/Gb/s/m^2$ for scenarios such as hotspots.
    \item Energy efficiency of $10$–$100$ times and a spectrum efficiency of $5$–$10$ times those of 5G.
\end{itemize}

The works in~\cite{liu2020vision,zhang20196g} provide a comparison of metrics, and KPIs among existing generations with 6G. 

Therefore, in essence, B5G/6G networks will primarily fulfill the targets of ultra-high speeds and complex connectivity with very low latency. Many potential techniques are suggested to tackle these performance requirements, ranging from physical requirements such as using Terahertz frequency bands, and improving network infrastructures, such as flexible heterogeneous networks~\cite{yang20196g} to software-based solutions. 

\subsection{Key 6G Technologies}

When considering the key driving B5G/6G technologies, we can categorize them into 1) technologies that enable 6G functionalities and 2) technologies associated with 6G services. The first one is associated with fulfilling the B5G/6G requirements that are expected. These may range from hardware to software solutions that enhance speed, reduce latency, improve overall service quality, security, privacy, etc. The second is about the technologies that will be possible to use at a large scale with the features provided by the future networks. We further discuss them as follows:

\subsubsection{B5G/6G Enabling Technologies}
   There are multitude of technologies that support in establishing the underlying network and infrastructure of B5G/6G. The work in \cite{gur2022integration} discuss about Information-Centric Networking (ICN), which can act as a prominent network technology focused on content-centric network architecture, rather than host-centric routing/operation. It can offer many benefits including improved storage and caching, high mobility support, and better support for realtime and context aware applications. The authors in \cite{saad2019vision} discuss trends in B5G/6G, including the emergence of smart surfaces, the increase of small data, the convergence of communication, sensing control, localization and computing. These will lead to the B5G/6G enabling technologies as discussed in~\cite{saad2019vision} are listed as follows:

\begin{itemize}
    \item \textbf{AI/ML} - Adding intelligence to B5G/6G networks will be primarily done through the use of AI. A substantial surge of AI research can be observed, starting from late 2000's~\cite{AISurveyEvolution}. Especially this can be done with ML models when considering their significant improvement over the recent years. With the ML approaches such as Deep Learning (DL) giving higher accurate results for unseen data, using them for B5G/6G will be trivial.
    
    \item \textbf{Above 6 GHz for B5G/6G} - The use of higher frequency bands above sub-6 GHz as a first step for B5G, and then exploring frequencies beyond mmWave, which resides in Terahertz (THz) band. This technology will be crucial to achieving the high speeds required for B5G/6G applications.

    \item \textbf{Communication with Large Intelligent Surfaces} - To facilitate these higher frequency bands, it is required to introduce intelligent surfaces that can optimally guide signals to reach the destination. Therefore, B5G/6G network communication will result from joint optimization of both devices and environment using Re-configurable Intelligent Surfaces (RISs)~\cite{dardari2020communicating}.
    \item \textbf{Edge AI} - AI-enabled capabilities will be supplemented by ``collective network intelligence'',  in which network intelligence is pushed to the edge to provide a distributed autonomy in B5G/6G, where AI and learning algorithms will be performed on edge devices. We will further discuss this in Section \ref{sec:solutions}.
    \item \textbf{Integrated Terrestrial, Airborne, and Satellite Networks} - To enhance the range of future networks, drones and terrestrial infrastructure will be used, and they will be connected with Low Earth Orbit (LEO) satellites for wide area coverage.
    \item \textbf{Energy Transfer and Harvesting} - B5G/6G network base stations will be able to provide basic power transfer for devices, particularly implants and sensors. This makes it a promising solution for powering future Internet-of-Things (IoT) devices~\cite{lopez2021massive}.
    \item \textbf{Beyond 6G} - A set of technologies at early stages at present will be matured with the arrival of B5G/6G. They will be helpful in the research and standardization efforts such as security, privacy, and long-distance networking capabilities for future networks.
    
\end{itemize}

Considering the intelligence that will be added as a core feature in B5G/6G, AI usage will not be limited to performance improvements. However, the vision for 6G wireless networks is to achieve an autonomous ecosystem with human-like intelligence and consciousness~\cite{zhang20196g}, with numerous ways to communicate and interact. Therefore, integrating AI and other technologies to the B5G/6G are crucial requirements to perform complex tasks that require a high level of intelligence, with fast decision making process and actions. 

\subsubsection{Technologies Associated with B5G/6G Services}
There are many technologies associated with B5G/6G services, which will be further discussed in Section \ref{sec:issues}, regarding privacy issues. To describe briefly, we see many novel approaches and applications such as virtual and mixed reality, haptics, nanorobots, holographic applications, etc. They will require much faster data rates as a base requirement. These applications demand a multi-dimensional, fully immersive experience with realistic graphic rendering in very high picture quality. They may also control millions or billions of small IoT sensor devices or robotic swarms in real-time. Also, currently, trendy technologies like blockchain, Terahertz, and visible light communications will be more mature and be used frequently with future networks.

\subsection{Visionary 6G Applications }
The introduction of 6G will boost the current technological society, industry, and human interaction. There are many applications associated with B5G/6G networks. The work in~\cite{chowdhury20206g} describes the following prospects and applications for 6G: super-smart society, Extended Reality (XR), Connected Robotics and Autonomous Systems, Wireless Brain-Computer Interactions, haptic communication, smart healthcare, biomedical communication, automation, manufacturing, Five Senses Information Transfer, as well as Internet of Everything. The authors in~\cite{saad2019vision} also describe four driving applications of 6G: multi-sensory XR applications, connected and autonomous robots, wireless brain-computer interactions, blockchain, and distributed ledger technologies. The next industrial revolution, industry 5.0, will involve co-working of robots with humans \cite{maddikunta2022industry}. Here, 6G will act as an enabling technology for industry 5.0, to provide the required bandwidth requirements, and other features like ultra-high reliability and optimized energy management strategies \cite{maddikunta2022industry}. The work in~\cite{de2021survey} also describes some of these applications mentioned above and others, including industry 5.0, autonomous vehicles, Unmanned Aerial Vehicle (UAV)-based mobility, UAV swarms~\cite{tahir2019swarms}, smart grid 2.0, holographic telepresence, and personalized body area networks. Smart healthcare is another domain which will lead to ubiquitous and continuous availability of personalized medical service, where future networks will be providing the telecommunication infrastructure \cite{aceto2020industry}. 

Therefore, one can identify these applications are available in different application areas such as entertainment, healthcare, transportation, energy, finance, and industrial automation. Consequently, the impact of 6G will be available in almost every segment of daily tasks, and B5G/6G will be an essential and integrated component of people's lives.

\section{Privacy Taxonomy} \label{sec:TM}

The term privacy is generally known to be the assurance that individuals get the control or influence of what details related to them may be collected and stored and by whom and to whom the information may be disclosed~\cite{stallings2014cryptography}. It is the capability that a person gets to seclude the information about themselves selectively. 

Understanding the categories of privacy is essential to identify potential privacy-related aspects in B5G/6G networks since the possible privacy threats may fall into these categories. The following section provides an overview of the taxonomy of privacy and its relation with B5G/6G networks.

Privacy consists of psychological and social background, as it is based on personal interests and their social influence. In~\cite{pedersen1999model}, the authors identify a set of privacy functions for six types of privacy, namely: solitude, isolation, anonymity, reserve, intimacy with friends, and intimacy with family. The privacy functions are autonomy, confiding, rejuvenation, contemplation, and creativity. 

The work presented in~\cite{finn2013seven} defines another set of seven types of privacy as follows:
\begin{itemize}
    \item \textbf{Privacy of the person} - right to keep body functions and body characteristics (e.g., genetic codes and biometrics) private.
    \item \textbf{Privacy of behaviour and action} - the capability to behave in public, semi-public or one's private space preserving privacy.
    \item \textbf{Privacy of communication} - avoid the interception of communications.
    \item \textbf{Privacy of data and image} - ensure that users' data is not automatically available and controlled by other individuals and organizations.
    \item \textbf{Privacy of thoughts and feelings} - the right not to share people's thoughts or feelings or to have those thoughts or feelings revealed.
    \item \textbf{Privacy of location and space} - individuals have the right to move about in public or semi-public space without being identified, tracked, or monitored.
    \item \textbf{Privacy of association} - people's right to associate with whomever they wish, without being observed.
\end{itemize} 

\subsection{Privacy in Actions on Data}
\begin{figure}[htb]
    \centering
    \includegraphics[width=0.6\textwidth]{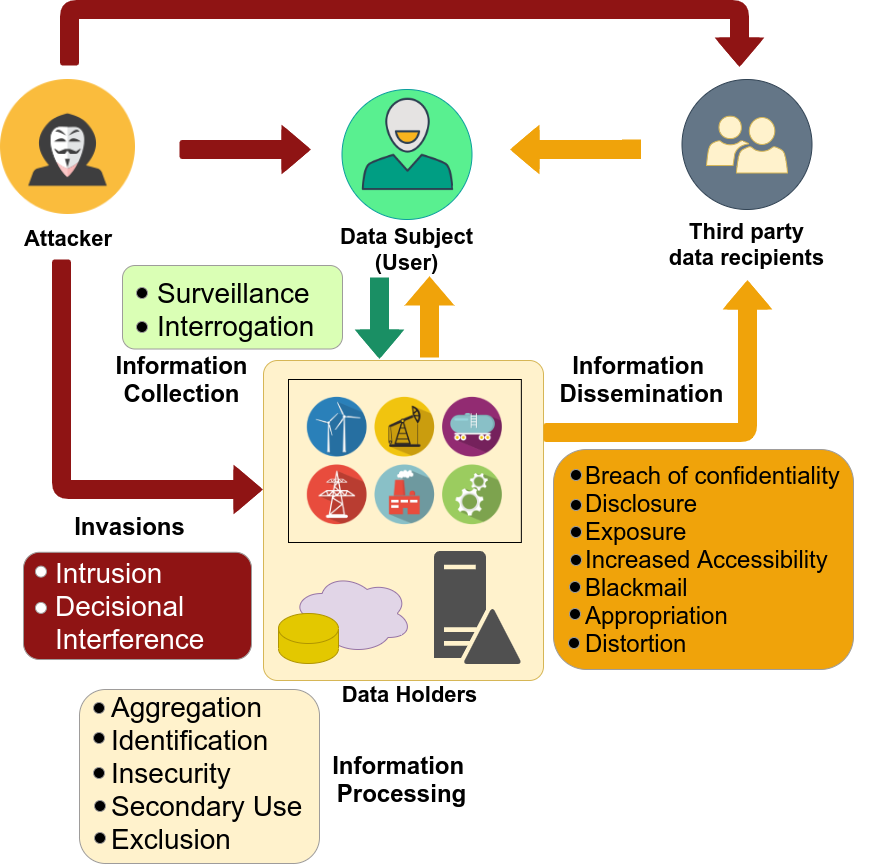} 
    \vspace{1mm}
    \caption{A Taxonomy of Privacy ~\cite{solove2005taxonomy} for Interactions of the Data Subject with External Parties}
    \label{Figure_Privacy_Taxonomy}
\end{figure}

Considering the users in the B5G/6G networks, their data holders, such as network components and third-party services, and potential adversaries, the required privacy actions should be different. The work in~\cite{solove2005taxonomy} divides the taxonomy of privacy in different actions occurring to the data subject, which is generally a B5G/6G user in our context. Figure~\ref{Figure_Privacy_Taxonomy} illustrates these actions and their components. We discuss these actions and their relation with the B5G/6G networks.

\subsubsection{Information Collection}
Third parties can collect user information for different reasons, including personalization and advertising for marketing purposes~\cite{aguirre2015unraveling}. This may happen via surveillance, through watching, listening to, or recording individuals' activities through various methods such as click-through rates~\cite{aguirre2015unraveling}. Another option is interrogation, through multiple means of questioning or probing for information.

This information can be collected from users directly via user devices by third-party services operated in the B5G/6G networks. The collection of information will be done through various sensor devices and personal devices such as mobile phones. A large quantity of data collection will be possible in B5G/6G compared with the current network capabilities due to the higher data rates provided in terabit speeds. Also, they will have intelligence-enabled in the collection points, which will be smart to capture only the essential information that could expose user privacy.

\subsubsection{Information Processing}
The use, update, and manipulation of already collected data can be considered as the information processing~\cite{solove2005taxonomy}. This can be used to get insights on the generators of data since, through processing, it may be possible to trace the origins and their links by connecting data. The work in~\cite{solove2005taxonomy} further divides the information processing into aggregation, identification, insecurity, secondary use, and exclusion.

The B5G/6G networks will be able to process data even from the edge devices with integrated intelligence. Also, from the architecture of B5G/6G defined in Fig.~\ref{Figure_6G_AI_architecture}, it can be expected a tremendous processing of data in the data mining and analytics layer of B5G/6G. For this, AI tools will be used to get insights into complex, big data, which could reveal patterns that expose user privacy. The Art. 5~\cite{gdprArt5} of the European Union (EU) data protection law General Data Protection Regulation (GDPR) defines that the personal data should process with  principles of lawfulness, fairness, and transparency, purpose limitation, data minimization, accuracy, integrity and confidentiality, and storage limitation.

\subsubsection{Information Dissemination}
This refers to the revealing of personal data or the threat of spreading of information~\cite{solove2005taxonomy}. The dissemination may include the breach of confidentiality, disclosure, exposure, an increase of accessibility, blackmail, appropriation, and distortion of information~\cite{solove2005taxonomy}.
These potential privacy threats come from a third party accessing the stored information. Therefore, these issues may be mitigated by adequately keeping information with proper access rights. The next problem arises when using third parties to store information, such as cloud storage providers.  They may be able to access this information due to various reasons. Examples are poor security measures, hacking, stolen media, deliberate or unintended access by employees, accidental publishing, etc. 

The B5G/6G networks may also be vulnerable to uncontrolled information dissemination due to a lack of regulations for future data types and technologies and loopholes in the existing laws. Also, there is currently no unified approach to solving information dissemination issues. 

\subsubsection{Invasion}

This involves threat caused to the user, and it does not necessarily involve information~\cite{solove2005taxonomy}. The authors in~\cite{solove2005taxonomy} categorise the invasions into two types, intrusion and decisional interference. Work in~\cite{denning1987intrusion} shows that intrusion can be detected from abnormal patterns of system usage.~\cite{ferrag2018security} identifies three intrusion detection methods: signature-based, anomaly-based, and hybrid systems for 4G and 5G cellular networks. The latter, decisional interference involves the governmental interference with people’s decisions~\cite{solove2005taxonomy}.

The invasions will be a definite threat to privacy in B5G/6G networks. The systems may not be designed by keeping privacy integrated into their systems, especailly when considering edge devices and networks. It is much possible due to their lack of computation capacity, which may lead to implementing only relatively weaker privacy preservation algorithms. Another possible invasion could be the attacks on AI in network management and applications in B5G/6G networks.

\subsection{Privacy Types for 6G}
When considering taxonomy in terms of privacy types for 6G networks,~\cite{porambage2021roadmap} categorizes privacy aspects into data, personal behaviour, action, image, communication, and location.  

\subsubsection{Data}
Data privacy represents the privacy of the stored data~\cite{liyanage20185g}. We observe users store their personal and sensitive data on mobile devices as mobile phones are easily accessible. Therefore, these devices need more measures to protect data to avoid leaking them to a third party without user consent. Due to the high data rates provided by future networks, an enormous volume of data will get collected at central storage within a short period. Therefore, attackers may find it easy to exploit privacy vulnerabilities in these single data nodes. Thus, privacy is much of a concern in terms of data. 

\subsubsection{Actions and Personal Behaviour}
People's actions and behaviour can play a crucial role in privacy with the influence of social media. Work in~\cite{garcia2019privacy} shows that a person's behaviour is predictable using only the information provided by people around them in a social network. This includes information for a third party to jeopardize user privacy. Even when a user leaves the social network, their ``shadow profile'' will still be available through their friends' data. The application of IoT widens the paths to track user behaviour through extended sensors in the environment.~\cite{massimo2018user} uses tourist data collected from IoT-enabled areas to simulate user behaviours under different contexts and scenarios and provide recommendations. In B5G/6G era, approaches such as Virtual Reality (VR), and smart sensing environments will gather more insights on personal behaviour and may even predict human thought patterns. Therefore, user privacy protection may need to be strictly considered in such cases.

\subsubsection{Image and Video}
With the increase of influence from social media platforms, users tend to add their personal details via images and videos. Though the users are given the freedom to select who will be seeing the post, there is information that may consist in the images that adversely affect privacy that users may not be aware of. Also, video surveillance allows third parties to track users and these parties may potentially use the surveillance data to compromise user privacy. The work in~\cite{rakhmawati2018image} discusses the existing image privacy protection techniques, including editing techniques: blurring, black-box, pixelation, masking, encryption, and scrambling. The authors also include other techniques such as face regions, false colour, and JPEG metadata embedding. It also shows that a balance among privacy, clarity, reversibility, security, and robustness must be maintained for public safety requirements such as identifying suspicious behaviour and knowing non-sensitive information like the number of people in a specific area.~\cite{cciftcci2017reliable} presents two reversible privacy protection schemes
implemented using false colours for JPEG images. The images are applied with encryption techniques to allow only authorized parties with authorization keys to reverse the false colouring. Authors in~\cite{yu2016iprivacy} propose a tool called "iPrivacy" as an automatic privacy recommendation system for image sharing from deep multi-task learning. However, ultra high resolution, multi dimensional images and videos will be very common in B5G/6G. Therefore, it will be possible to expose more unintentional details from these digital media. Thus, more safeguarding of privacy of them is essential in future.

\subsubsection{Communication}
Privacy in different modes of communication is another classic yet important aspect since communication with rich media has increased drastically over the development of platforms in recent years~\cite{gurtov2016secure}. There will be different types of communication providers provide a wide range of services through voice, text, video, and other novel sensory modes for users in B5G/6G as options to choose. Therefore, the users will have more possibility for privacy leakages. Through IoT communication, sensor data might carry vital health signatures of individuals. The work in~\cite{babun2021survey} shows that IoT platforms, with their cloud connections, pose vulnerabilities that attackers may exploit to extract useful information. It also indicates that over-privileged IoT devices and services may also collect user data themselves that may cause breaches in privacy~\cite{ahmad2018towards}.

\subsubsection{Location}
The location privacy is important for an individual since it includes details of physical locations that the person has travelled and which may also reveal the many related information including personal behaviour, financial status, habits, beliefs, interests, and even political preferences~\cite{primault2018long}. They are highly valuable data for a third party and location data include moving objects, spatial coordinates, current time, and other unique features~\cite{yin2017location}. 
Many recent works show the importance of location privacy~\cite{liu2018location,primault2018long,yin2017location}. Location-Based Services (LBS) have ever-increasing popularity, and they consist of components: a positioning system, users, networks, LBS server content provider, and a location privacy server~\cite{liu2018location}. These components are interconnected together to provide the LBS. The authors show that an adversary can attack any of the components in the network, the location privacy server, the LBS server, or from the user side.~\cite{primault2018long} provides many existing Location Privacy Protection Mechanisms (LPPM) considering privacy, utility, and performance aspects. In addition, IoT systems are also vulnerable to location-based privacy issues~\cite{yin2017location}, since they may include components that can sense location. Future B5G/6G are expected to incorporate with gadget free environment, fully automated travel and transport options, where mandatory location privacy preservation will be needed.

\section{6G Privacy Challenges and Issues} \label{sec:issues}

\begin{figure}[ht]
\centering
    \includegraphics[width=0.95\linewidth]{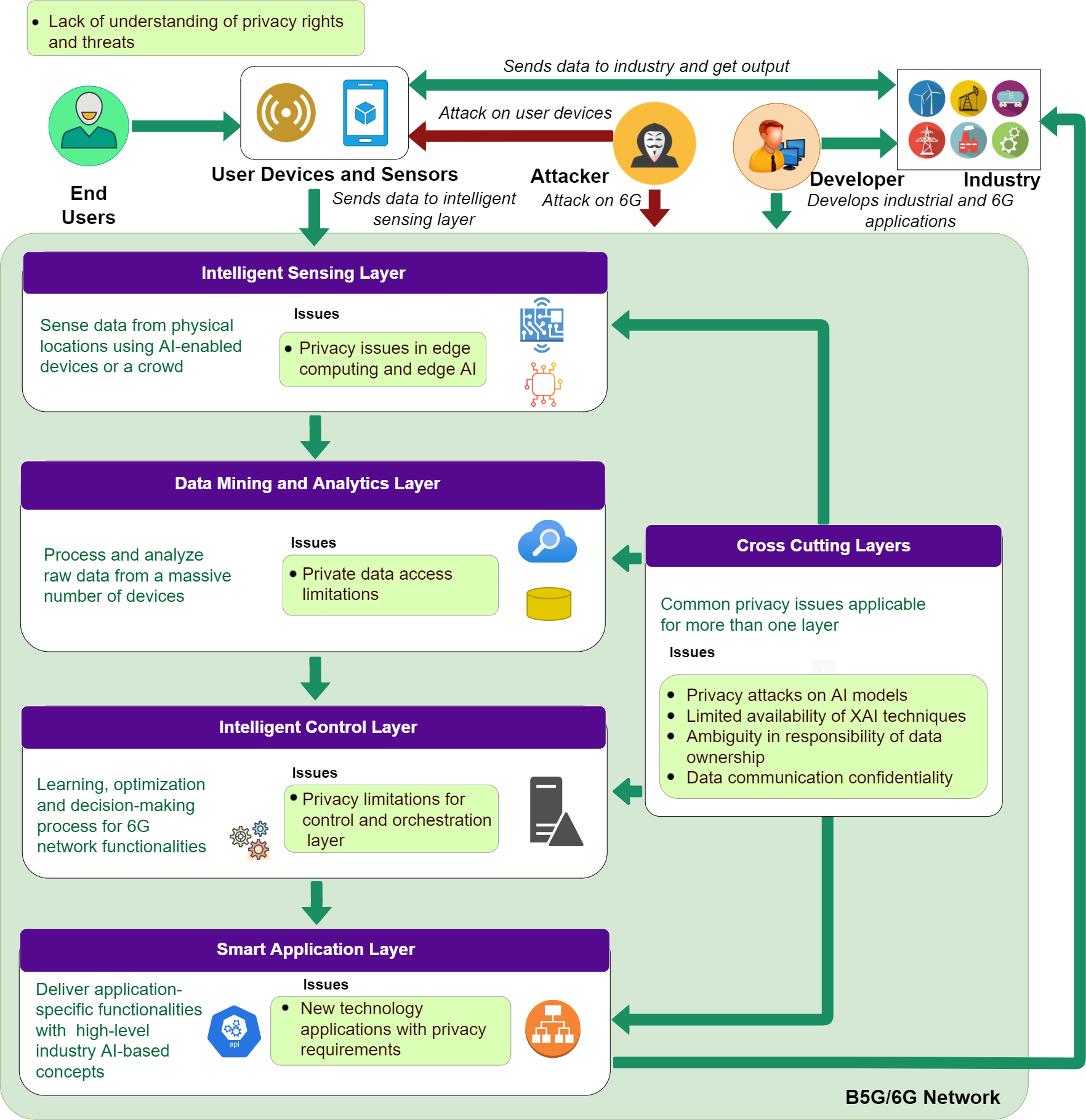} 
    \vspace{1mm}
    \caption{Classification of Privacy Issues based on B5G/6G Vision Architectural Layers.  End users generate data using their devices and sensors. The data is processed in the intelligent sensing layer, where AI-enabled sensing and edge devices may perform local operations and analytics on data. The data will then get forwarded to the data mining and analytics layer, where further processing and analysis of data will be conducted. The intelligent control layer may control and coordinate this collected data toward the smart application layer. Each layer consists of identified privacy issues. Cross-cutting layers indicate common issues applicable to multiple layers in this architecture. }
    \label{issues_archi}
\end{figure}

To formulate a set of privacy challenges, we considered different  potential privacy issues in proposed B5G/6G architecture discussed in Section \ref{sec:6GNetwork}. The Figure~\ref{issues_archi} shows an overview of the distribution of these challenges and issues.  This is a visionary 6G layered architecture \cite{yang2020artificial} that consists of four layers: intelligent sensing layer, data mining and analytics layer, intelligent control layer, and smart application layer. We identified that each of these layers could have potential privacy issues. Further, some issues are spread over multiple layers. Thus we denote them as cross-cutting layer issues. In the figure, we have briefly described the functionality of each layer.  There are several actors involved in the architecture, end-users, developers, and attackers.  The end-users generate data using their user devices. In the B5G/6G era, there will be billions of sensor devices that will track user actions, movements, and the surrounding environment. The developers can utilize this generated data to derive solutions for industry or B5G/6G-related requirements.  Meanwhile, attackers attempt to obtain this data illegally through adversarial methods by attacking either the network or the user devices. However, there is a possibility for them to attack industries in a wide variety of methods as well. 

Considering the data flow in Figure~\ref{issues_archi}, first, the end-users interact with their user devices, such as smartphones or sensors, to generate private or personal data. Industries may store this data for their requirements and it may affect user privacy. Then, this data is forwarded to the intelligent sensing layer in the B5G/6G enabling technology architecture. Next, the data is processed and/or forwarded to the data mining and application layer. There, they are stored and analyzed to harness more data related to user privacy. After that, the data is handled by the intelligent control layer. The smart application layer then obtains the control layer's data to interact with the external industry or services. Industries or services may forward this data to users to interact with. Several issues related to all layers are summarized in cross-cutting layers.  It is identified these issues have a significant impact on all the layers mentioned since those issues are clearly possible to arise in each of them. We discuss each of these privacy issues in all layers in more detail in the following subsections. Each privacy issue is arranged as follows: We first give an introduction to the issue on each privacy issue identified. Then, we provide a discussion on related works available on the topic. The summary and our opinion on the issue are finally discussed. 

\subsection{New technology Applications with New Privacy Requirements} \label{tech_app_iss}

B5G/6G network capabilities provide an arena for many associated new technologies to the consumer market.  One such example is the emergence of the digital twin, which is used in many industrial applications, including smart manufacturing. It will improve the environmental and social impact of manufacturing via the virtualization of physical entities \cite{li2022digital}.  With the improvements in digital twin and extended reality, the concept of the metaverse is also becoming another popular topic. The metaverse can be considered a fully-immersive virtual world. However, metaverse incorporates many privacy requirements, such as ensuring the behaviours of people are not tracked by third parties, the private and confidential data kept in the virtual environment is not exposed or stolen, and the biometric and other critical profile data used to access metaverse is not compromised. Another example would be autonomous driving, and the advancement of vehicle-to-vehicle communication with B5G/6G. The sensitive data such as location could be able to track by third parties and the identity of the owners, their travel preferences and habits will be revealed easily. Therefore, we identify that it is essential to discuss about potential privacy threats this eco-system of technologies bring with B5G/6G.

The work in~\cite{nguyen2021security} mainly discusses B5G/6G associated technologies of extended reality/digital twin, tactile interaction, space-air-sea communications, smart medical nano-robots, autonomous driving, and holographic telepresence. It shows that the XR/digital twin suffers from the possibility of leakage of biometric data and physical movements.
In addition,~\cite{9419108} indicates that in XR, a variety of data types can be collected. This could include real scene information, biometric data such as gait, eye or head movements, body appearance, domicile information, heart rate, inferred emotional states, and potentially many more.
Also,~\cite{9419108} shows that domicile data, for instance, may include a record of household objects to build a psychological profile about individuals. Tactile interaction is also vulnerable to exposing biometric data as discussed in~\cite{nguyen2021security}. The authors also show communication in the space-air-sea makes location tracking and identity exposure possible.  Another area discussed in~\cite{nguyen2021security} is the nano-robots for medical applications, where these nano-robots may expose individuals' health information.  For autonomous driving, the paper~\cite{collingwood2017privacy} discusses that it is possible to track individuals' location and their future location interests can be predicted using data from these vehicles. The work in~\cite{glancy2012privacy} proposes two models of autonomous vehicles, self-contained and interconnected vehicles, with the latter being more vulnerable to privacy leakages, including the collection of vehicle sensor information through vehicle to vehicle communication.
Also,~\cite{bloom2017self} provides details on four different aspects of privacy-invasive technologies:
\begin{itemize}
    \item The ubiquitous capture of data in the public
    \item Physical surveillance by a privately owned company
    \item The ability to scale without additional infrastructure
    \item The difficulty to notice and choose data practices for physical sensors that capture data about non-users
\end{itemize}

The study in~\cite{karnouskos2017privacy} shows autonomous vehicles track the environment through their sensors and collects detailed data that users otherwise might not be shared. Holographic telepresence can expose personal behaviour, habits, and biometric data as shown by~\cite{nguyen2021security}. 

Considering novel modes of data communication, the work in~\cite{wang2020security} provides technologies of molecular communication and mentions malicious behaviour, encryption issues, and authentication privacy issues are possible with the technology. The work also discuss quantum communication privacy issues like encryption. 

Blockchain will be another main technology associated with the B5G/6G networks and its applications. The blockchain-related privacy concerns includes the issues in authentication, and access control~\cite{wang2020security}. Further issues on the blockchain are discussed in~\cite{feng2019survey}, where the authors claim de-anonymization and transaction pattern exposure as privacy threats. The work in~\cite{halpin2017introduction} shows that the distributed ledgers may encounter privacy issues as users all share the same data. The authors of~\cite{nguyen2021security,wang2020security} give a set of technologies associated with 6G networks. The Table~\ref{tbl:techIssues} summarizes the potential privacy issues of these provided new technologies related to B5G networks.

\textit{From the related work, we summarized a set of privacy issues for associated technology changes that will be built on top of B5G/6G. We observe some technologies, such as autonomous driving and blockchain, already consist of a significant study of potential privacy issues. 
This could be due to the availability of these concepts for a long time and evolution from pre-5G era. However, others may require further investigation on privacy since they are new concepts in early stages. The applications of these associated technologies are tremendously valuable for the public, but their impact on privacy should be regulated.
}

\begin{center}
\scriptsize
\begin{longtable}[htb]{|p{3cm}|p{6.5cm}|p{2cm}|}
\caption{B5G Technologies and Their Privacy Issues} \label{tbl:techIssues}
\renewcommand{\arraystretch}{1}
\\
\hline

\multicolumn{1}{|c|}{\cellcolor[HTML]{CBCEFB}{\color[HTML]{000000}
\textbf{Technology}}} & \multicolumn{1}{|c|}{\cellcolor[HTML]{CBCEFB}{\color[HTML]{000000} \textbf{Privacy Issues}}}
& \multicolumn{1}{|c|}{\cellcolor[HTML]{CBCEFB}{\color[HTML]{000000} \textbf{References}}}
\\ \hline 
\hline
\begin{tabular}[l]{@{}l@{}}Extended reality\\ digital twin\end{tabular} 

& \begin{tabular}[c]{@{}l@{}} Exposure of biometric data, \\physical movements, \\ domicile information, emotional states\end{tabular} &~\cite{nguyen2021security}~\cite{9419108}                                                                                  \\ \hline
Tactile interaction                                                                      & Expose biometric data  &~\cite{nguyen2021security}                                                                                          \\ \hline
Space-air-sea communications                                                             & \begin{tabular}[c]{@{}l@{}}Signalling-based location tracking, \\ expose identity \end{tabular}     &~\cite{nguyen2021security}            \\ \hline
Smart medical Nano-robots                                                               & Expose body health information                                   &~\cite{nguyen2021security}                                              \\ \hline

Autonomous driving                                       &
\begin{tabular}[c]{@{}l@{}}location tracking, compromized credentials, \\internal vehicle sensor information,
\\ubiqutous data capture of public data and\\ non users
\end{tabular}              & 
\begin{tabular}[c]{@{}l@{}}
~\cite{nguyen2021security}~\cite{collingwood2017privacy}~\cite{karnouskos2017privacy}\\~\cite{bloom2017self}~\cite{glancy2012privacy}
\end{tabular}
\\ \hline
Holographic telepresence                                                                 & \begin{tabular}[c]{@{}l@{}}Expose personal behaviour, habits, \\ biometric data\end{tabular}    &~\cite{nguyen2021security}               \\ \hline

Molecular communication                                                                  & \begin{tabular}[c]{@{}l@{}}Malicious behaviour, encryption, \\ authentication\end{tabular}
 &~\cite{wang2020security}\\ \hline
Quantum communication                                                                    & Encryption, communication    
 &~\cite{wang2020security}                                                                        \\ \hline
Blockchain                                                                               & \begin{tabular}[c]{@{}l@{}}Authentication, access control, \\de-anonymization,\\ Transaction pattern exposure      \end{tabular}         &~\cite{wang2020security}~\cite{halpin2017introduction}~\cite{feng2019survey}\                                     \\ \hline
THz                                                                                      & \begin{tabular}[c]{@{}l@{}} Authentication, malicious behaviour,  \\user location, device orientation, mobility and\\ traffic patterns, surrounding environment \\and blockages 

\end{tabular}                                       &~\cite{wang2020security}~\cite{singh2020thz} \\ \hline
Visible light communication                                                              & Communication, malicious behaviour              &~\cite{wang2020security}                                                               \\ \hline
\end{longtable}
\end{center}

\subsection{Privacy Preservation Limitations for B5G/6G Control and Orchestration Layer} \label{control_iss}
After considering the associated technologies and their ecosystems, we focused on the B5G/6G architecture itself for potential privacy issues. The B5G/6G control architecture is composed of many novel features, including AI-based zero-touch network orchestration, optimization and management~\cite{zhang20196g}. These features are essential in fulfilling the ultra high-speed requirements of future networks and serve billions of highly interconnected devices.  Intent-based networking (IBN) is also an emerging area where automated network management 
solutions are driven by a set of specifications called intents~\cite{zeydan2020recent}. The intents affect the operation of network management. However, due to this, it may be a potential interest of attackers. Closed loop automation is another step in automating network management, where the process monitors and assesses and automatically acts in network occurrences such as congestion and faults~\cite{closedLoop}. If this monitoring process is compromised, adversaries may be able to cause various privacy related attacks, and go unnoticed. As the control architecture of networks evolves, it is important to consider the privacy strengths and potential weaknesses of future architectures. 

We identify B5G networks to have privacy issues related to existing 5G networks and privacy concerns associated with the AI. To achieve fully automated network management, the work in~\cite{benzaid2020ai} propose an AI-driven ZSM architecture for B5G/6G. The authors also show that ZSM could cause privacy issues through AI models used in the ZSM by adding adversarial examples for training and test data. Also, the authors show that safe shared learning through data sharing between multiple mobile operators is essential for speeding up the accuracy of ML models. Nevertheless, it is limited due to potential privacy leakages in sharing. Similarly, IBN incorporates AI driven approaches for understanding of intents\cite{zeydan2020recent}. Therefore, the privacy issues related with AI may affect the IBN decision making process. The intent-based interfaces in ZSM can carry information about the desires of the application, including the peer connections, network traffic regulation and advertising services, which may expose important information~\cite{benzaid2020zsm}. The closed-loop automation is also at risk of getting influenced by attackers by observing, creating, or hiding information in the network channel~\cite{benzaid2020zsm}. Therefore, if this network channel handles sensitive or private data, owners are at risk of the data being observed or stolen.

\textit{The work on B5G/6G architecture is currently at their early stages and consist of concepts such as ZSM, IBN and closed-loop automation. These works promise intelligence on top of the existing architecture of 5G to manage network operations efficiently. We see a great interest in AI in future network control. However, having AI integration opens these networks to issues related to AI privacy with these future networks. Therefore, it is critical to investigate privacy requirements further when using these in the automation of networks.
}

\subsection{Privacy Attacks on AI Models} \label{attacks_ai_iss}
As discussed in the previous two privacy issues, AI models will be used extensively for B5G/6G. For another example, zero-touch 6G networks will heavily use AI as a key component for its automated decision making process \cite{benzaid2020ai}. Therefore, it is crucial to identify privacy attacks that could be possible on AI models. The systems that use AI techniques such as machine learning could be subjected to attacks named adversarial attacks, where an attacker gets details on machine learning models and the data used to train the models~\cite{thuraisingham2020can}. The solution proposed to this issue is called adversarial machine learning~\cite{thuraisingham2020can}. Another issue arises from AI models themselves. The machine learning models can learn from big data and predict patterns and trends~\cite{thuraisingham2020can}. If these trends could reveal sensitive information on individuals, then it is a privacy issue, which an AI model makes. 

\begin{figure}[ht]
    \centering
    \includegraphics[width=0.6\linewidth]{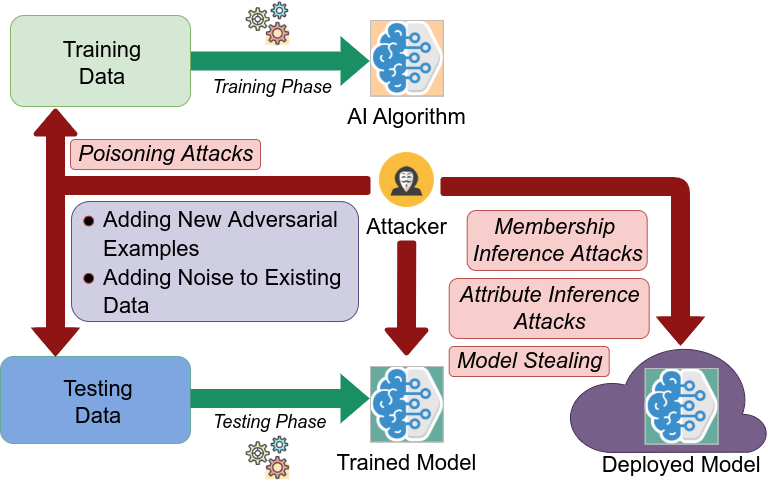} 
    \vspace{1mm}
    \caption{Categories of Machine Learning Attacks in Training and Testing Phase}
    \label{ml_attacks}
\end{figure}

We can find many works associated with AI model vulnerabilities for different categories of AI. These categories include machine learning, reinforcement learning, deep learning, vision, etc. In~\cite{sun2020machine}, the authors provide a classification of different types of attacks on ML models, which is presented in Figure~\ref{ml_attacks}, where ML is a subset of AI. We can observe that ML attacks can be launched in both training and testing phases from the figure. The training phase can poison the input data to make the model less accurate or vulnerable to privacy attacks. An adversarial test poisoning attack is a poisoning attack that happens in the testing phase.  The authors in~\cite{saha2020hidden} discuss that attackers may poison data to create backdoors by embedding patterns on the AI model. These backdoors may trigger when a certain types of data is entered, causing model performance drop. This may also cause privacy leak as the attacker will identify the type of data that triggered the backdoor.  Also, during the testing phase, an adversary can make reverse attacks where they reverse-analyze the model. A membership inference attack is used to infer whether a particular member is included in the training data~\cite{sun2020machine}, which violates the privacy of that member. The work of~\cite{liu2020privacy} shows that DL is vulnerable to poisoning attacks in the training phase and in the testing phase, it is possible to encounter model extraction, model inversion, and adversarial attacks. Among them, the model extraction and model inversion attacks invade the models' privacy.  Here, the model extraction attacks focuses on model information and the model inversion attack tries to extract training data. Another type of attack is the attribute inference attack. In this attack, attackers target sanitized data released by the data owner to infer the attributes of the data~\cite{zhao2019adversarial}, rather than only inferring the membership.  Therefore, an attacker of an attribute inference attack may attempt to infer a user's private attributes such as location, gender, sexual orientation, and/or political view via leveraging its public data~\cite{jia2018attriguard}. Another one is a model stealing attack, where an adversary attempt to create a counterfeit of the functionality of a victim ML model by exploiting its black box queries such as input and outputs, which could therefore be regarded as prediction poisoning~\cite{orekondy2019prediction}. Having AI being attacked with prediction poisoning may mislead the end users who may use it for privacy-related decision-making, leading them to decisions that compromise their privacy. 

Considering the other AI categories,~\cite{sakuma2008privacy} compares several distributed RL models and shows that privacy is not guaranteed in these models since data is shared. In~\cite{pan2019you}, the authors show that Deep Reinforcement Learning (DRL) can be used to make privacy-sensitive information leakages, proposing algorithms to infer floor plans from trained Grid World navigation DRL agents with LiDAR perception.~\cite{silva2020using} discusses NLP can be used to extract privacy violations and Personally Identifiable Information (PII), which makes it a tool for an adversary to exploit privacy-related information in documents. In computer vision, the work in~\cite{das2017assisting} show that privacy concerns are surfacing since cameras can process and transfer PII at unknown extents for those who get affected. 

Since AI can be used for adversarial tasks, privacy violations could be made using AI models as a tool. The work in~\cite{sun2020machine} shows that AI is a ``double-edged sword'',  which can be beneficial in protecting user privacy if appropriately used. However, privacy violations can be made via AI. They summarise a list of such violations. These violations include eavesdropping network traffic\cite{conti2015analyzing}, load monitoring~\cite{zhu2018big}, de-anonymize authors in academic comments~\cite{payer2014you} etc. using supervised and unsupervised learning models. The next breakthrough of AI is expected to give it emotional capabilities \cite{zhang2021study}, which makes these models possible to understand how users in B5G/6G would react emotionally and might make them to reveal their private data for attackers.

\textit{We see there are various privacy issues in the existing AI categories.  They require a significant further investigation of privacy. There are numerous types of attacks possible on AI models. Some of them may expose the data of subjects used to train the models and some may steal the trained model.  Also, we see AI can be used as a tool for adversaries to launch attacks against privacy, such as revealing the identities of actual users from already anonymized data. Therefore, these points must be considered: 1) when adopting these models for B5G/6G-related tasks, the issues with AI models are needed to be addressed before design and implementation; 2) the possible methods to detect AI threats to user privacy need further evaluation.}

\subsection{Privacy Issues in Edge Computing and Edge AI} \label{edge_iss}
Edge computing is a paradigm aimed at solving IoT and localized computing requirements by bringing computing resources close to the 'edge' near the end-users. 
This is achieved through adding computing nodes close to user devices, thereby reducing the overhead to cloud computing~\cite{yu2017survey}. The 
5G brings edge computing in to customers with more control to their data, but lack of consumer trust is an issue that prevent its benefits~\cite{porambage2019sec}. It is indubitably favourable for computing in the edge, rather than communicating directly with the cloud for resource-constrained devices, since it can save their energy usage, improve the QoS and reduce the network traffic. However, the privacy concerns here are very important to consider prior to apply it in B5G/6G. There is a frequent availability of these edge devices in many physical locations where attackers might get easy access. Also, a malicious edge computing device or a compromised device can eavesdrop or steal user data processed through it, such as sensitive health data. With the addition of many billions of more edge devices in B5G/6G, edge computing and edge AI will also increasingly face similar issues due to possible less privacy protection measures in resource constrained environments. The AI algorithms are vulnerable to all the privacy attacks discussed, such as poisoning, membership inference, and attribute inference attacks.

\begin{figure}[ht]
    \centering
    \includegraphics[width=0.7\linewidth]{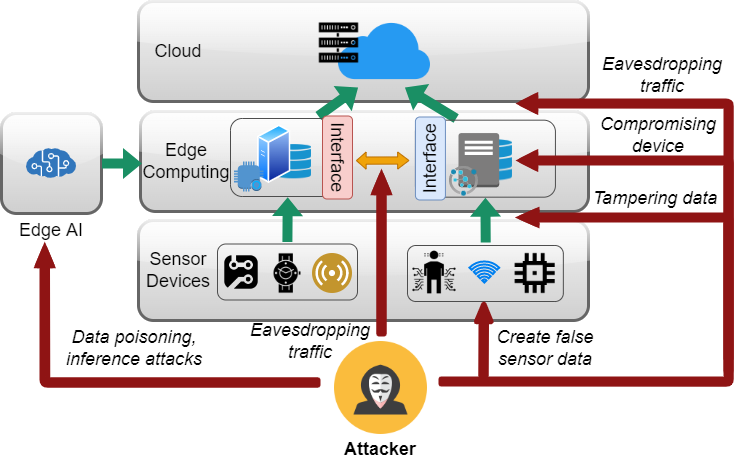}
    \caption{Edge Computing Overview and Attack Scenarios on Sensor Devices and the Edge Layer}
    \label{Edge}
\end{figure}

There is a clear possibility of exploiting privacy-sensitive information at the edge~\cite{yu2017survey}. The work in~\cite{zhang2018data} summarizes a set of privacy challenges for edge computing, based on the components in edge computing: core infrastructure, edge servers, edge network, and mobile edge devices. The issues like privacy leakage or data tampering can occur in the core infrastructure. The authors mention that edge servers are vulnerable to privacy leakages and privilege escalation that may cause unauthorized parties to access data. Edge networks and edge devices are also vulnerable to being compromised. The authors propose lightweight data encryption methods and fine-grained data sharing systems, distributed access control, resource-constrained, and efficient privacy-preserving mechanisms to mitigate these issues. However, implementing these mechanisms raises the problem of embedding privacy defensive features for a vast range of devices~\cite{ranaweera2021survey} from various manufacturers, running different types of algorithms, with varying capabilities in processing and energy consumption. This makes unified privacy-preserving approaches difficult due to heterogeneity of edge devices.  The work in~\cite{ranaweera2022realizing} discusses privacy issues of Multi-access Edge Computing (MEC) based IoT applications. They include contact tracing which exposes location of participants, and possibility of leaking health data extracted from medical IoT devices.  Considering the edge AI, due to privacy concerns, it may be impossible for data collecting from different edge systems and share them with each other~\cite{liu2021bringing} for applications such as model training tasks.

\textit{The edge computing paradigm supports B5G/6G improving the service performances by bringing computing resources near users. However, there are privacy concerns for these edge devices that may process sensitive data and be easy targets for adversaries. Many of the related works show that there could be vulnerabilities for edge computing and edge AI in different components. We highlight the issue of privacy mainly lies in the lack of proper privacy-preserving mechanisms. The lack of AI-based approaches for identifying privacy issues for edge computing is another issue that needs improvement.} 

\subsection{Limited Availability and Vulnerabilities of Explainable AI (XAI) Techniques} \label{xai_iss}
We recently observed a substantial interest in AI research, especially in the fields of deep learning, reinforcement learning, computer vision, etc. AI is commonplace in daily activities, and recently,  a wide range of tools are available in web-based ML as a Service (MLaaS) platforms~\cite{ribeiro2015mlaas}. Despite the significant development, there is a lack of understanding about these AI models' decision-making process. These are, therefore, generally regarded as ``black boxes''~\cite{samek2019towards}.
As discussed in Section \ref{sec:6GNetwork}, 6G will add intelligence as a new core element integrated to it. This will bring back the question of justifying the decision taken by the AI models, as highly-sensitive private information will be transmitted through the network and may get processed by these AI models. In case of privacy violation by AI models, to investigate the roots of the issues, XAI techniques can be used. The concept of XAI, therefore, emerges again to explain and justify the underlying processes of these complex, mostly non-linear models when making decisions, though the field itself dates back to forty years ago~\cite{xu2019explainable}. We see many recent works related to XAI that attempt to describe the nature of the models in many fields, including healthcare and industrial practices~\cite{xu2019explainable,pawar2020incorporating,li2020survey}. However, the area of XAI for deep neural networks has emerged recently, and it is currently in its initial stage of development. Hence, there is a great challenge for B5G/6G in providing a reasonable explanation for the black-box approach in models such as deep neural networks.  Furthermore, it may be required to identify possible threats and vulnarabilities of XAI itself, since there could be possible approaches to attack on XAI decision or use XAI as a tool for launching privacy attacks in B5G/6G against AI. 

Considering the popularity of XAI, a survey in~\cite{adadi2018peeking} shows there has been a significant increase in interest and a trend for XAI in recent years. However, the work also claims that many works lack formalism in terms of problem formulation and ambiguity in definitions. The human's role in XAI is not sufficiently addressed. The authors conclude by mentioning that considerable effort is required to tackle future challenges and issues in XAI. Also, the authors of~\cite{adadi2018peeking} mention that there exists a tradeoff between accuracy and interpretability. The most accurate models like deep neural networks and boosted trees are usually not explainable. However, simple models such as linear or logistic regression are much interpretable, yet their accuracy is relatively low. Another work in~\cite{das2020opportunities} also mentions that the current field of XAI is ``still evolving'', and one should be careful when developing and selecting XAI methods. Their survey on evaluation shows that currently used evaluation techniques are immature and focus only on human-in-the-loop evaluations. In XAI visualizations,~\cite{das2020opportunities} discuss two flaws: 1) the inability of human attention to deduce XAI explanation maps for decision making, and 2) unavailability of a quantitative measure of completeness and correctness of the explanation map.~\cite{gunning2019xai} shows several active issues and challenges for XAI: ambiguity in considering whether the XAI method should be starting from computers or starting from people, accuracy versus interpretability, use of abstractions for simplifications and, explanation of competencies versus explanation of the decisions.  When considering the attacks on XAI methods, the authors in \cite{kuppa2020black} designed a black box attack on gradient based XAI methods to mislead explainability method without affecting the AI model classifier output. Privacy attacks can be launched with the support of XAI as a tool. The work in \cite{kuppa2021adversarial} uses an XAI technique called counterfactual explanation methods to launch membership inference attacks that link users and identify their passwords from leaked datasets. The authors also worked on model extraction and poisoning attacks on real-world datasets. Similarly, the work in \cite{zhao2021exploiting} investigates privacy attacks on image data that reconstructs private image data from model explanations. Therefore, it is important to identify the exploitation of XAI for attacks and take possible privacy protection measures to mitigate them. A right balance between explainability and privacy is a crucial requirement. 

\textit{The field of XAI is still evolving with a rapid pace, and there are some works mentioning flaws in existing XAI approaches.   These issues include lack of formalism, ambiguity, the tradeoff between accuracy and interpretability, and potential privacy leakages through XAI.   For B5G/6G networks, XAI-based approaches will be highly required for proving and justifying the users' privacy is preserved by the AI models used throughout the network. Therefore, a significant improvement on the techniques should be made to make them applicable in these future networks.}

Table \ref{tab:privacyAttacks} provides a summary of different types of privacy attacks that can affect AI, edge AI and XAI. Their impact on 6G privacy and the relevance on each AI type is presented. 

\begin{centering}
\scriptsize
\begin{longtable}[!htbp]{|m{20mm}|m{40mm}|m{40mm}|m{6mm}|m{6mm}|m{6mm}|}
\caption{Privacy Attacks on AI, Edge AI, and XAI, their Impact and Relevance }
    \label{tab:privacyAttacks}
    \\
\hline
\rowcolor[HTML]{CBCEFB} 
\cellcolor[HTML]{CBCEFB}                         & \cellcolor[HTML]{CBCEFB}                                                                                                                                                                                & \cellcolor[HTML]{CBCEFB}                                                                                                                         & \multicolumn{3}{c|}{\cellcolor[HTML]{CBCEFB}Relevance}                                                        \\ \cline{4-6} 
\rowcolor[HTML]{CBCEFB} 
\multicolumn{1}{|c|}{\multirow{-2}{*}{\cellcolor[HTML]{CBCEFB}Attack}} & \multicolumn{1}{|c|}{\multirow{-2}{*}{\cellcolor[HTML]{CBCEFB}Description}}                                                                                                                                            & \multicolumn{1}{|c|}{\multirow{-2}{*}{\cellcolor[HTML]{CBCEFB}Impact on 6G privacy}}                                                                                   & \multicolumn{1}{c|}{\cellcolor[HTML]{CBCEFB}AI} & \multicolumn{1}{c|}{\cellcolor[HTML]{CBCEFB}Edge AI} & \multicolumn{1}{c|}{\cellcolor[HTML]{CBCEFB}XAI}  \\ \hline
Poisoning Attack                                 & Include adversarial inputs from an attacker in either training or test phase that affects the model integrity~\cite{liu2020privacy, orekondy2019prediction, kuppa2021adversarial}. & Attackers may poison an AI model in B5G/6G to embed backdoors~\cite{saha2020hidden} that may leak privacy-sensitive data. & High                      & Very High                       & High \\ \hline

Membership Inference Attack
& Infer whether a particular data is included in the model's training data~\cite{sun2020machine}. & The membership state can be critical as it can target an individual, whether they are participating or using the services in the B5G/6G applications. & High                      & High                        & High \\ \hline

Attribute Inference Attack
& An attempt to recover specific properties in the training set such as gender and location~\cite{jia2018attriguard}. & Revealing specific properties of datasets can directly affect the data generators as privacy-sensitive properties can reveal the identity or other private information. & High                      & High                       & High \\ \hline

Model Stealing Attack
& Adversary attempts to create a counterfeit of the functionality of an ML model through black box queries~\cite{orekondy2019prediction}. & The attackers may reconstruct data or identify features in the training set  that is used to train the original model, which may be used for privacy-critical services in B5G/6G. & High                      & Very High  & High \\ \hline
\end{longtable}
\end{centering}
\color{black}

\subsection{Ambiguity in Responsibility of Data Ownership} \label{ownership_iss}

With the expansion of services associated with B5G/6G, many entities will collect, process, and store personal data.  Data ownership is getting complicated with this increased number of data controllers. Therefore, problems can arise when claiming the data ownership. During a privacy leakage, the parties who process and store the data may be unwilling to take responsibility for it.  There is also a risk that these parties may misuse data, claiming they own it. Eventually, the consumer could be the victim of any such scenario if the responsibility is not handled correctly. In B5G/6G, massive amount of big data will be collected with or without the consent of data producers. The development multi-dimensional abundant sensor environment and rapid communication will facilitate this. Then, they will be processed, analysed with edge AI and third parties who provide new technologies and services emerged with B5G/6G. Further these data will be stored and sold to other parties. The data will be transmitted through the 6G zero-touch network, where AI models may process the data. But a question arises on who will be responsible in case of a privacy breach. Furthermore, it will be difficult to trace where the breach occurs, as there are many intermediate parties involved in the communication. 

One of the issues in creating ambiguity is duplication. The work in~\cite{tjong2018data} raises the problem of the capability of data duplication without any change in the integrity of the original file, which could only be prohibited through legal means. The authors mention this may make ``keeper'' of the data the copy's owner unless contractually prohibited. Another ambiguity occurs when data owners have lack understanding of the laws. For example, in Australia, non-personal information such as business or commercial information is governed by the law of contract~\cite{wiseman2018rethinking}. Therefore, one might misunderstand their personal business information is private to them. Having multiple stakeholders is another problem, as shown in~\cite{galvin2020developments}, which mentions that multiple stakeholders complicate the idea of ``ownership'', even without the data subject's consent.

Even though the ownership is unambiguous, theft can be another issue for actual owners in the B5G/6G. Data, AI models, or other intellectual property can be stolen without the owner's consent. The adversaries may claim they own the model since it may be difficult to detect the owner once the data or models are slightly modified. Therefore, this makes the situation more complicated, with many owners claiming the ownership of data or intellectual property. There are existing methods for AI models to add unique watermarks in a model's decision interface. However, the work in~\cite{maini2021dataset} show this approach has downsides due to the reduction of model accuracy and therefore, it may require retraining. 
 
\textit{With the increased involvement of multiple parties to collect, process, and store personal data in B5G/6G, there will be an issue in who owns data, causing ambiguity among users and service providers. We see many issues in related work: data duplication, lack of understanding of legal terms related to ownership, stealing and modification of data that leads to claims of multiple ownership. In B5G/6G networks, the issue of data ownership can be expected to rise due to having more intermediate services with multiple connectivities and new ways to collect personal and sensitive data further. Ambiguity in data ownership can increase due to several factors such as duplication, ignorance, having multiple owners, and theft. Therefore, we think this should be addressed as an issue for privacy.
}

\subsection{Data Communication Confidentiality Issues} \label{confidential_iss}

With trillions of interconnected devices, the B5G/6G will have to fulfill scalability requirement of the network. Furthermore, with limited power of most of these devices, it is practically difficult to maintain advance privacy protection. Furthermore, there are different modes of communications such as Visible Light Communication (VLC), body-area communication, space internet, deep sea communication, that many  incompatibilities could occur when taking privacy measures in each.
During communication, there is a possibility that a third party reveals data communicated between the sender and the receiver.  To mitigate this issue, confidential communication can be achieved using conventional End to End (E2E) data privacy solutions such as encryption. With encryption, only the intended receiver can understand the content.  Since privacy is a fundamental problem, these techniques have evolved from early generations of wireless communications. However, with the involvement of several intermediate devices in B5G/6G networks, users will have to inevitably transmit their data from one device to another. Therefore, data transmission through many devices needs to have mechanisms to prevent unauthorized parties from accessing data.  These devices may have different capabilities and there may be intermediate data processing requirements as well. At the same time, upcoming networks may require to sense the type of service the users are using at the moment. It will be an important requirement to provide an intelligent, and differentiated QoS~\cite{5gHuawei}.  Furthermore, with the introduction of quantum computing, it may be possible to easily break current encryption schemes. Therefore, it is important to consider quantum resistant cryptographic mechanisms, however, their implementation in resource constrained devices is challenging. 

When considering IoT communication, the work in~\cite{coroller2018position} mentions that the IoT paradigm lacks E2E privacy solutions. This happens since many IoT devices are resource constraint, and their cost of implementation can be high, as discussed in the privacy issue of cost on privacy enhancements. The authors in~\cite{coroller2018position} also propose a control architecture for IoT using distributed event-based publish-subscribe pattern. This mechanism uses a set of brokers to relay messages from the publisher to the subscriber. Their work reflects that all the brokers in the system cannot be fully trusted, such that there could be privacy leaks. In the case of wireless sensor networks, the existing work for the establishment of E2E encryption may require high computation requirements in decryption~\cite{cui2018data}. The conventional E2E techniques may cause increased storage spaces, thus costs in cloud services, due to factors such as the inability of deduplication and compression from the encrypting side~\cite{10.1145/2664168.2664176}. Short-range techniques such as D2D communication will be extensively used in B5G/6G networks to facilitate fast data exchange between physically proximate devices~\cite{haus2017security}. However, they may get subjected to potential privacy leakages such as location spoofing, eavesdropping, and man-in-the-middle attacks~\cite{haus2017security}.

\textit{Privacy in data communication is a fundamental requirement. However, due to cost and performance capabilities limitations, establishing secured E2E connections is challenging in platforms such as IoT. Even though attempts have been made to secure it, there could be possible privacy leakages and attacks on the communication. Therefore, guaranteeing E2E privacy in communication is crucial, especially with relatively new fields and techniques such as D2D communication. There exists a significant challenge in introducing lightweight solutions in this sector due to resource limitations.}

\subsection{Private Data Access Limitations} \label{data_acc_iss}

Though a huge amount of data is generated each day, access to this data is limited due to privacy issues. With the B5G/6G, the data generation will be much higher as they will facilitate scalable connections for a large number of devices than the present. But, eventually, this data may get discarded without any proper use. Especially, health data generated by sensors are wasted, which may otherwise help save future lives through improved AI models for disease prediction. 

Most of the current machine learning models get trained on publicly available datasets, which may not represent the current situation and do not have up-to-date information. If these models could get access to private data without breaking the privacy of data owners, we will be able to achieve a more significant leap of accuracy with improved AI models. Therefore, it will be an open issue to address and make data available without breaking privacy, especially in the era of B5G/6G that need better AI models.

Usually, when publishing data for public use, anonymization techniques are used to hide PIIs of individuals~\cite{goswami2017privacy}. However, the problem lies in whether it is possible to infer individuals' identity despite these anonymization methods. many works show that this is possible. For example, the work in~\cite{sweeney2013matching} matches anonymized patient-level health data with newspaper stories and infer patients' identity. Sweeney et al.~\cite{sweeney2013identifying} identify critical information such as genomic information, details on medications, diseases of individual profiles in the Personal Genome Project by linking profile data with public records such as voting lists. In~\cite{acquisti2009predicting}, the authors predict social security numbers of individuals through correlation of data from multiple sources, including data brokers and social networks. Therefore, we see correlations from other datasets applied to anonymized data can lead to identify personal information on data subjects.

On the other hand, B5G/6G requires a large amount of timely and quality data sets for its AI models, as high accuracy models can be made through supervised learning approaches. In ZSM architecture, the authors of~\cite{benzaid2020ai} show this requirement, mentioning the lack of 5G specific high-quality, timely, and high volume datasets due to privacy issues. The work also show that the currently available generated data is synthetic and lacking completeness.

However, to fulfill this data requirement, there are potential hurdles to overcome.
In particular,~\cite{majeed2020anonymization} discusses many issues on Privacy-Preserving Data Publishing (PPDP) techniques. Some of them are listed as follows:
\begin{itemize}
    \item Less resilience of existing approaches in groups privacy preservation
    \item Imbalanced datasets anonymization
    \item Tradeoff between anonymization and data utility
    \item Differences of user privacy preferences for exposure of their data
    \item Adversaries background knowledge on data subjects
\end{itemize}

\textit{Having up-to-date data for AI models makes them more accurate in their predictions. This will be important in 5G/6G as it heavily relies on AI-based decision-making processes and applications. However, we observe there are many issues in the existing approaches to anonymize data since it is possible to infer the identity of the data subjects through correlation. At the same time, future networks will lack enough data due to these privacy issues. There will be a great opportunity for the progress of AI if more abundance of data is available for them to train on. If we can use private data without violating privacy, we can expect a significant development leap for B5G/6G.
}

\subsection{Lack of Understanding of Privacy Rights and Threats in General Public} \label{public_iss}

The definitions of privacy terms and rights seem far-fetched for the general public as they could be technical and require context to be understood. In the B5G/6G era, we discussed that there will be many new technologies that come with novel modes that could collect user data, such as body movements, biometrics, and even thoughts. The gadget-free communication is an emerging concept for B5G/6G where each object can sense, gather and process the information and can be able to take context based decisions~\cite{kumar2017gadget}. This will enable digital services at convenience without explicit gadgets or even wearables~\cite{kumar2017gadget}. However, this may also make ways to collect data without consent. People may lack intuition about what approaches potentially leak their data to an undesired party using these new sensing methods and it can make privacy considerations more challenging in these circumstances. For example, though people could understand their personal information is somehow collected, they may be clueless as to which range these subtle data could be used to analyse, classify, predict, or track each aspect of their lives. With the future networks, this will be increasing even further as the complexity of the interconnectivity increases. Users may get some privacy terms and conditions when using services, but, in practice, they may not go through the privacy agreements they make when signing up for third-party applications and provide their personal and sensitive information. This creates an opportunity for third parties to collect this information and misuse them. Also, due to the uneven distribution of privacy rights, legal backgrounds, education levels, cultural influence, and many other reasons, people may have varying concerns about their privacy, making the future situation even more complicated. 

One of the issues people may face is the unclear terms that can cause ambiguity. The survey on genetic information in~\cite{clayton2018systematic} shows that many people are concerned about revealing their medical and genetic information. Still, they lack the understanding of the difference between privacy, confidentiality, control, and security and often seem to conflate these ideas. Also, a lack of knowledge on what actions are performed on their data by third parties is another concern. In cloud computing, the work in~\cite{ghorbel2017privacy} shows that most cloud customers do not have a clear idea about practices that data actors can perform with their private data.
Also,~\cite{galvin2020developments} mentions that most consumers are unaware of the ways a service can collect the data and how third-parties use them. Due to length and complexity, even if they are aware of the implications, they often do not read these policies before consenting. 

\textit{In summary, the B5G/6G networks will provide capabilities to facilitate services related to a new range of technologies that can collect users' information, potentially in every aspect of their lives. However, users may lack understanding of their privacy rights and potential threats though they will enjoy the new experience of these technologies. Therefore, we observe some issues related to the ambiguity of the privacy-related terms and unawareness in general consumers about which actions are performed on their data by third parties. Educational systems may need to consider providing sufficient knowledge on user privacy in this digital age. The requirement of convincing the general user on privacy issues and making them aware is the responsibility of entities who use their data. Therefore, we see this issue as challenging since proper mechanisms to make it possible are currently not adopted by everyone.}

\section{Privacy Solutions} \label{sec:solutions}

As discussed in the previous section, the privacy issues in B5G/6G networks exist, and they have to be addressed by the period we start using the next generation of wireless networks.

We selected the solutions considering different approaches to address issues inherited in B5G/6G. We have identified three main issues that were mentioned in Section \ref{sec:issues}:

\begin{enumerate}
\item The significant role of AI in B5G/6G.
\item Transmission of big and small data via the network.
\item The requirements of regulation of communication by the authorities especially with the new possibilities of privacy threats, vulnerabilities and attacks to the general public.
\end{enumerate}

Considering our first approaches of AI, we discuss four major solutions for B5G/6G. These solutions are outlined in Section \ref{decentralized_ai_sol} - Decentralized AI, Section \ref{edge_ai_sol} - Edge AI,  Section \ref{int_mgt_sol} - Intelligent Resource Management and Section \ref{xai_sol} - XAI for Privacy. The decentralized AI will mainly prevent privacy leakages that are happening from current centralized AI approaches where data is not directly shared through the B5G/6G network. Instead they use techniques like sharing only the model parameters. It will prevent possible privacy leakages in data transmission, and third parties will not obtain user data. Instead, it will remain within the user's device. To make the AI training process and implementation more realistic with less latency, we can use edge AI. Edge AI can protect privacy in B5G/6G since AI decisions can be offloaded to local secure, trustworthy environments. To manage the network, we can use AI for B5G/6G that can identify privacy threats, monitor the network, and make rapid, critical decisions in minimum time. The novel approaches of XAI can be used to get reasonable explanations of the behaviour of these AI tools used in B5G/6G and applications. Also, they will help to identify if an AI process has any privacy vulnerability.

The second set of solutions is under the data privacy requirements. For that, we consider the data and its properties like Personal Identifiable Information (PII), mutability, ownership, resilience against attacks, and utilizing data in a privacy preserved manner in B5G/6G. The solution in Section \ref{pii_sol} - Privacy Measures for PII presents the possible techniques that can be taken to ensure the privacy of personal identifiers in data. Especially with new technologies introduced, we discuss many possible ways that will be available to capture sensitive data from users that contain PII. Thus, protecting the privacy of the data is essential. Another approach is the solution in Section \ref{blockchain_sol}, to use blockchain, where the blockchain stores data in robust storage, making it difficult to tamper. This will enable keeping true records of data ownership clearly. Furthermore, the data should be encrypted with resilient, quantum-resistant, and lightweight encryption techniques to prevent it from being eavesdropped on or deciphering its potentially sensitive contents, as mentioned in the solution in Section \ref{light_quantum_enc_sol}. One major disadvantage of the classic encrypted data is that we cannot learn about the data unless it is decrypted. But, insights from up-to-date data are essential as well for use cases such as training the B5G/6G AI algorithms. Therefore, solutions in Section \ref{homomorphic_sol} - Homomorphic Encryption, and in Section \ref{ppdp_sol} - Privacy Preserving Data Publishing Techniques can help by making the data available, meanwhile preserving privacy.

The third consideration on privacy solutions is the regulations, where privacy could be integrated as an essential component for the design of B5G/6G communication and the associated technology ecosystem. This would reduce the privacy leakages from new technologies as the laws and regulations will direct the industries to follow the solution in Section \ref{design_sol} - Privacy Protection by Design. Further, the end consumers may not need to have advanced expertise in privacy when using the facilities provided by B5G/6G in their day-to-day life if such privacy laws and regulations are in place, as discussed in Section \ref{gov_reg_sol}. 

Apart from those solutions listed above, we consider location privacy preservation as another important area to be considered in privacy preservation. Furthermore, we mention using personalized privacy, which can also be considered as an alternative approach to solving privacy issues by giving more controllability to the user to customize their privacy as required. Thus, we analyze privacy in different aspects of B5G/6G, providing potential solutions for all issues discussed.

The following subsections will discuss these solutions we present in detail. First, we give an introduction on the proposed solution. Then, we mention how the solution can be applied to the issues of the Section \ref{sec:issues}. Next, the associated research done under the proposed solution is discussed. Finally, we summarize the key points of the solution.

\subsection{Privacy Preserving Decentralized AI} \label{decentralized_ai_sol}

The AI is one of the most impactful aspects of B5G/6G networks for automated decision making. Many applications and features thus include AI as an essential requirement. Recently, there has been an increasing interest in decentralized learning in AI. It is an approach for an AI model to learn from different sources, which may be separated from each other. 

The data generators in B5G/6G can face a significant problem of sharing their private data with third parties. In the context of ML, more accurate models are made if up-to-date data is available in large quantities. We discussed this issue in Section \ref{data_acc_iss} - Private Data Access Limitations. However, an emerging solution for this issue is decentralized AI. With decentralized AI approaches, data may not need to be moved from user devices. The idea is to locally train a model on the data generator side and then send this model to the third party. The third-party can perform the aggregation operation of all the local models, creating a more accurate model in general. Therefore, both the data generator and the organizations will be benefited in terms of privacy. Even though the data cannot be accessed directly for training by third parties, they still get the required outcome from the private data.
 
Another issue that decentralized AI can solve is providing better AI models for new technology applications. Services that use new technologies can use this approach to improve their AI models and provide a better service in B5G/6G without disrupting privacy. Especially with IoT and privacy requirements in sensor environments, decentralized AI is emerging as a solution, as the data does not require to move through an untrustworthy network. Especially, IoT communication can be compromised due to weaker security protocols. It is, therefore, safer to keep data within the device. Hence, such examples show decentralized AI solutions help to address the issue \ref{tech_app_iss} - New Technology Applications with New Privacy Requirements. 

Furthermore, the issue \ref{attacks_ai_iss} - Privacy Attacks on AI Models are mitigated through this approach, as the model training process is decentralized, and it will be difficult for an attacker to target a certain single point like a server with data for significant privacy leakage. This approach also helps reduce the costs of processing at a centralized server, often offloading the majority of that requirement from the service to the client.

There exist different approaches of decentralized learning, which may be applicable in different scenarios, depending on the requirement. The work in~\cite{zantedeschi2020fully} provides existing decentralized learning methods for ML: federated multi-task learning and decentralized learning. 

\begin{figure}[ht]
    \centering
    \includegraphics[width=0.6\linewidth]{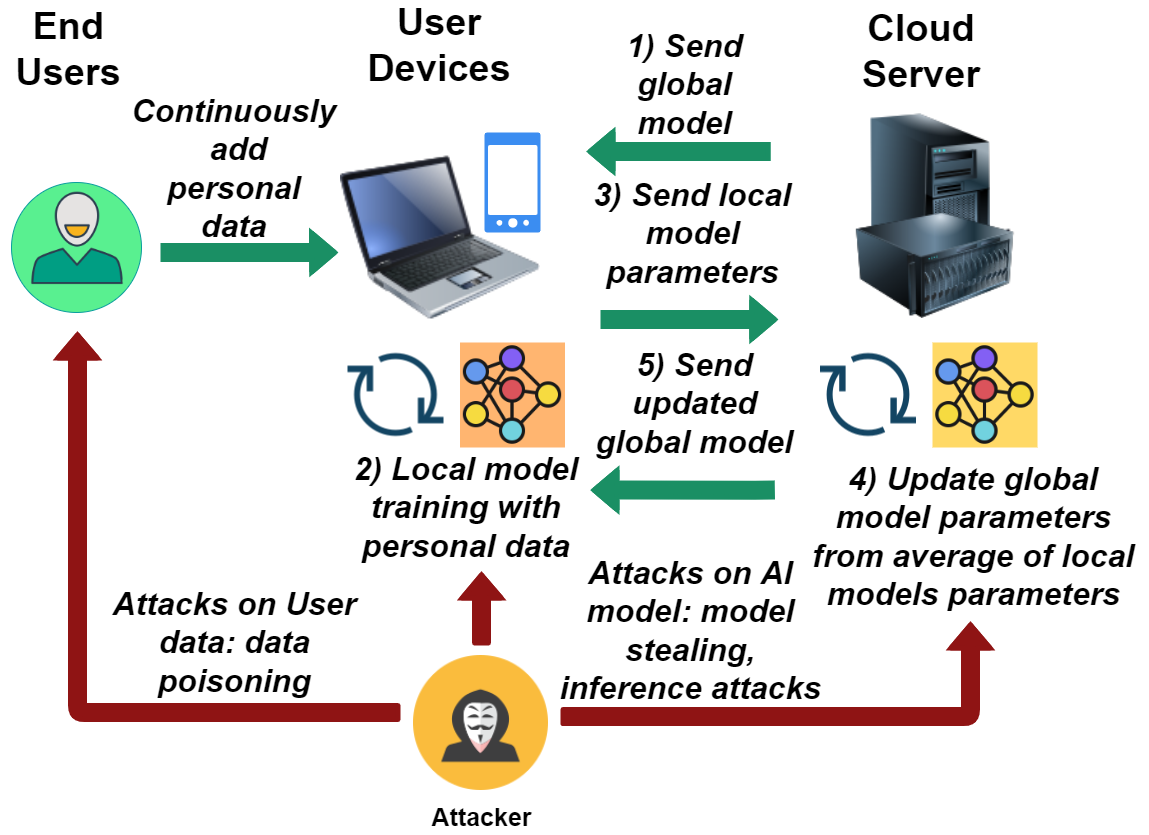} 
    \vspace{1mm}
    \caption{Process Flow of Federated Learning and Possible Attacks on Privacy}
    \label{FL_process_flow}
\end{figure}

\paragraph{Federated Learning}
FL is a new form of AI that processes learning at the edge, using decentralized data at these devices. In~\cite{mothukuri2021survey} the basic flow for FL is provided as (Figure~\ref{FL_process_flow} further illustrates these steps):
\begin{itemize}
    \item Model selection - pre-trained global model is shared with all the clients  
    \item Local model training - pre-trained models are trained locally with individual training data
    \item Aggregation of local models - Trained model updates are sent to a central server to update and improve the global ML model
\end{itemize}

The methodology of implementing FL solution in B5G/6G can be summarised as follows: First, the participants in the network will support the model training process via locally training an initial model. Then they can securely send the details to an aggregator through the network. Next, the aggregators can finally accumulate the models together using a suitable aggregation algorithm. This aggregated model can be used as the initial model by all nodes for the next training iteration. Unlike the traditional ML processes, only the model weights or gradients can be shared in the case of Federated Learning (FL) without transferring data directly. The model aggregators can also be decentralized, making it difficult to launch an attack on a single aggregator, such as an ML server.

The authors in~\cite{mothukuri2021survey} also show there are different approaches of FL techniques. Some of them are given as:
\begin{itemize}
    \item Centralized vs. clustered network topology - centralized FL still depends on a central server to manage the responsibility and collect data models. Clustered FL is designed to address the heterogeneity of clients, where an intermediate model is created among a group of clients.
    
    \item Data partition - this is divided into three sections: 1) horizontal partitioning, where homogeneous clients of the same domain contribute to train the global model, 2) vertical partitioning, where heterogeneous clients contribute to train the model; and 3) transfer FL - knowledge is transferred from a trained heterogeneous model to personalized models.    
    
    \item Data availability - based on the data availability and number of clients nodes, there are two categories: cross-silo FL and cross-device FL.
\end{itemize}

The work in~\cite{mothukuri2021survey} also highlights several privacy threats in the existing FL models, including membership inference attacks, unintentional data leakage, reconstruction through inference,  and Generative Adversarial Network (GAN)-based inference attacks. Authors also provide techniques to mitigate these issues using secure multi-party computation, differential privacy, and adversarial training. Another work in~\cite{jeon2021privacy} proposes a protocol that controls the communications of participants in distributed aggregation rounds for FL to minimize privacy leakage. They also guarantee privacy against an honest-but-curious adversary. Authors in \cite{mcenroe2022survey} propose FL can be used for UAV systems to improve privacy of AI models in UAV communications.

\paragraph{Swarm Learning}
In~\cite{warnat2021swarm}, the authors introduce swarm learning, which is a decentralized form of ML that combines edge computing, blockchain-based peer-to-peer (P2P) networking, and coordination maintaining confidentiality without a central coordinator. 
Similar work in~\cite{zantedeschi2020fully} proposes P2P exchanges without a central coordinator, where users with their personal datasets collaborate to learn together. For privacy preservation, it is suggested using differential privacy~\cite{dwork2006differential} (which will be discussed in a later section) to guarantee that personal data cannot be inferred from user information. Here, the swarm edge nodes build models individually, improving themselves by sharing parameters in the network. Members are trustworthy and governed by the blockchain.

\paragraph{Other Approaches}
A decentralized learning technique named Gossip Learning is a different approach to the above two methods. Here, the models perform random walk over the network, updating at every node they visit and merging with other models they meet~\cite{ormandi2013gossip}. The work in~\cite{hegedHus2021decentralized} compares Gossip Learning with FL and show that it is fully decentralized with no requirement of a parameter server. It also compares the performance and concludes that it is a highly competitive alternative to the FL due to its decentralized nature, but with some tradeoffs such as a slower convergence rate compared with FL.

For RL applications, a decentralized solution is available, which can be used especially in robotics, where a learning problem is decomposed into several sub-problems where the resources are managed independently. However, they work towards a common goal~\cite{leottau2018decentralized}. The work in~\cite{han2020enabling} mentions decentralized RL has emerged  in areas such as underwater energy harvesting. Authors in ~\cite{leottau2016decentralized} use decentralized RL in mobile robots. However, the privacy implications of this learning category still need to be investigated further.

Though using different types of decentralized AI techniques looks as a great research aspect, we can see there are trade-offs of each of these methods as discussed. Therefore, it may be difficult to choose the best option for real-world implementations in practice. Also, suppose such an AI approach is changed to a different one. In that case, the upgrading costs might increase significantly since the decentralized models could be spread in millions of devices unless these devices and platforms offer a flexible softwarized approach for replacing the existing models.  Table \ref{tab:decentralizedAI} below summarises decentralized AI approaches where privacy issues in each decentralized learning method and possible solutions are provided.

\begin{centering}
\scriptsize
\begin{longtable}[!htbp]{|p{3cm}|p{6cm}|p{6cm}|}
    \caption{Decentralized AI Types, Privacy Issues, and Potential Solutions}
    \label{tab:decentralizedAI}
    \\
\hline
    \rowcolor[HTML]{CBCEFB} 
\multicolumn{1}{|c|}{\textbf{Learning Method}} & \multicolumn{1}{|c|}{\textbf{Privacy Issues}} &\multicolumn{1}{|c|}{\textbf{Potential Solutions}}   \\\hline \hline
Federated Learning   & Possibility of compromising the central aggregator, semi-honest or malicious aggregators that can steal client information.
\vskip 0cm
Vulnerable to privacy attacks, including inference attacks, reconstruction attacks, and unintentional data leakage~\cite{mothukuri2021survey}.
  &  
Can use techniques including secure multi-party computation, differential privacy, and adversarial training to mitigate the issues~\cite{mothukuri2021survey}.
\vskip 0cm
Use of decentralized aggregator-based architectures~\cite{jeon2021privacy}.
 
\\\hline

Swarm Learning   & Fully decentralized ML may leak privacy of data to third parties.
\vskip 0cm
Peer-to-peer networking and coordination can have members that may not be trustworthy.
  &  
Use of differential privacy to ensure the privacy of personal information when transmitting models among peers~\cite{dwork2006differential}.
\vskip 0cm
Blockchain is used to ensure trust among members~\cite{warnat2021swarm}.

\\\hline

Gossip Learning   & Malicious nodes in gossip learning resulting in less accurate models and privacy leakages when merging.
\vskip 0cm
Slow convergence of models in gossip learning~\cite{hegedHus2021decentralized}.
  &  
Secure aggregation of models when merging.
\vskip 0cm
Enhanced model convergence with faster merging algorithms.

\\\hline

Decentralized RL   & Lack of privacy-related investigations in decentralized RL solutions.
  &  
Further investigation of techniques for privacy of agents in decentralized RL such as blockchain and pseudo identities~\cite{nguyen2020privacy}~\cite{zhu2022swarm}.
\\\hline
\end{longtable}
\end{centering}

\textit{There are many decentralized learning approaches available such as federated learning, swarm learning, gossip learning, and decentralized RL for use in B5G/6G networks. the existing work on these may need to improve regarding privacy in their areas. Therefore, we have to consider the trade-off of these methods when selecting an approach. The costs of upgrading or switching to a different AI approach should be considered as well. 
From the related works, it is clear that FL has progressed a lot over the recent years, and new areas such as swarm learning with blockchain technology are also emerging as alternatives to address existing issues in FL. }

\subsection{Edge AI} \label{edge_ai_sol}
The idea of edge AI is the implementation of AI algorithms on edge devices. This technique is helpful for B5G/6G networks because, With the ever-increasing edge connectivity, it will create huge network traffic from billions of these small devices if a central server performs AI computations on their data. Therefore, to mitigate this issue, it is inevitable to bring AI functions to the edge by offloading intelligence from the cloud to these devices to facilitate ``bringing code to data, not data to code''~\cite{kumar2020federated}.

The solution is implemented in the edge or IoT environments within B5G/6G environments. The AI algorithms are developed to run near the original data sources, such as sensor nodes. The data generators first capture data at varying rates. Then, this data is sent to the edge AI, where it can perform operations such as intelligent data pre-processing, initial model training, and predictions. Here, significant speed in performing these operations can be achieved as data is processed at the vicinity of the origin.  

The idea of bringing AI close to the edge is aimed at achieving much faster data processing through edge AI and obtaining decisions quickly. This is clearly important for B5G/6G as faster decision-making processes can help to bring down the latency in communication. Otherwise, data may need to be transferred to a remote server, which may significantly increase AI processing time. However, one major benefit of edge AI in terms of privacy is that privacy-sensitive data can be processed in local AI models without needing the data to be transferred to a remote server over an untrustworthy network. These models can be protected locally from attacks rather than trusting a remote server. Thus, the data generators will get more control over the AI models their data is processed. These reasons show edge AI helps to solve the issue \ref{edge_iss} - Privacy Issues in Edge Computing and Edge AI. There are several issues existing in the edge AI itself, which will be discussed later in this section, with potential approaches to solve them. 

The associated works further discuss the advantages offered by edge AI. The authors in~\cite{loven2019edgeai} discuss the benefits provided by AI in the edge. The implementation includes privacy-preserving regularization and models in the context of privacy. Introducing Edge computing for AI includes fine-grained control and management of personal data ownership or decentralizing trust with blockchain~\cite{loven2019edgeai}. Another privacy preserved implementation of edge AI is in~\cite{rahman2020towards}, where the authors propose a privacy-preserving AI task composition framework for pushing AI tasks to the edge networks based on homomorphic encryption.~\cite{kumar2020federated} suggests a federated k-means clustering for edge networks with privacy preservation since the data does not leave the edge device.  The authors in \cite{ding2022roadmap} discuss opportunities of edge AI to achieve privacy through locality of data, early detection of attacks, and intelligent attack migration closer to the user. 

However, considering the downsides, having edge AI computations may cause performance degradation on these resource-constrained devices. Therefore, for performance improvements, there are three approaches proposed by~\cite{lee2018techology}: 1) introducing power-efficient hardware processors for edge AI, and 2) Efficient software management frameworks and tools with the latest AI implementations. Implementing lightweight AI for the edge may be a requirement since we have to consider the costs of performance and energy in the edge devices. 

The architecture of edge AI models is also important to consider in performance and separation of functions.
Indeed,~\cite{tsigkanos2019architectural} proposes a layered architecture for edge AI highlighting privacy concerns. It consists of two layers: 1) the application layer with user goals and business logic, and 2) the edge support layer with privacy governance facilities such as anonymization, policy enforcement, and control. 

When considering other costs, attacks on these AI models could cause privacy losses. Especially with decentralized AI approaches, we discussed in \ref{decentralized_ai_sol} the interest of attackers on edge AI might increase, launching more sophisticated attacks. The survey in \cite{alwarafy2020survey} discusses possible countermeasures to the attacks, including securing software, firmware, and hardware, use of reliable routing protocols, and cryptographic schemes. We will also further discuss such cryptographic approaches later as privacy solutions, as they can be applied to many areas in B5G/6G. 

\textit{The application of AI in the edge provides many benefits, including ensuring privacy preserved intelligence, low latency for the edge devices response, and reduction of network traffic. Much recent research work is in progress to provide solutions based on AI and suitable architectures to drive the change in the direction of edge AI. This is closely connected with the decentralized AI approach we discussed, where decentralized local AI models will run on these edge nodes. Therefore, we believe edge AI can be a significant part of IoT, helping B5G/6G networks offload AI functions to the edge and providing users with a smoother experience when dealing with edge services.}

\subsection{Intelligent Management with Privacy} \label{int_mgt_sol}

Software service architectures often use automated resource allocation and service management for load balancing and network management tasks. This can be applied for B5G/6G networks as well. The use of intelligence to automate resource and service management will be an essential requirement for future networks since optimum decisions have to be taken considering various dynamic factors at high speeds. Therefore, in the case of privacy, the services can add such intelligence to monitor the service allocation and status to identify potential threats that could endanger user privacy.

The intelligent integration of management and orchestration process can be done by the design of AI models to perform supervision to these tasks. The existing management functions can be separated into different domains, and AI tools can be integrated into these domains for numerous operations, including optimization of performance, identifying vulnerabilities and threats to the network, and getting analytics on the usage patterns.

AI-based management operations in B5G/6G can automatically monitor the network for potential anomalies and privacy violations using the designed AI algorithms. This will bring down the costs of privacy preservation mechanisms, as manual inspection will be lesser. Therefore, using intelligence-based control, potential privacy threats could be identified beforehand and managed automatically. For example, automated privacy-enhanced controlling can contribute to fixing the privacy issues related to edge devices since possible privacy leakages can be detected through anomaly detecting AI models in real-time. Automated mechanisms are placed to scan the network for any potential privacy vulnerabilities in the deployments or orchestration. Periodic inspections can be done on each management domain by these intelligent-based managing AI agents in B5G/6G. Therefore, this solution helps to address those issues we mentioned in Section \ref{control_iss} - Privacy Preservation Limitations for B5G/6G Control and Orchestration Layer.

The intelligent management also contributes to preventing privacy issues in Section \ref{edge_iss} - Privacy Issues in Edge Computing and Edge AI as follows: In this issue, we mentioned many privacy attacks on edge AI, such as eavesdropping traffic, data tampering, data poisoning, and inference. Management of edge devices with AI intrusion detection, traffic monitoring, and anomaly detection can support mitigating such issues. AI management operations can automatically terminate the connection with the detected malicious or compromised devices with the B5G/6G network in such situations. Therefore, intelligence-based automatic management and orchestration can enormously support the B5G/6G network operations for protecting privacy. 

Intelligent resource management operations can use AI techniques such as sophisticated, high accuracy DL for decision making. The work in~\cite{ahmed2019deep} discusses DL techniques used in related work in wireless resource management problems, such as channel and power allocation, throughput maximization, and spectrum sharing. They observed two categories of DL used in this context: DRL and supervised DL. For edge node management,~\cite{xu2020joint} defines an optimization problem for resource utilization and load balancing while maintaining privacy and proposes a balanced service offloading method. The authors in~\cite{ratnayake2020novel} propose an approach for realtime network intrusion detection using XGBoost with LSTM, which can be used in B5G/6G for realtime network monitoring and alerts. The work in~\cite{mukherjee2020energy} proposes an energy-efficient resource allocation for massive IoT systems using a clustering algorithm to categorize subsets of systems. 

Considering service management, the ZSM architecture proposed in~\cite{benzaid2020ai} splits services to different Management Domains (MD), where MDs interface with each other using an inter-domain integration fabric. The separation of functions in MDs improves privacy since they can have their data services and communicate using interfaces without fully exposing the implementations. Within the scope of each MD, they can use intelligent automation of orchestration, control, and assurance of resources and services. They also provide E2E services for multiple domains with different administrative entities.~\cite{zhang20196g} mentions that a combination of AI with existing technologies of SDN, NFV, and Network Slicing (NS) can reach zero-touch network orchestration, optimization, and management for 6G networks. The work in \cite{liyanage2022survey} discusses the need for ZSM and service management for reasons such as increasing network complexity, generation of new business-oriented services, performance requirements, and the requirement of efficient management for B5G/6G networks. The authors also highlight several privacy issues related to it, especially the AI/ML adversarial attacks, and discuss the importance of defending against them. 

Having high accurate AI can be an expensive task since we have limitations on accuracy with capabilities of hardware and software used to make these models. There can be possible privacy leakages due to low-accuracy AI algorithms. Also, with offloading of services, there will be costs involved. Moreover, there is a substantial cost in designing and implementing new architectures and algorithms.

\textit{The intelligent-based resource and services management can be done for B5G/6G networks to preserve privacy in a rapid decision-making environment. We found related intelligence application approaches such as DL-based resource management, clustering-based IoT nodes resource allocation, and domain-based ZSM management using intelligence. We see that intelligence-based solutions maintain privacy, and their design leads to privacy preservation in these related works. Therefore, significant work is done in resource, service management, and optimization. However, we mentioned some challenges due to various costs, such as architectural design and implementation and the accuracy of AI models. Using intelligence to directly manage privacy can be proposed as future work in this area. The architectural updates, use of privacy-preservation algorithms, and combining AI are critical factors that drive the change to improve this area.}

\subsection{XAI for Privacy} \label{xai_sol}
When using intelligence-based solutions, a reasonable explanation needs to be provided using XAI for justification of AI-based decisions for B5G/6G, as discussed in Section~\ref{sec:issues}. It is critical to privacy in future networks since users have rights, and any privacy violations may cause legal actions. The decisions may depend on how explainable and reasonable the AI decision is. 

\begin{figure}[htb]
    \centering
    \includegraphics[width=0.7\textwidth]{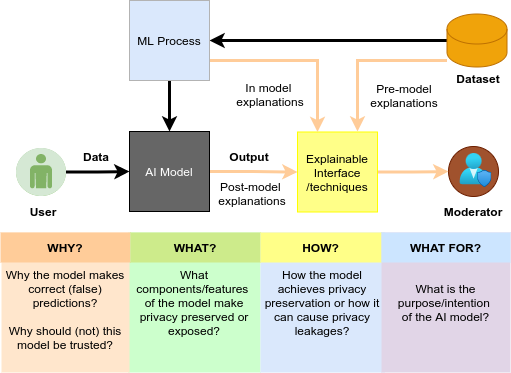} 
    \vspace{1mm}
    \caption{Application of XAI for Privacy in B5G/6G AI models and Questions to Address}
    \label{xaiPrivacy}
\end{figure}

There are numerous classifications of XAI techniques. One such is the pre-model, in-model and post-model explanations~\cite{singh2020explainable}. The XAI algorithms can be developed to get explanations with the dataset used to train ML models. This is called pre-model explanations. These explanations may provide insights into how the ML model can alter its decisions due to dataset properties such as amount, relevancy, and diversity. Another possibility for evaluating AI models is in-model explanations, which are related to the ML model creation process. The post-model explanations can provide meaningful insights into what exactly a model has learned after the training process. Figure~\ref{xaiPrivacy} provides an overview of how XAI can be applied for trustworthy AI in terms of privacy for B5G/6G networks. Here, a moderator or authorized personnel can access and evaluate the explanations of the AI models available in the B5G/6G network via an explainable interface. The availability of such an approach can primarily support identifying any issues with these AI models after deployment. They can provide answers to different questions related to the privacy of these models, as indicated in Figure~\ref{xaiPrivacy}.

There has been an increasing trend of discovering novel approaches in XAI in recent years. The work in~\cite{das2020opportunities} illustrates different XAI methods discovered from 2007 to 2020. Therefore, we can assume that the trend for XAI will continue to rise in upcoming years as well, due to the increased requirement of getting explanations of the black box AI models for privacy reasons. With the increased usage, there will be attacks by exploiting vulnerabilities in the XAI. However, as we discuss in related works, defense mechanisms for XAI are also emerging in recent research works. Therefore, the increased new approaches to XAI and new defense techniques for XAI can help mitigate the discussed privacy issue in Section \ref{xai_iss} - Limited Availability and Vulnerabilities of XAI Techniques. Further investigation of XAI techniques helps bring down the limitations of availability in explanations for AI, especially considering the issues on XAI itself, such as the lack of formalism and standardization of XAI techniques.

 Considering how XAI can contribute to enhancing the privacy of B5G/6G, the use of XAI may help define the levels of privacy violations. The privacy levels may help to support decision-making processes in determining the privacy tradeoffs. Other than that, the application of XAI can support convincing various parties as a validation mechanism to agree upon privacy standards, which can help mitigate discrepancies and privacy differences in different regions. 

To apply XAI in the privacy domain, we must identify which areas we want to consider in current XAI works. In~\cite{kuppa2020black}, the authors divide the explainability space into three regions for the predictions/data in the context of the security domain: 1) explanations for the predictions/data itself, 2) explanations for covering security and privacy properties, and 3) explanations covering the threat model.

One of the main concerns in ensuring privacy is identifying how a particular AI model guarantees privacy. The authors in~\cite{wolf2019explainability} introduce the concept of ``explainability scenarios'', where it focuses on what type of explanation people need to understand on AI systems, rather than what an AI system is capable of explaining. 

We can observe a great interest in the field of XAI starting from recent years. The survey in~\cite{das2020opportunities} shows many related works and categorizes them based on scope, methodology, and usage. Their work show more recent work have been added in recent years.   The work in \cite{srivastava2022xai} shows that XAI has become more popular over the recent years, where the related works have kept increasing. Many novel XAI approaches are investigated, and numerous types of applications using the existing XAI techniques are included in these research works. Further, they show open-source packages for XAI have significantly improved over recent years. Considering the defense against XAI attacks, the authors in \cite{vigano2020explainable} propose a new paradigm in security research named Explainable Security (XSec), where they discuss six aspects called ``Six Ws'' of XSec. They include questions based on who receives the explanation, what is explained, where the explanation is, when the explanation is given, why the stakeholders require XSec, and how to explain it. These questions and paradigms in XSec can be used as a roadmap for finding techniques to enhance the defense in XAI against malicious attacks.  The work in~\cite{dazeley2021explainable} discusses that explainable RL (XRL) also should be a part of the mainstream XAI. The authors also aim to introduce a Causal XRL Framework to merge ideas of XRL from existing work. 

One of the main costs of XAI could be its implementation cost for XAI algorithms since it involves the engineering of XAI solutions and development costs of interfaces for presenting explainability.

\textit{XAI is an emerging field that can be used to justify AI-based decision-making for B5G/6G. It is undoubtedly helpful regarding privacy-based AI applications since users and authorities may need reasons for how the AI models guarantee their privacy. In recent years, we have seen an increasing interest in XAI-based techniques from the associated work, which may continuously improve further. It positively impacts B5G/6G networks since AI will empower them, and thus, user privacy may depend on their explainability of actions.}

\subsection{Privacy Measures for Personally Identifiable Information} \label{pii_sol}

For a person, their PIIs play a key role in privacy as they directly provide a way to distinguish them easily. Therefore, the protection of these PIIs is an essential requirement for B5G/6G. For instance, information such as digital image sharing is continuously popular with new mobile phones and advanced capturing devices, which may expose the PIIs of users. Therefore, it is suitable to consider privacy awareness in these devices and services when processing them. Also, privacy concerns are rapidly escalating, considering the increasing popularity of biometric authentications, geolocation, and ever-growing personal data exposure in social media. New technologies will generate new modes of PII, such as holographic imaging and Brain-Computer Interface (BCI) wave patterns, which need attention when considering privacy.

Due to rising concerns about PII, many privacy preservation methods were proposed in the literature. PII can be protected via different means, such as encryption techniques, perturbation mechanisms, and secure storage. In B5G/6G networks, a large quantity of data can be generated over a short period of time. It can consist of PII, which can be used to trace individuals. Especially, new modes of technologies can be protected with the discussed privacy-preserving mechanisms to prevent potential privacy leakages. The improved privacy measures for new technologies and sensing modes will reduce the capability of attackers to exploit their weaknesses of them. It will bring down the privacy issues discussed in Section \ref{tech_app_iss} - New technologies and applications for privacy requirements. Identifying PII and making suitable protection mechanisms for them will significantly improve the anonymity of data. Also, it may contribute to solving the issue in Section \ref{data_acc_iss} - Private Data Access Limitations as the anonymized or modified data can be shared among third parties, where PII cannot be determined exactly. Thus, more data will be available for obtaining insights on usage trends and training AI models; meanwhile, data generators or owners' privacy is preserved. Having enough privacy measures for PII requires only hiding crucial information on users. It may help organizations release other non-critical information publicly, thus contributing to improvements in up-to-date big data sets. In B5G/6G, PIIs can be protected through the use of different approaches, technologies, and tools. It may depend on the application requirements. In the following paragraphs, we discuss different approaches to protecting PII. 

The authorities should update users and organizations with guidelines on how it should be done to achieve privacy in PII. The authors in~\cite{onik2019personal} provide a set of recommendations and review existing work on the privacy of PII: a blockchain-based personal data sharing for privacy in data processing, privacy by design (this is further discussed as a privacy solution),  privacy-preserving data aggregation at the owner side, privacy awareness, and differentiation between privacy and security. 

When considering the privacy of PII in digital media, the work in~\cite{abur2021personal} uses AES encryption to cover digital images to hide PIIs when transiting users' attributes to a federated cloud environment. In~\cite{ziad2016cryptoimg}, the authors implemented a library for privacy-preserving image processing operations using homomorphic encryption. The survey in~\cite{ribaric2016identification} provides a comprehensive de-identification method in multi-media documents for non-biometric identifiers such as text, hairstyle, license plates, as well as physiological, behavioural, and soft-biometric identifiers. 

In terms of future technologies associated PIIs, the works~\cite{de2019security,bye2019ethical} discuss privacy implications on mixed reality, which is an emerging technology where devices blend virtual and physical elements to create a new concept of reality. As shown by~\cite{bye2019ethical} mixed reality collects biometric data such as eye-tracking, facial
tracking, gait detection, emotional sentiment analysis, galvanic skin response, Electroencephalography (EEG), Electromyography (EMG), and Electrocardiography (ECG), which extends the list of existing PII. The work in~\cite{onik2018privacy} brings the idea of Privacy of Things (PoT), where the authors developed a privacy monitoring system for collecting, analyzing, and detecting incoming and outgoing traffic from IoT devices. 

With the use of encryption and other privacy-preserving techniques, there could be potential performance costs due to PII protection measures. Another cost is to make the public aware of the importance of PII privacy.
Also, investigating future PII privacy measures for future technologies is another possible expansion we can expect.

\textit{In this solution, we focused on existing methodologies for PII privacy protection. A set of recommendations can be provided to ensure PII privacy for users in industrial applications in B5G/6G. Also, the current digital media should be handeled with prior identification of PII and providing proper protection. However, we see some costs for protecting PII, such as cost in reduction of performance due to privacy-protecting techniques, and financial costs of making the public aware of PII protection. The future PII collecting concepts may need further investigation of potential privacy threats. The work on PIIs are rapidly growing as they are crucial for privacy, and also, they are one of the most valuable yet very vulnerable type of data.} 

\subsection{Blockchain-based Solutions} \label{blockchain_sol}

Blockchain is a decentralized and distributed public ledger technology in a peer-to-peer network~\cite{feng2019survey}. In simple terms, it is a list of linked records, or blocks, which are connected with links that make it challenging to change any block after creation~\cite{al2019blockchain}. The blocks that store required transactions are timestamped, and encrypted~\cite{al2019blockchain}. As hash values for blocks are unique, fraud can be effectively prevented since modifications to a block in the chain change the hash value immediately. A new block can be added to the chain if most nodes in the network agree. We do this via a consensus method to verify the legitimacy of transactions in a block. It also verifies the validity of the block itself,~\cite{nofer2017blockchain}. Even some or all blocks of a blockchain in the network are changed, it is virtually impossible to tamper in practice. These transactions are duplicated and distributed across the entire network of computer systems.

Though the implementations of blockchain have existed since 2008 with the famous Bitcoin concept~\cite{nakamoto2008bitcoin}, it has attracted great interest in the research community in recent years because of its decentralized and secured nature~\cite{feng2019survey}.
The application of blockchains on privacy has also made significant progress in research. Therefore, potential solutions exist, such that we can get their adoptions for B5G/6G networks. 

We can observe from research literature that many new technologies can get support from blockchain to enhance their data privacy and provide data confidentiality, integrity and availability. Therefore, blockchain can support to mitigate the privacy issues mentioned in Section \ref{tech_app_iss} - New Technology Applications with New Privacy Requirements. We already mentioned some examples of such applications in swarm learning~\cite{warnat2021swarm}~\cite{dwork2006differential} and PIIs~\cite{onik2019personal} in previous solutions on how they use blockchain for usecases such as privacy preserved tamper-proof storage or to reach a privacy preserved consensus. More such solutions are discussed here as well. Also, blockchain helps remove the issue discussed in Section \ref{ownership_iss} - Ambiguity in the responsibility of data ownership, as blockchain maintains robust records of transaction details. The records are nearly impossible to modify by the design of the blockchain. Thus, it can be used to keep records of guaranteed ownership and transfer ownership if needed.

The blockchain can be implemented in the B5G/6G services and applications as tamper-proof storage of data. Different consensus mechanisms can be introduced to perform operations on data in a privacy-preserved manner. The access control can be provided if necessary to allow only trusted entities to enter and perform the consensus. Examples of such implementations from related works are given below.

We notice direct industrial and technological applications in the blockchain are significantly rising. In~\cite{dorri2017blockchain}, the authors discuss a potential solution using blockchain for smart vehicles to enhance privacy in communication and privacy-sensitive data storage. The authors in \cite{ch2020security} discuss the application of blockchain to Virtual Circuit (VC) based devices such as UAVs, drones, and other IoT devices. To improve privacy, they use Elliptic curve cryptography and Secure Hash Algorithms (SHA) based on secured data storage in a public blockchain. The work in \cite{kalla2022emerging} shows the blockhained 6G with features such as pseudonymity, secure cryptographic techniques, and smart contract-based access control has the potential to provide a privacy-compliant ecosystem.  In \cite{LU201980}, the authors mention blockchains and smart contracts can be used to collaborate industrial automation processes in 5G environments, which can be extended to 6G. The survey in~\cite{al2019blockchain} shows blockchain can provide digital identities, distributed security, smart contracts, and micro-controls. They also show that it is beneficial to use blockchain for many domains, including financial, healthcare, logistics, manufacturing, energy, agriculture, robotics, entertainment, construction, and telecommunication. Authors claim that in telecommunications domain, we can achieve:
\begin{itemize}
    \item Enhancement in telecommunications service management
    \item Improvement in traceability and transparency
    \item Enabling efficient contract management
    \item More cost-effective governance process
\end{itemize}
Currently, the user's digital identity is spread among multiple parties, including government organizations, social media, and other private or public entities. Therefore, the user has little control over their data. Blockchain can also be used to create digital identity/self-sovereign or self-managed identity. Here, the user gets the ownership and control of their digital identity. Citizens can then use this identity to authenticate when using digital services~\cite{rivera2017digital}. Blockchain can also be used with IoT to eliminate a requirement of a trusted third party, by improving the scalability and reducing the friction of the business processes \cite{viriyasitavat2019blockchain}, which will be very helpful in B5G/6G QoS, guaranteeing privacy.

As technologies adopt blockchain for their privacy, recent work ensure privacy in the blockchain itself. The survey in~\cite{feng2019survey} shows methodologies for identity privacy preservation techniques. These include mixing service transaction's relationships in the blockchain, use of ring signature to sign the message on behalf of ``ring'' of members to hide the identity, Zero Knowledge Proof (ZKP) to utilize cryptography to prove a given statement without additional information leakage. They also discuss transaction privacy preservation methods: non-interactive ZKP, homomorphic encryption, and Pedersen commitment scheme, which is one of the implementations of the homomorphic commitment scheme.

The costs for blockchain include maintenance costs for these networks, especially if the blockchain is private. There are issues in high energy consumption and computation costs \cite{bamakan2020survey} for laborious calculations such as proof of work, which is essentially a consensus mechanism used in blockchain.

\textit{There is a very significant development in blockchain, which can be applicable for privacy preservations of many applications in B5G/6G networks. Continuously improving research is in progress to ensure the privacy of blockchain as well. However, there could be maintenance costs and energy consumption with blockchain technology. Yet, more works are continuously being done on addressing these issues. We see B5G/6G networks can use this tool to ensure privacy, including the use with AI applications, network communications, and data storage.}

\subsection{Lightweight and Quantum Resistant Encryption Mechanisms} \label{light_quantum_enc_sol}

Encryption is a technique to prevent unauthorized parties from accessing information by applying the mathematical function and converting it to an incomprehensible format. There will be a key to decrypt or recover the original information during that process. This will only be converted back by an entity with this valid key, which is only revealed by the party who performed the encryption. Encryption is often used in communication in networks to ensure data privacy primarily; since if a third party receives the data, it will not be helpful as they are impossible to understand.

Since there is an apparent computational cost in encryption, it will be incorporated with a delay. As per the limitations of hardware and cost considerations such as computation capacity in network infrastructures such as IoT or cloud computing environments, there are proposed solutions for enhancing privacy through encryption mechanisms through less complex computations. It may help achieve better preservation of privacy and, at the same time, increase overall performance and quality of experience for the end-user. Therefore, B5G/6G networks need to get empowered by lightweight yet powerful encryption mechanisms. Similarly,  the quantum resistant encryption techniques are necessary to ensure the data privacy is preserved in quantum computing environments facilitated by B5G/6G as the current approaches of encryption fail in them. For example, the work in \cite{shor1994algorithms} shows
quantum algorithms can be used for factoring and the computation of discrete logarithms efficiently, making the public key cryptography techniques used today insecure.

With the increased use of edge devices in B5G/6G, we discussed that they could face several attacks, such as poisoning, inference, and eavesdropping of traffic. One of the main reasons for these issues is the lack of proper protection mechanisms for data communication due to resource limitations. Classic encryption techniques may be too expensive to run on these devices in terms of computing cost. Therefore, new lightweight encryption methods can support the fulfillment of privacy requirements for edge devices. As these lightweight encryption techniques may run smoothly on resource-constrained devices, existing issues in Section \ref{edge_iss} - Privacy Issues in Edge Computing and Edge AI can be solved by running these encryption algorithms within the devices. It will enable privacy protection of data during transmission or storage, as the attackers cannot poison, infer or eavesdrop on encrypted data. Therefore, E2E data communication confidentiality can be preserved during the transmission of data with encryption. Also, lightweight encryption significantly supports bringing down the costs in privacy enhancements, processing requirements, and energy consumption aspects. In addition, lightweight encryption may significantly reduce the workload on cloud computing environments as well since the decryption process may also be much simpler and more efficient than general encryption mechanisms.

Quantum-resistant algorithms also play a significant role in ensuring communication privacy among multiple entities using the applications of quantum computing in B5G/6G.  We identify that the issues in Section \ref{confidential_iss} - Data communication confidentiality can be solved with the aid of lightweight and quantum-resistant encryption techniques. This is because these encryption techniques would prevent data leakage through any device, such as resource constraint devices, general user devices, or quantum computing devices. Thus, they further ensure confidentiality for B5G/6G services and applications. 

In essence, the encryption will make data look meaningless after performing a reversible mathematical operation on the data during the encryption process. The lightweight encryption advantage for B5G/6G is that they will perform the encryption with algorithms that require much less computation cost than classic algorithms. Such encryption algorithms may provide similar privacy protection to a more complex algorithm at a lower cost. Similarly, quantum-resistant algorithms will be used in B5G/6G, which is resilient against quantum attacks. The encrypted data will have no pattern and might look like noise to an external observer. Even if the communication channel is compromised, an adversary may not be able to derive any information from it. Thus, their implementation will solve the privacy issues mentioned above. Now we present several examples available in related works on lightweight and quantum resistant encryption.

The work in~\cite{singh2017advanced} discusses lightweight cryptographic algorithms for IoT devices. They classify a set of lightweight cryptographic primitives into several categories: lightweight block cipher, lightweight hash function, high-performance system, lightweight stream cipher, and low resource devices.~\cite{bansod2014implementation} presents a lightweight, compact encryption system based on bit permutation instruction group operation. The work in~\cite{shabisha2021security} presents lightweight symmetric-key-based operations for resource constrained environments to preserve anonymity and untraceability with a use case of a third-party mobile relay-based emergency detection system.~\cite{al2016lightweight} proposes a lightweight encryption scheme for smart home applications, supporting public key management through identity-based encryption, without certificate handling. A review of lightweight encryption schemes in~\cite{naru2017recent} focused on many different schemes, and authors show every technique has some advantages and disadvantages in IoT, such as the requirement of more storage but fewer computations vice versa. The work in~\cite{baharon2015new} proposes a homomorphic encryption-based lightweight scheme for mobile-cloud computing. Considering the quantum resistant algorithms, the survey in \cite{perlner2009quantum} discusses the threat of quantum attacks and discusses a set of public key cryptographic schemes that are believed to be used as quantum resistant algorithms. The work in \cite{cheng2017securing} mentions the algorithms such as RSA, and Elliptic Curve Cryptography (ECC) used in IoT environments are insecure against quantum attacks. They also discuss on ongoing post-quantum cryptography research projects for protecting IoT.

Lightweight encryption schemes might have a limited scope of security, which could cause privacy leakages. Also, we can see costs of network traffic due to a large number of parameters requests from a cloud server in some of the lightweight schemes since they may offload complex calculations to a cloud server. These schemes may also incorporate computation costs for encryption in low-resource IoT devices. The complexity and non-intuitive nature of quantum resistant algorithms may make the integration of them with the applications difficult.

\textit{There are many active types of research going on for lightweight encryption schemes. We observe that most of them focus on resource-constrained IoT devices related applications. Some of them may come with privacy costs due to limited scope, resulting in more network traffic. However, they can be equally valuable for resourceful environments such as mobile or cloud computing due to their lesser utilization of energy and improved performance. When considering the quantum resistant encryption, the future networks will pose a great threat of quantum attacks as most current encryption schemes fail. Thus, more research is ongoing for finding better quantum resistant algorithms.}

\subsection{Homomorphic Encryption} \label{homomorphic_sol}

To fulfill the privacy requirements of data in B5G/6G environments, safe encyrption techniques are needed. However, the drawback of most encryption techniques is they will cause data cannot be used for any useful task without decryption. With the emerging need for preserving privacy meanwhile having the possibility of analysing data in B5G/6G, homomorphic encryption provides a promising solution. It allows mathematical operations on data to be carried out on a ciphertext, which is an encrypted form of the input data/the plain text, instead of on the actual data itself\cite{rocha2018overview}. 
It consists of three~\cite{rocha2018overview} forms: 
\begin{itemize}
    \item Somewhat homomorphic - supports mathematical operations for addition and multiplication, yet limited to a certain number of operations due to noise accumulation.
    \item Partially homomorphic - Supports any number of operations, yet only limited to a specific type of operation.
    \item Fully homomorphic - Support both addition and multiplication operation any number of times.
\end{itemize}

\begin{figure}[htb]
    \centering
    \includegraphics[width=0.55\linewidth]{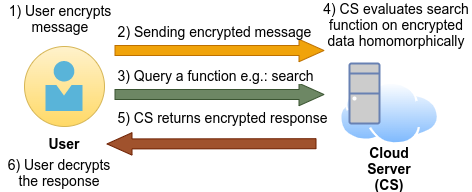} 
    \vspace{1mm}
    \caption{Process Flow of Homomorphic Encryption}
    \label{Figure_homomorphic_encryption}
\end{figure}

Even though the first fully homomorphic scheme was introduced in 2009, improvements are continuously carried out to make it practical to use in every platform~\cite{acar2018survey}.
Figure~\ref{Figure_homomorphic_encryption} illustrates the basic overview of homomorphic encryption.

In the context of B5G/6G privacy, as mentioned, one of the main drawbacks of encryption is that the utility of data from encryption is lost unless it is decrypted. However, B5G/6G services may still need to get insights from data and perform operations on them. Due to privacy issues, it is unsafe to send plain data as well. Homomorphic encryption helps to solve this private data access issue discussed in Section \ref{data_acc_iss} since private organizations can allow their data to be encrypted and outsourced to commercial cloud environments for processing while encrypted. The process of homomorphic encryption can be performed for data, and the applications such as ML model training can be made on this encrypted data, unlike classic encryption methodologies. This will clearly be useful for B5G/6G since such encrypted data will preserve privacy; meanwhile, this data could be used for real-world applications. Therefore, we can consider it as a viable solution for this privacy issue in B5G/6G network services.  

many works exist related to homomorphic encryption. The survey in~\cite{acar2018survey} provides a detailed description of such several homomorphic encryption methods in all three types mentioned. Their evaluation shows that the security, speed, and simplicity have increased over time, but they can improve further. Authors also note that fully homomorphic encryption can provide solutions for functional encryption, which controls access over data while allowing computation based on the features of identity/attribute. Yet, the related work in this area are limited. The work in~\cite{aono2017privacy} uses additively homomorphic encryption in combination with asynchronous stochastic gradient descent algorithm for deep learning to build system with: 1) no information is leaked to the server, and 2) accuracy is kept intact. In~\cite{geng2019homomorphic}, the work compares homomorphic algorithms Hill, RSA, Paillier, and ElGamal and discuss applications of these algorithms in a cloud environment. The authors conclude that the current public key encryption efficiency is not high, and speeds need to be improved. 

Homomorphic encryption is relatively slow in operation and increases computational cost. Unfortunately, fully homomorphic encryption techniques are currently more susceptible to attacks as they do not guarantee the reliability of its secret key used for encryption~\cite{acar2018survey}.

\textit{We see homomorphic encryption has extended applicability in future networks for privacy preservation, due to its capability of processing data without decryption. This area has great attention in the research community, and a significant progress. We consider this a potential solution for ensuring data privacy for B5G/6G networks, especially addressing the private data access limitations.}

\subsection{Privacy-preserving Data Publishing Techniques} \label{ppdp_sol}

Data can be modified for privacy during storage and usage in big data AI models in B5G/6G. Different data sanitation methods are available to remove or add noise to user data, including sensitive personal information or PII.  For example, the work in \cite{lin2020privacy} uses a technique of ant colony optimization adopting multiobjective sanitization and transaction deletion to protect sensitive data in 6G IoT networks. They also use expert knowledge in defining the confidential events or information for deletion to reduce the side effects of the sanitization process.  Therefore, essential facts to consider in data sanitization and publishing techniques are protecting user privacy; meanwhile, ensuring the original data not to be mutated too much for publishing. Not limited to data publishing, privacy-preserving techniques help enhance privacy in other data-related usages, such as storing data in an unsecured environment. The actual privacy of owners of those data will not be affected even if the data is exposed. Therefore, discovering more techniques on privacy-preserving data publishing may help B5G/6G networks to improve their guarantees on privacy for their users' data. 

Privacy preserved data publishing techniques in B5G/6G will make the data anonymized or untraceable. For example, noise addition through a technique called differential privacy is an approach for this, where an attacker will not be able to distinguish whether the data received is real or a random output. Therefore, data can be freely shared without much privacy concerns. This can help in private data access issues discussed in Section \ref{data_acc_iss}, where up-to-date data for AI model training and data science applications are lacking. Since third parties cannot reveal a user's identity, more data can be published by private or public repositories, given that privacy is guaranteed through these methods. It also helps reduce the issue in Section \ref{ownership_iss} - Ambiguity of privacy responsibility in data ownership since owning data by any party may not cause privacy issues for data contributors.

The implementation of these techniques can be done in the data preprocessing stages in the B5G/6G network. After the data is captured by devices such as sensors, they can directly be subjected to data modification techniques. Otherwise, the data can be securely transferred to a particular destination, and before storing, privacy-preserving operations can be performed.

There are many methods of privacy preservation for data in the associated work. Two approaches for data modification are shown by~\cite{binjubeir2019comprehensive}: 1) data perturbation techniques and 2) anonymization. There are several anonymization techniques currently available as shown in~\cite{goswami2017privacy}, based on:
\begin{itemize}
    \item Data nature - tabular, data sets, graphical data
    \item Anonymization approaches - generalization, suppression, perturbation, permutation
    \item Objectives of anonymization - k-anonymity, L-diversity, objectives based on presumptions on attacker's knowledge
\end{itemize}
The works~\cite{goswami2017privacy,wang2017utility} classify anonymization techniques based on 1) syntactic approaches and, 2) differential privacy. We also describe them briefly as follows. 

\paragraph{Syntactic Anonymization}
The syntactic anonymization modifies the input data set to achieve generalization. As shown in~\cite{clifton2013syntactic}, most of the syntactic models are based on generalizing table entries.~\cite{goswami2017privacy} shows several techniques for syntactic anonymization, namely: 1)~k-anonymity, which is the warranty that in a series of $k$ groups, the probability of identifying that person is less than $1/k$, 2)~L-diversity, which makes the maximum probability of recognizing the sensitive user information to be $1/L$ for $L$ different groups, and 3)~T-closeness, where the dissemination distinction between sensitive data and its values within groups does not exceed a value $T$. The work in~\cite{choudhury2020syntactic} uses syntactic anonymization for federated learning based on $(k,k^n)$-anonymity model.

\paragraph{Differential Privacy}
The differential privacy is a relatively recent introduction of methods to privacy preservation, which first appeared in 2006~\cite{hassan2019differential}. This is a technique used to maintain a trade-off between privacy and accuracy by adding a desirable amount of noise to data~\cite{hassan2019differential}. In~\cite{dwork2014algorithmic} it is shown that using differential privacy techniques saves the user privacy meanwhile having no additional cost, and the accuracy further increases with further increase of the number of samples. Therefore, this is well applicable for the scenario of big data. In~\cite{hassan2019differential}, the authors discuss privacy attacks on cyber-physical systems, namely: disclosure attack, linking attack, differencing attack, and correlation attack. They also discuss differential privacy applications and design requirements for smart grids, transportation systems, healthcare and medical systems, and industrial IoT systems.

In differential privacy, as data gets subjected to noise addition, its accuracy might get reduced, especially, when the data set is small. Loss of actual user information in anonymization techniques is another issue that comes as a cost for enhancing privacy.

\textit{There are different approaches to privacy-preserved data publishing, which will be mandatory in B5G/6G due to its support for increased growth in big data. We discussed mainly syntactic anonymization, and differential privacy. We observe the current trend for privacy-enhancing data publishing techniques moving towards differential privacy, yet older techniques are also applied according to the use case.}

\subsection{Privacy by Design and Privacy by Default} \label{design_sol}

With the expanding network capabilities of B5G/6G from technologies that capture and use personal data, general users will get subjected to an increased risk of privacy leakages in their lives. Therefore, privacy by design is an important aspect that should be considered in every step of the designing process life cycle of B5G/6G services, and their capability should be evaluated. Moreover, by default, these services or products should safeguard privacy requirements, even without any manual input from users. It is taking actions to protect beforehand, not after a privacy breach happened~\cite{cavoukian2009privacy}. This means that, rather than having to come up with complicated and time-consuming 
``patches'' later on, it is necessary to detect and assess potential data protection issues when creating new technology and to incorporate privacy protections into the overall design~\cite{schaar2010privacy}.

As we discussed in Section \ref{tech_app_iss} - New Technology Applications with New Privacy Requirements, there will be numerous technologies added to operate with B5G/6G. These technologies may potentially lead to various privacy leakages as the industries, developers, and the general public may lack understanding of these technologies and their impact on privacy. Therefore, vulnerabilities can go undetected. To mitigate this issue,  Privacy by design can help since designing such new technological applications focusing on privacy beforehand will eliminate most privacy threats. Careful inspection of vulnerabilities, threats, and potential attack scenarios can enormously support mitigating privacy leakages and attacks. It also provides a level of guarantee for privacy by default. Therefore, privacy is ensured even though the general public may pay less attention to privacy rights and threats. Thus it helps the issues mentioned in Section \ref{public_iss} - Lack of Understanding of Privacy Rights and Threats in the General Public using these services of B5G/6G. 

The privacy by design techniques should be implemented by including them to the overall design of the B5G/6G applications and services. It may require standardizations and regulations to make this approach to be accepted by the organizations. A minimum set of requirements would be feasible to be included for the industries to follow such that privacy by design practices will be implemented based on the scope and financial capabilities of the organizations. Examples of the importance and approaches of privacy by design from related works are discussed below.

As the self-decision-making capability of AI increases, it will get the opportunity to impact on the B5G/6G network and end users. Therefore, the AI design process should be done with privacy concerns as a key requirement. The authors of~\cite{van2017privacy} raise the concern that if the autonomy of AI increases, it may be difficult to maintain transparency. They mention that the European Committee on Civil Liberties, Justice, and Home Affairs has given high-level indications for future regulations to stress the responsibility of AI designers and developers. They should ensure the AI products are safe, secure, fit for purpose, and follow procedures for data processing compliant with existing legislation, confidentiality, anonymity, fair treatment, and due process.

Considering the design of associated technologies with B5G/6G, we can provide a set of guidelines to ensure privacy in designing. For instance, the work in~\cite{sion2019architectural} presents an architectural view for data protection by design for an e-health application, which is compatible with the EU GDPR. Work presented by European Union Agency for Network and Information Security in~\cite{d2015privacy} and~\cite{guideprivacydesign} discusses some privacy by design strategies as follows:

\begin{itemize}
    \item Minimize the amount of personal data as possible
    \item Hide personal data from plain view
    \item Separate personal data in a distributed fashion
    \item Processing of personal data should be done at the highest level of aggregation
    \item Inform data subjects with transparency
    \item Data subjects should be supported for the processing of personal data
    \item Enforce privacy policy compatible with legal requirements
    \item Demonstrate compliance with privacy policy into force and any applicable legal requirements
\end{itemize}

\begin{figure*}[ht]
    \centering
    \includegraphics[width=0.95\linewidth]{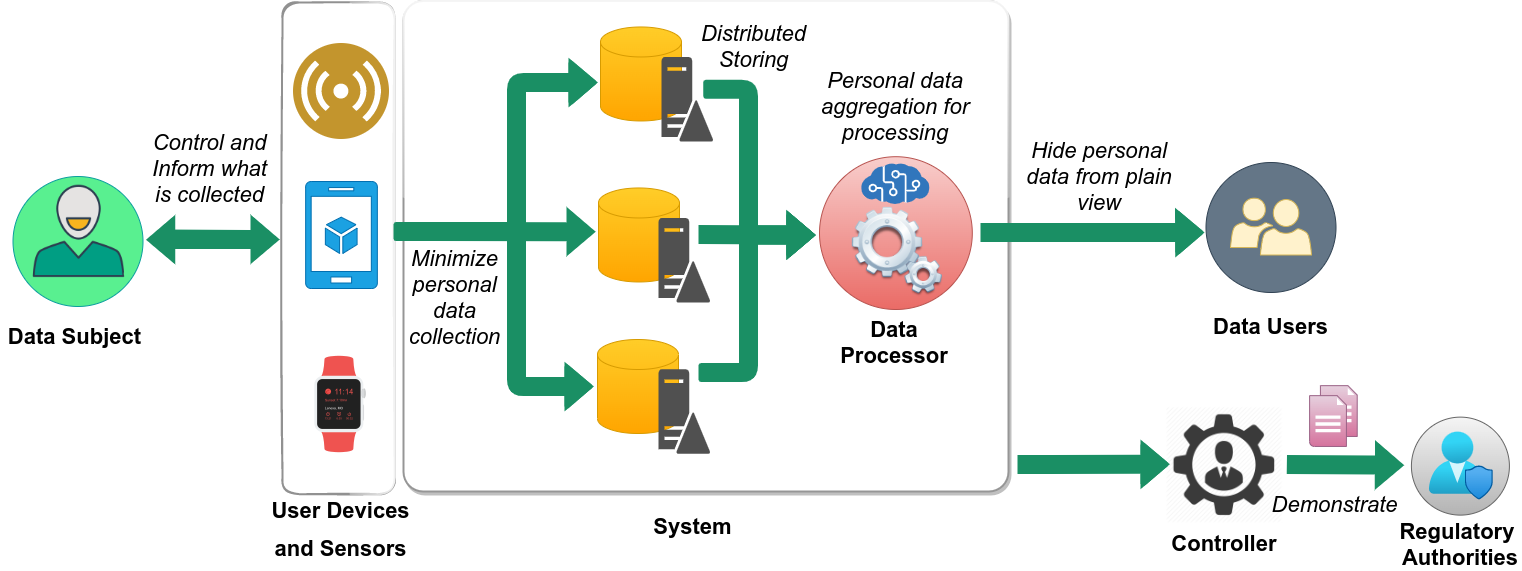} 
    \vspace{1mm}
    \caption{Privacy by Design Strategies Implementation for Privacy Enhancement}
    \label{privacy_by_Design}
\end{figure*}
Figure \ref{privacy_by_Design} illustrates how these design strategies could be applied.

Having privacy enabled in design will automatically provide enhancements of security aspects.
In fact,~\cite{onik2019personal} shows that privacy by design provides two levels of security for IoT systems: 1) check of data fitness to the context when collecting information 2) capability of a system to assess the scope of data sharing to the internet and other associated risks. Similarly, in~\cite{lodge2018iot}, the authors develop an IoT environment to help developers to engage with data protection at design phase. Authors in \cite{siriwardhana2021role} provide privacy by design as a potential solution in the usecases of e-health and telemedicine, where privacy sensitive health data is aggregated and privacy protection is a legal requirement.

There is a design cost for privacy by default implementations, coupled with an additional cost of privacy preservation techniques for implementation. Also, when updating the privacy by design schemes, there will be costs for any design changes if new standards are imposed.

\textit{Adopting privacy by design for B5G/6G is an essential step in assuring guarantees to privacy beforehand, rather than fixing when a privacy leakage occurs. We see an emergence in the privacy by design topic in terms of AI privacy-aware design, guidelines for design, and discussions on how privacy design helps mitigate risks. Therefore, future network architecture components, associated services, and technologies can fundamentally include privacy-protected design principles by default. However, privacy by design could be expensive to design and implement, and changes will also carry heavy costs. Hence, more work should be done to identify these costs and propose methodologies to mitigate them in the designs.}

\subsection{Regulation of Government, Industry, and Consumer} \label{gov_reg_sol}

Many parties in B5G/6G networks handle user data, and they could be linked with the government authorities as a mediator to oversee the actions of this data by the industry. Also, governments directly collect user data with the aid of wireless communication. For instance, using online apps by the government for contact tracing in the COVID-19 pandemic~\cite{li2020covid} was such an initiative. These apps may collect vast amounts of user data with high privacy-related information such as user location, identity, and medical records. If such a government system is hacked, citizens will have critical information leakages. Also, industries collect and provide data to third parties, even without the consent of users~\cite{tuttle2018facebook}. Therefore, the involvement of regulations helps ensure user privacy since all parties who utilize privacy-related data will require to adhere to them.

We discussed that privacy leakages could occur for data owners or generators due to their lack of understanding of the importance of protecting their own privacy. Adversaries may also gain the advantage of this by exploiting the lack of knowledge of the general public to social engineer attacks or gain the advantage of weaker privacy protection mechanisms made by users. Imposing proper regulations on privacy for B5G/6G service providers can make the services provide a minimum set of privacy standards when providing the services. Then, even if the users lack the knowledge of privacy, their private data can be protected.  Therefore, establishing proper regulations will support solving the privacy issue in Section \ref{public_iss} related to a lack of understanding of privacy rights and threats to the general public. An example of such an approach is making lightweight two-factor authentication for IoT devices \cite{gope2018lightweight} a default requirement when accessing private data to verify if data is accessed by the data owner or an authorized person.

Furthermore, regulations will promote privacy best practices for the service providers. New innovations in privacy preservation techniques may emerge due to these regulations as well, since the industries may require to obtain data analytics meanwhile preserving privacy to satisfy the standards. Also, it may help to solve privacy differences based on location if these regulations can tackle the general privacy requirements despite the physical location. Thus, imposing these regulations can help solve the issue of legal disputes among entities.  

We can identify different types of regulatory approaches in the literature to address the issues. In~\cite{liyanage20185g}, the authors categorize the regulatory approach into three directions:
\begin{itemize}
    \item Government regulation
    \item Industry self-regulation
    \item Consumer or market regulation
\end{itemize}

Industry and government regulations influence macro-level general privacy concerns. Their work categorizes privacy concerns to different themes regarding general privacy concerns such as: consumer attributes and privacy, macro-environment surrounding consumer privacy, technology-mediated E-Commerce privacy. The specific privacy concerns include vendor-related attributes and consumer privacy, consumer-vendor interaction, and consumer-vendor trust relationship. The work ~\cite{liyanage20185g} finally presents possible future directions for consumer privacy: through technical mediation of E-commerce privacy, individual-centered situational privacy determinants, and having a decision-environment in situational privacy.
Other work in introducing privacy regulations include the work in~\cite{lederman2016private}, which proposes a framework for Intelligent Transportation Systems (ITS) from existing non-ITS technologies. They consider that ITS-specific privacy should address the four dimensions 1) who is collecting the information, 2) the potential criminal implications resulting from the collected data, 3) whether PII is being collected, and 4) whether there is a possibility of secondary collected data. 

From a consumer perspective,~\cite{zhang2021whether} shows that consumers are generally aware of having privacy rights but have insufficient knowledge and resources to exercise these rights properly. This work also identifies consumers rely on to access privacy violations based on the moral values of trust, transparency, control, and access.~\cite{bandara2020privacy} presents privacy concerns in e-commerce where they discuss that consumers are generally have limited knowledge on security and privacy and rely on laws and safety mechanisms. Therefore, introducing regulations for privacy will ultimately mitigate users from breaching their privacy. 

These regulations may not be applicable everywhere, yet they can cause a high extra cost in implementation in only a specific region. Also, many industries may suffer financial losses if these regulations are very strictly imposed.

\textit{With the enhanced capabilities of wireless networks, we see a requirement to regulate government, industry, and consumer privacy. The existing work suggest such approaches for each of these parties, showing where the privacy concerns arise and future directions. Also, recent work show that consumers have a general understanding of privacy, yet they lack detailed knowledge of laws and mechanisms. Therefore, the regulations imposed on the governments, industry, and consumers play a crucial role in protecting user privacy since this will ensure legal assurance for the users and their data. However, we suggest this field should need further recent work and investigation sufficing with updated regulations for the B5G/6G networks.
}

\subsection{Other solutions}

In addition to the mentioned privacy solutions, we found other solutions that are already available, which can be used for addressing some of the related privacy issues in B5G/6G networks. We summarize them as follows:

\paragraph{Location Privacy Considerations}
Location privacy is an important aspect to consider since it is a type of PII. The establishment of location privacy measures helps solve privacy issues related to edge computing since most of the location trackers are placed in IoT platforms and mobile devices. Associated work on the discussion of location-based privacy can be applied to ensure privacy in B5G/6G networks. 

For instance, the work in~\cite{shivaprasad2016privacy} discusses location-based services for privacy preservation. The authors classify and discuss them, including 1) the cryptographic method of Private Information Retrieval and 2) non-cryptographic approaches such as spatial obfuscation, mix zone, k-anonymity, and dummies. They suggest future research methods need to consider user-based protection and real-time protection by considering different user habits, interests, or preferences. That means they propose a personalized approach for location privacy preservation. The trajectory of locations is important since they directly include tracking user movements. For this,~\cite{shaham2020privacy} proposes a metric called "transition entropy" that considers the privacy of users in trajectories and an algorithm to improve the transition entropy for a given dummy-generation algorithm. Here, dummy-generation algorithms generate false location data to make it challenging to identify the user to an adversary such as an untrusted server. However, hiding actual locations may cause difficulties for location-based personalization services to deliver accurate service. 

Therefore, we can see different approaches, including encrypted and un-encrypted methods for location privacy protection. Also, to ensure further privacy, techniques such perturbation addition can be done for the location to avoid tracking by an adversary. We can see personalized privacy can be utilized for future location privacy preferences. This can be applicable in scenarios such as contact tracing, where the work in~\cite{sandeepa2020social} uses GPS location information for COVID-19 contact tracing from users to collect their location data, only when the user is willing to contribute.

\paragraph{Personalized Privacy}
Personalized privacy is an approach to achieving privacy levels for users specialized for their requirements. Having personalized privacy may help resolve privacy issues based on different locations since no matter the location, privacy is configured based on user choices. 

Instead of going to a unified privacy metric, an alternative suggestion could be to rely on a personalized approach in some scenarios applicable at the consumer level since privacy preferences can differ from one individual to another. In~\cite{xiong2019personalized}, the authors show the existing privacy-preserving methods adopt a unified approach for sensing data where effective privacy metrics are lacking. They propose a personalized privacy protection framework based on game theory and data encryption for mobile crowdsensing in Industrial IoT. Similarly, the work in~\cite{wang2018personalized} proposes a personalized privacy-preserving task allocation framework for mobile crowdsensing, meanwhile providing location privacy. The authors in~\cite{colnago2020informing} evaluate the user perceptions on Personalized Privacy Assistants (PPA) for IoT, which help users to discover and control data collection practices of IoT resources. They interviewed 17 participants and recommended PPA solutions that address users' differing automation preferences and reduce notifications, which can overload users. 

Therefore, we see that a potential solution for the existing privacy issue of difficulty in the definition of levels for privacy when using personalization where it can be applied. However, the cost of creating platforms for personalized privacy and coordination of privacy choices of a large number of users could be high. Also, it might neither be practical nor accepted as reliable in complex decision-making situations like the government level. However, we suggest the potential for personalized privacy metrics can be investigated as a future research direction related to privacy for B5G/6G.

\subsection{Summary of Privacy Solutions}
The solutions presented above are possible approaches to address the privacy issues discussed in Section \ref{sec:issues}. Table \ref{tbl:solutionsForIssues} provides the summary of issues that the solutions have mostly addressed. In the table, we list each privacy solution and show how many issues it is addressing. We have already provided a detailed explanation of how these issues are addressed, in the description of each of these solutions.

From Table \ref{tbl:solutionsForIssues}, it is evident that we have addressed all the issues since the provided solutions can provide privacy preservation to one or multiple issues mentioned. Similarly, multiple solutions can be available for a particular privacy issue, and one can choose the best option or combinations of these solutions to address their privacy issue in B5G/6G.  
\begin{center}
\scriptsize
\begin{longtable}[!ht]{|l|c|l|l|l|l|l|l|l|l|l|}
\caption{Privacy Solutions that mostly Address Privacy Issues in B5G/6G}
\label{tbl:solutionsForIssues}
\\
\hline
\rowcolor[HTML]{CBCEFB} 
\cellcolor[HTML]{CBCEFB}                                   & \multicolumn{9}{c|}{\cellcolor[HTML]{CBCEFB}\textbf{Addressed Privacy Issues}}                                                                                                                                                                                                                      \\ \cline{2-10} 
\rowcolor[HTML]{CBCEFB} 
\multirow{-2}{*}{\cellcolor[HTML]{CBCEFB}\textbf{Privacy Solution}} & \multicolumn{1}{c|}{\cellcolor[HTML]{CBCEFB}\textit{IA}} & \multicolumn{1}{l|}{\cellcolor[HTML]{CBCEFB}\textit{IB}} & \multicolumn{1}{l|}{\cellcolor[HTML]{CBCEFB}\textit{IC}}  & \multicolumn{1}{l|}{\cellcolor[HTML]{CBCEFB}\textit{ID}} & \multicolumn{1}{l|}{\cellcolor[HTML]{CBCEFB}\textit{IE}} & \multicolumn{1}{l|}{\cellcolor[HTML]{CBCEFB}\textit{IF}} & \multicolumn{1}{l|}{\cellcolor[HTML]{CBCEFB}\textit{IG}} & \multicolumn{1}{l|}{\cellcolor[HTML]{CBCEFB}\textit{IH}} & \textit{IJ} \\ \hline
\hline
Privacy Preserving Decentralized AI                         & \multicolumn{1}{c|}{\checkmark}                         & \multicolumn{1}{l|}{}                           & \multicolumn{1}{l|}{\checkmark}                           
& 
\multicolumn{1}{l|}{}                           & 
\multicolumn{1}{l|}{}                           & 

\multicolumn{1}{l|}{}                           & \multicolumn{1}{l|}{}                           & \multicolumn{1}{l|}{\checkmark}                         & 

\multicolumn{1}{l|}{}                            
\\ \hline 
Edge AI                                                    & \multicolumn{1}{c|}{}                           & \multicolumn{1}{l|}{}                           & \multicolumn{1}{l|}{}                           & 
\multicolumn{1}{l|}{\checkmark}                         &\multicolumn{1}{l|}{}                           & 
& \multicolumn{1}{l|}{}                           & \multicolumn{1}{l|}{}                           & \multicolumn{1}{l|}{}                           
\\ \hline
Intelligent Management with Privacy               & \multicolumn{1}{c|}{}                           & \multicolumn{1}{l|}{\checkmark}                         & \multicolumn{1}{l|}{}                           &
\multicolumn{1}{l|}{\checkmark}          
& 
\multicolumn{1}{l|}{}                           & 
\multicolumn{1}{l|}{}                           & \multicolumn{1}{l|}{}                           & \multicolumn{1}{l|}{}                           & 
\multicolumn{1}{l|}{}                           
\\
\hline
XAI for Privacy                                            & \multicolumn{1}{c|}{}                           & \multicolumn{1}{l|}{}                           & \multicolumn{1}{l|}{}                           & 
\multicolumn{1}{l|}{}                           & \multicolumn{1}{l|}{\checkmark}                           & 
\multicolumn{1}{l|}{}                           & \multicolumn{1}{l|}{}                           & \multicolumn{1}{l|}{}                           & 
\multicolumn{1}{l|}{}                                  
\\ \hline
Privacy Measures for PII   & \multicolumn{1}{c|}{\checkmark}                         & \multicolumn{1}{l|}{}                           & \multicolumn{1}{l|}{}                           &
\multicolumn{1}{l|}{}                           &
\multicolumn{1}{l|}{}                           & \multicolumn{1}{l|}{}                           &
\multicolumn{1}{l|}{}                           & 
\multicolumn{1}{l|}{\checkmark}                 &

\multicolumn{1}{l|}{}                           
\\ \hline
Blockchain Based Solutions                                 & \multicolumn{1}{c|}{\checkmark}                         & \multicolumn{1}{l|}{}                           & \multicolumn{1}{l|}{}                           & 
\multicolumn{1}{l|}{}                           &
\multicolumn{1}{l|}{}                           & 
\multicolumn{1}{l|}{\checkmark}                         & \multicolumn{1}{l|}{}                           & \multicolumn{1}{l|}{}                           &
\multicolumn{1}{l|}{}                           
\\ 
\hline
Lightweight and Quantum Resistant Encryption Mechanisms                          & \multicolumn{1}{c|}{}                           & \multicolumn{1}{l|}{}                           & \multicolumn{1}{l|}{}                           & 
 \multicolumn{1}{l|}{\checkmark}                   
& 
\multicolumn{1}{l|}{}                           & 
\multicolumn{1}{l|}{}                           & \multicolumn{1}{l|}{\checkmark}                         & \multicolumn{1}{l|}{}                           & 
\multicolumn{1}{l|}{}                           
\\ \hline
Homomorphic Encryption                                     & \multicolumn{1}{c|}{}                           & \multicolumn{1}{l|}{}                           & \multicolumn{1}{l|}{}                           & 
\multicolumn{1}{l|}{}                           & 
\multicolumn{1}{l|}{}                           & 
\multicolumn{1}{l|}{}                           & \multicolumn{1}{l|}{}                           & \multicolumn{1}{l|}{\checkmark}                 &
\multicolumn{1}{l|}{}                           
\\ \hline
Privacy-preserving Data Publishing Techniques              & \multicolumn{1}{c|}{}                           & \multicolumn{1}{l|}{}                           & \multicolumn{1}{l|}{}                           & 
\multicolumn{1}{l|}{}                           & 
\multicolumn{1}{l|}{}                           & 
\multicolumn{1}{l|}{\checkmark}                         & \multicolumn{1}{l|}{}                           & \multicolumn{1}{l|}{\checkmark}                     & 
\multicolumn{1}{l|}{}                           
\\ \hline
Privacy by Design and Privacy by Default                              & \multicolumn{1}{c|}{\checkmark}                         & \multicolumn{1}{l|}{}                           & \multicolumn{1}{l|}{}                   
& \multicolumn{1}{l|}{}                           &
\multicolumn{1}{l|}{}                           & 
\multicolumn{1}{l|}{}                           & \multicolumn{1}{l|}{}                           & \multicolumn{1}{l|}{}                           & 
\multicolumn{1}{l|}{\checkmark}                         
\\ \hline

Regulation Government, Industry, and Consumer                    & \multicolumn{1}{c|}{}                           & \multicolumn{1}{l|}{}                           & \multicolumn{1}{l|}{}                           & 
\multicolumn{1}{l|}{}                           & 
\multicolumn{1}{l|}{}                           & 
\multicolumn{1}{l|}{}                           & \multicolumn{1}{l|}{}                           & \multicolumn{1}{l|}{}                           & 
\multicolumn{1}{l|}{\checkmark}                          
\\ \hline
Other Solutions                                            & \multicolumn{1}{c|}{}                           & \multicolumn{1}{l|}{}                           & \multicolumn{1}{l|}{}                           & 
\multicolumn{1}{l|}{}                           & 
\multicolumn{1}{l|}{}                           & 
\multicolumn{1}{l|}{\checkmark}                         & \multicolumn{1}{l|}{}                           & \multicolumn{1}{l|}{}                           &
\multicolumn{1}{l|}{}                           
\\ \hline
\end{longtable}
\begin{tabular}{ll}
\begin{tabular}[c]{@{}l@{}}\\IA - New Technology Applications with New Privacy Requirements\\IB - Privacy Preservation Limitations for B5G/6G\\ Control and Orchestration Layer\\IC - Privacy Attacks on AI Models\\ ID - Privacy Issues in Edge Computing and Edge AI\\ IE - Limited Availability and Vulnerabilities of XAI techniques\\ \end{tabular} & \begin{tabular}[c]{@{}l@{}} IF - Ambiguity in Responsibility of Data Ownership\\ IG - Data Communication Confidentiality\\ IH - Private Data Access Limitations\\ IJ - Lack of Understanding of Privacy Rights and Threats in\\ General Public \end{tabular}
\end{tabular}
\end{center}
Table~\ref{tbl:privacySolutionsSummary} summarizes the privacy solutions, how they can be applied in the B5G/6G context, and their potential limitations.

\begin{center}
\scriptsize
\begin{longtable}{|p{2cm} | p{7cm} | p{5.5cm}|}
\caption{Privacy Solutions, Methodology and Limitations} \label{tbl:privacySolutionsSummary}
\\
\hline
\rowcolor[HTML]{CBCEFB} 
\multicolumn{1}{|c|}{\textbf{Privacy Solution}} & \multicolumn{1}{|c|}{\textbf{Methodology}} &\multicolumn{1}{|c|}{\textbf{Costs of Privacy Preservation}}  \\\hline \hline
\vskip 0cm
Privacy Preserving Decentralized AI                       & 
    \vskip 0cm
    Use of Federated Learning, and Swarm Learning~\cite{zantedeschi2020fully,mothukuri2021survey,warnat2021swarm} 
    \vskip 0cm
    Possible use of other approaches such as Gossip Learning~\cite{ormandi2013gossip,hegedHus2021decentralized}, and decentralized Reinforcement Learningl~\cite{leottau2018decentralized,han2020enabling}
&
    \vskip 0cm
    Difficulty in selection of which learning technique to be used due to trade-offs,
    \vskip 0cm
    Cost in modification of AI algorithm when needed 

\\ \hline
\vskip 0cm
Edge AI                                              & \vskip 0cm
Decentralized AI algorithms in edge platforms~\cite{loven2019edgeai}
\vskip 0cm
Privacy preserving AI models~\cite{rahman2020towards,kumar2020federated}
\vskip 0cm
Storage and process data locally reducing cloud usage
\vskip 0cm
Separation of functionalities, and using privacy enhanced architectures~\cite{tsigkanos2019architectural}  
& 
\vskip 0cm
Performance degradation of resource constrained edge devices 
\vskip 0cm
Attacks on AI models causing privacy losses
\vskip 0cm

\\ \hline
\vskip 0cm
Intelligent Management with Privacy                      & 
\vskip 0cm
 Deep Learning for wireless resource management~\cite{ahmed2019deep}
\vskip 0cm
 Privacy preserved service offloading~\cite{xu2020joint}
\vskip 0cm
Use of clustering for massive IoT networks to identify subsets~\cite{mukherjee2020energy}
\vskip 0cm
Separation of tasks using ZSM Management domains~\cite{benzaid2020ai}
\vskip 0cm
Combination of AI with SDN, NFV, and NS~\cite{zhang20196g}
& 
\vskip 0cm
Privacy leakages due to low accuracy AI algorithms
\vskip 0cm
Costs of offloading
\vskip 0cm
Implementation cost of new architectures and algorithms
\vskip 0cm

\\ \hline
\vskip 0cm
XAI for Privacy & 
\vskip 0cm
Apply XAI for data, predictions, and threat model~\cite{kuppa2020black}
\vskip 0cm
Use of explainability scenarios~\cite{wolf2019explainability}
\vskip 0cm
Utilize open source packages for XAI model development~\cite{das2020opportunities}
& \vskip 0cm
Implementation cost of XAI algorithms
\vskip 0cm
Development costs of interfaces for explainability

\\ \hline
\vskip 0cm
Privacy Measures for Personally Identifiable Information &
\vskip 0cm
 Providing guidelines on PII privacy~\cite{onik2019personal}
\vskip 0cm
 Encryption techniques, and design principles with privacy~\cite{abur2021personal,ziad2016cryptoimg}
\vskip 0cm
 Data aggregation at the owner side~\cite{onik2019personal}
\vskip 0cm
 Differentiate between privacy and security~\cite{onik2019personal}
\vskip 0cm
 Investigation of future technologies using new types of PIIs  
& 
\vskip 0cm
Performance costs for encryption and other techniques
\vskip 0cm
 Cost for making awareness in the public 
\vskip 0cm
Research costs for detection of future technologies and possible new PII privacy measures

\\ \hline
\vskip 0cm
Blockchain Based Solutions                      
& \vskip 0cm
Using blockchain for privacy sensitive data storage~\cite{dorri2017blockchain}
\vskip 0cm
Privacy protected digital identities~\cite{al2019blockchain}
\vskip 0cm
Creating smart contracts without privacy leakage\cite{LU201980}
\vskip 0cm
Use of ring signatures for hiding identity of members~\cite{feng2019survey} 
\vskip 0cm
Zero Knowledge Proof~\cite{feng2019survey}
& 
\vskip 0cm
Maintenance costs for blockchain networks 
\vskip 0cm
Energy, and computation costs \cite{bamakan2020survey} for calculations such as proof of work
\vskip 0cm

\\ \hline
\vskip 0cm
Lightweight and Quantum Resistant Encryption Mechanisms               
& 
\vskip 0cm
Use of lightweight block ciphers and hash functions~\cite{singh2017advanced}
\vskip 0cm
Lightweight symmetric key operations~\cite{shabisha2021security}
\vskip 0cm
Public key management Identity-based encryption without certificates~\cite{al2016lightweight} 
\vskip 0cm
Investigating new algorithms for quantum resistant encryption\cite{perlner2009quantum,cheng2017securing} 
& 
\vskip 0cm
Security limitations causing privacy costs for lightweight encryption schemes
\vskip 0cm
Costs of network traffic due to large number of parameters requests from a cloud server
\vskip 0cm
Computation costs for encryption
\vskip 0cm
Non intuitive and complexity of encryption algorithms

\\ \hline
\vskip 0cm
Homomorphic Encryption                                   & 
\vskip 0cm
    Encryption of messages and use homomorphic functions to evaluate the encrypted data without breaching privacy~\cite{aono2017privacy}   
    \vskip 0cm
& 
\vskip 0cm
    Slower in operation and increased computational cost
    \vskip 0cm
    Susceptible to attacks for fully homomorphic encryption as it does not guarantee the reliability of secret key~\cite{acar2018survey}

\\ \hline
\vskip 0cm
Privacy-preserving Data Publishing Techniques            & 
\vskip 0cm
Anonymization techniques based on syntactic approaches~\cite{clifton2013syntactic,goswami2017privacy,choudhury2020syntactic}
\vskip 0cm
Differential privacy by adding random noise to data~\cite{hassan2019differential,dwork2014algorithmic}
\vskip 0cm
& 
\vskip 0cm
Accuracy loss due to noise addition in differential privacy
\vskip 0cm
Loss of actual user information in anonymization techniques

\\ \hline
\vskip 0cm
Privacy by Design and Privacy by Default                       & 
\vskip 0cm
Minimization and hiding plain view of personal data usage~\cite{d2015privacy,guideprivacydesign}
\vskip 0cm
Distributed personal data storage~\cite{d2015privacy,guideprivacydesign}
\vskip 0cm
Processing of personal data at the highest level of aggregation~\cite{d2015privacy,guideprivacydesign} 
\vskip 0cm
Facilitation of transparency for data subjects~\cite{d2015privacy,guideprivacydesign}
\vskip 0cm
Enforce and demonstrate privacy policy compatibility with legal requirements~\cite{d2015privacy,guideprivacydesign}
& 
\vskip 0cm
Design cost for privacy by default implementations
\vskip 0cm
Additional cost of privacy preservation techniques for implementation
\vskip 0cm
Updating costs for any implemented design according if new standards are imposed
\vskip 0cm

\\ \hline
\vskip 0cm
Regulation of Government, Industry, and Consumer                 & 
\vskip 0cm
    Technical mediation for E-commerce privacy~\cite{liyanage20185g}
    \vskip 0cm
    Individual-centered situational privacy~\cite{liyanage20185g}
    \vskip 0cm
    Having a decision-environment for the situational privacy~\cite{liyanage20185g} 
    \vskip 0cm
    Regulations on who is collecting data, criminal implications from data, if data contains PII, possibility of secondary collected data~\cite{lederman2016private}
    \vskip 0cm
    Make consumers aware on privacy regulations,~\cite{zhang2021whether,bandara2020privacy}
& 
\vskip 0cm
Regulations may not be applicable everywhere, yet cause a significant extra cost in implementation 
\vskip 0cm
Industries may suffer financial losses due to strict privacy regulations
\vskip 0cm

\\ \hline
\vskip 0cm
Location  Privacy  Considerations                        & 
\vskip 0cm
Using cryptographic and non-cryptographic mechanisms~\cite{shivaprasad2016privacy}
\vskip 0cm
False location generation algorithms to prevent tracking movements~\cite{shaham2020privacy}
                                        & 
\vskip 0cm
    Location based personalization services may face difficulty to deliver accurate service 
\vskip 0cm

             \\ \hline
\vskip 0cm
Personalized Privacy                                     &
\vskip 0cm
    Use of consumer preferences to adjust the level of privacy they require in using services and products~\cite{wang2018personalized}
& 
\vskip 0cm
    Cost of creating platforms for personalized privacy and coordination of privacy choices of a large number of users  
    
\\ \hline
\end{longtable}
\end{center}

\section{6G Privacy Projects and Standardization: } \label{sec:projects}

Several B5G research initiatives are underway, bringing the academic and industry partners together from all around the world. We describe a number of projects in this section and include a description of their major goals and planned deliverables.. 

\subsection{ Research Projects}
1) AI4EU~\cite{AI4EU} - the project is aiming to build a comprehensive European AI-on-demand platform to lower barriers to innovation, boost technology transfer, and catalyse the growth of start-ups and SMEs in all sectors through open calls and other actions. The platform built as part of AI4EU acts as a broker, developer, and one-stop-shop providing and showcasing services, expertise, algorithms, software frameworks, development tools, components, modules, data, computing resources, prototyping functions, and access to funding. Training is also available for different user communities such as engineers or civic leaders to obtain skills and certifications. The AI4EU platform aims to establish a world reference, built upon and interoperable with existing AI and data components and platforms. 

2) XMANAI~\cite{XMANAI}- EU-funded project that focuses on explainable AI. Researchers working on the XMANAI project plan to carve out a ``human-centric'', trustful approach that will be tested in real-world manufacturing cases. XMANAI aims to demonstrate (using four real-life manufacturing cases) how it will help the manufacturing value chain shift towards the amplification of the AI era. It is done by coupling (hybrid and graph) AI 
``glass box'' models that are explainable to a 
``human-in-the-loop'' and produce value-based explanations. This is done with complex AI assets management-sharing-security technologies to multiply the latent data value in a trusted manner and targeted manufacturing apps to solve concrete manufacturing problems with high impact. XMANAI pilots are carried out in collaboration with CNHi of Italy (generating a virtual representation/digital twin, of the plant based on 3d-2d models and production, logistic, maintenance data of the lines), Ford (real-time representation of production and traceability), UNIMETRIK (intelligent measurement software that warns that the point sets defined for the measurement strategy are adequate) and Whirlpool (platform capable to ensure a reliable sales forecasting for the D2C channel).

3) SPATIAL~\cite{SPATIAL} - EU-funded project set up to tackle the identified gaps on data and black-box AI. It is done by designing and developing resilient accountable metrics, privacy-preserving methods, verification tools, and system solutions that will serve as critical building blocks for trustworthy AI in ICT systems and cybersecurity. The project tackles uncertainties in AI that directly impact privacy, resilience, and accountability. Similar to some of the issues identified in Section \ref{sec:issues} of this paper, the SPATIAL project identifies possible attacks using XAI techniques and potential misjudgment of XAI techniques. Therefore it aims to propose resilient accountability metrics and embed them into the existing ``black-box'' AI algorithms. Another aim of the SPATIAL project is to propose detection mechanisms to identify data biases and carry out explanatory studies about different data quality trade-offs for AI-based systems. 

4) STAR~\cite{STAR}- is a project that links AI and digital manufacturing experts towards enabling the deployment of standard-based secure, safe, reliable, and trusted human-centric AI systems in manufacturing environments. STAR researches how AI systems can acquire knowledge to make timely and safe decisions in dynamic and unpredictable environments. It will also research technologies that enable AI systems to confront sophisticated adversaries and to remain robust against security attacks. Partners working on this project consider several AI-powered scenarios and systems, including active learning systems, simulated reality systems that accelerate Reinforcement Learning (RL) in human-robot collaboration, XAI systems, human centric digital twins, advanced reinforcement learning techniques for optimal navigation of mobile robots, and for the detection of safety zones in industrial plants and cyber-defense mechanisms for sophisticated poisoning and evasion attacks against deep neural networks operating over industrial data. These technologies will be validated in challenging scenarios of quality management in manufacturing lines, human-robot collaboration, and AI-based agile manufacturing. STAR aims to eliminate security and safety barriers against deploying sophisticated AI systems in production lines.

5) 6G Flagship~\cite{6GFLAGSHIP} - is a research project funded by the Academy of Finland that aims to realize 5G networks from the 5G standard to the commercialisation stage and the development of the new 6G standard for future digital societies. The main goal of 6G Flagship is to develop the fundamental technology needed to enable 6G. The 6G Flagship research program has published the world’s first 6G white paper~\cite{6GflagshipWhitepaper} which opened the floor for defining the 2030 wireless era. The authors of that paper identified several interesting security challenges and research questions, i.e., how to enhance information security, privacy and reliability via the physical layer technologies and whether this can be done using quantum key distribution. Additionally, the 6G Flagship project will target wireless connectivity, distributed intelligent computing, and privacy to develop essential technology components of 6G mobile networks. Finally, the 6G ﬂagship project will also carry out the large pilots with a test network with the support of both industry and academia.

6) INSPIRE-5Gplus~\cite{INSPIRE5GPLUS} -  The project's objective is to advance beyond 5G network security and privacy. INSPIRE-5Gplus is committed to enhancing security across multiple areas, including overall vision, use cases, architecture, integration with network management, assets, and models. INSPIRE-5Gplus addresses critical security challenges across a range of vertical applications, from autonomous and connected vehicles to critical industry 4.0. INSPIRE-5Gplus will develop and implement a fully automated end-to-end smart network and service security management framework that enables protection, trustworthiness, and accountability in managing 5G network infrastructures across multiple domains. INSPIRE-5Gplus's conceptual architecture is divided into Security Management Domains (SMD) to facilitate the separation of security management concerns. Each SMD is accountable for intelligent security automation of its scope's resources and services. The E2E service SMD is a unique SMD dedicated to managing end-to-end services' security. Using orchestration, the E2E service SMD manages communication between domains. Each SMD, including the E2E service SMD, comprises a collection of functional modules that work in a closed-loop fashion to provide software-defined security orchestration and management that enforces and controls security policies for network resources and services in real-time.

7) Deamon~\cite{DAEMON} - The DAEMON Network Intelligence (NI) - native architecture introduces intelligence directly into the user plane, creating a hierarchy of NI instances for network management that removes the restrictions of closed-loop models. The project concept identifies three levels of NI, depending on their operation timescale: NI at Orchestrators, NI at Non-Real-Time Controllers, and NI at Real-Time Controllers. As an inherent part of its proposed NI-native architecture, DAEMON sets out a structured approach to system-wide NI coordination by introducing an original NI Orchestration layer for selecting the best NI algorithm to be executed within each NI instance. It offers a predefined range of options, coordinating all NI instances in the system, in order to ensure gracious operation of all live mechanisms operating at different timescales and in different micro-domains. DAEMON aims to advance current solutions for managing resources at the Edge by designing dedicated NI algorithms that implement a holistic orchestration approach where diverse Edge functionalities are operated at different timescales and locations whilst acting as a single integrated system via a multi-time scale loop. In addition to this, the project studies the design of loss functions that are based on expert knowledge and builds on recent advances in machine learning for the automated solution of the loss metric mismatch by means of an active bridge that continuously receives feedback about the relevant service metrics. 

 The Table \ref{tbl:projectssum} provides a summary of each project discussed, that highlights the main objectives of each projects, its applications or outcomes and the implemented use-cases that demonstrate the applicability of these projects.

\begin{center}
\scriptsize
\begin{longtable}[htb]{|p{1.5cm}|p{4.3cm}|p{4.3cm}|p{4.3cm}|}
\caption{Summary of B5G/6G Research Projects} \label{tbl:projectssum}
    \\
\hline
    \rowcolor[HTML]{CBCEFB} 
\multicolumn{1}{|c|}{\textbf{Project}} & \multicolumn{1}{c|}{\textbf{Objectives}} & \multicolumn{1}{c}{\textbf{Applications/Outcomes}} & \multicolumn{1}{|c|}{\textbf{Use-cases/Implementation}}    
\\ \hline

AI4EU~\cite{AI4EU} &  To develop a European AI-on-demand platform and ecosystem to facilitate knowledge transfer from research to business. &     Industrial applications with AI on-demand services. & A scalable AI-as-a Service service for the deep edge, an AI in energy service for small and medium-sized enterprises, a matchmaking service for connecting businesses, earth observation, experimentation, planning and scheduling services. \\\hline
XMANAI~\cite{XMANAI} 
& Utilize XAI models in the manufacturing value chain to provide models explainable to ``human in the loop'' and to produce value-based explanations. & Catalogue of XAI algorithms applicable for different manufacturing problems and a platform to collaborate among main actors in AI systems. & Four main industrial  use-cases with AI for production optimization, product demand planning, quality optimization and smart measurement planning.

\\ \hline
SPATIAL~\cite{SPATIAL} &
To create resilient accountability metrics, privacy-preserving techniques, verification tools, and system improvements that will serve as essential elements of a trustworthy AI architecture. & Provide systematic verification and validation software/hardware mechanisms, enhance the resilience of AI in decentralized, uncontrolled environments, guidelines for trustworthy AI solutions, improve the socio-legal awareness, and a communication framework for accountable and transparent understanding of AI. & Four use-cases for privacy preserving AI on the edge and beyond, improving explainability, resilience and performance of cybersecurity analysis of 5G/4G/IoT networks, accountable AI in emergency e-call systems, and resilient cybersecurity analytics.

\\ \hline
STAR~\cite{STAR} &
Enable the deployment of  standard-based secure, safe, reliable, and trusted human-centric AI systems robust to security attacks in manufacturing environments. & Specifications for safe and human centric AI in manufacturing, security and data governance for manufacturing AI systems, safe transparent and reliable human robot collaboration, digital twins, digital innovation hub for secure and safe AI in manufacturing, and standardization. & Human-cobot collaboration for robust quality inspections, human-centered AI for agile manufacturing 4.0, and human behaviour prediction and safe zone detection for routing.
\\ \hline
6G Flagship~\cite{6GFLAGSHIP} &
Commercialization of 5G networks and the creation of the next 6G standards. It also targets wireless connectivity, distributed intelligent computing, and privacy for 6G technologies. & Understanding on 6G vertical applications: health, energy, automotive, and industry. & Implementation and showcase of radio technologies, and 6G test network development.
\\ \hline
INSPIRE-5Gplus~\cite{INSPIRE5GPLUS} &
To advance B5G network security and privacy to address challenges across multiple domains. introduce new architecture for security management domains for separation of security management requirements. & Enhancing security across multiple areas in B5G, including overall vision, use cases, architecture, integration with network management, assets, and models. & Devise and implement a fully automated end-to-end smart network and service security management framework.
\\ \hline
Deamon~\cite{DAEMON} -  DAEMON Network Intelligence &
Design a B5G architecture that is native to network intelligence and integrate network intelligence-assisted features to introduce intelligence directly into the user plane. & Designing a network intelligence native architecture for B5G systems, developing special network intelligence assisted network functionalities, and establishing fundamental guidelines. & Development of network intelligence framework and tools, real-time control and VNF intelligence, and intelligent orchestration management.
\\ \hline
\end{longtable}
\end{center}

\subsection{Standarization related to 6G privacy}
Standardization is essential for establishing the technological criteria for B5G networks and selecting the best technologies for 6G network implementation. As a result, standards shape the worldwide telecommunications industry. Standardizing 6G has been assigned to several Standards Developing Organizations (SDOs). Standardization activities in the field of 6G privacy are summarized in the Table~\ref{tab:standards}.
\begin{centering}
\scriptsize
\begin{longtable}[!htbp]{|p{1cm}|p{1.5cm}|p{3.5cm}|p{8.5cm}|}
    \caption{Standards relevant to privacy}
    \label{tab:standards}
    \\
\hline
    \rowcolor[HTML]{CBCEFB} 
\multicolumn{1}{|c|}{\textbf{Org. }} & \multicolumn{1}{|c|}{\textbf{Identifier}} &\multicolumn{1}{|c|}{\textbf{Standard Title}}     
&\multicolumn{1}{|c|}{\textbf{Description}}     \\\hline \hline
    NIST   &   NISTIR 8228~\cite{NISTIR8228}&   Considerations for Managing IoT Cybersecurity and Privacy Risks & The purpose of this publication is to help federal agencies and other organizations better understand and manage the cybersecurity and privacy risks associated with their IoT devices throughout the devices’ lifecycles. This publication is the introductory document providing the foundation for a planned series of publications on more specific aspects of this topic.\\\hline
    
    NIST   &   NIST SP 800-53 Rev. 4~\cite{NISTSP80053}&   Security and Privacy Controls for Federal Information Systems and Organizations&This publication provides a catalog of security and privacy controls for federal information systems and organizations and a process for selecting controls to protect organizational operations (including mission, functions, image, and reputation), organizational assets, individuals, other organizations, and the state from a diverse set of threats including hostile cyber attacks, natural disasters, structural failures, and human errors (both intentional and unintentional). The security and privacy controls are customizable and implemented as part of an organization-wide process that manages information security and privacy risk. \\\hline
    
    NIST	&   SP 800-53A Rev. 4~\cite{NISTSP80053A} &   Assessing Security and Privacy Controls in Federal Information Systems and Organizations&This publication provides a set of procedures for conducting assessments of security controls and privacy controls employed within federal information systems and organizations.\\\hline
    
    NIST	&   SP 800-53 Rev. 5~\cite{NISTSP80053rev5}	&   Security and Privacy Controls for Information Systems and Organizations& This publication provides a catalog of security and privacy controls for information systems and organizations to protect organizational operations and assets, individuals, other organizations, and the state from a diverse set of threats and risks, including hostile attacks, human errors, natural disasters, structural failures, foreign intelligence entities, and privacy risks. \\\hline
    
    NIST	&   SP 1500-4r2~\cite{NISTSP1500}&   NIST Big Data Interoperability Framework: Volume 4, Security and Privacy Version 3&This publication considers new aspects of security and privacy concerning Big Data, reviews security and privacy use cases, proposes security and privacy taxonomies, presents details of the Security and Privacy Fabric of the NIST Big Data Reference Architecture (NBDRA), and begins mapping the security and privacy use cases to the NBDRA.\\\hline

    NIST	&   SP 800-37 Rev. 2~\cite{NISTSP800-37}	&   Risk Management Framework for Information Systems and Organizations: A System Life Cycle Approach for Security and Privacy&This publication describes the Risk Management Framework (RMF) and provides guidelines for applying the RMF to information systems and organizations. The RMF provides a disciplined, structured, and flexible process for managing security and privacy risk, including information security categorization; control selection, implementation, and assessment; system and common control authorizations; and continuous monitoring.  \\\hline

    ETSI	&   ETSI TR 103 305-5 V1.1.1 (2018-09)~\cite{ETSITR1033055}	&   CYBER; Critical Security Controls for Effective Cyber Defence; Part 5: Privacy enhancement&The ETSI TR 103 305-5 document is describing privacy-enhancing implementations using the Critical Security Controls. These presently include a privacy impact assessment and use of the Controls to help meet provisions of the EU GDPR.\\\hline

    ETSI	&   ETSI TS 103 485 V1.1.1 (2020-08)~\cite{ETSITS103485}&   CYBER; Mechanisms for privacy assurance and verification& The document defines the means to enable assurance of privacy, using the conventional Confidentiality, Integrity, Availability (CIA) paradigm and with reference to the functional capabilities for privacy protection described in Common Criteria for Information Technology Security Evaluation. The document addresses privacy assurance within the context of Identity Management following the model described in ETSI TS 103 486. The present document addresses the cases where both transient and persistent identifiers are used, where identifiers are used in isolation, and where identifiers are used in combination.\\\hline

    ETSI	&   ETSI TR 103 304 V1.1.1 (2016-07)	\cite{ETSITR103304}&   CYBER; PII Protection in mobile and cloud services&The document proposes several scenarios focusing on today's ICT and develops an analysis of possible threats to PII in mobile and cloud-based services. It also presents technical challenges and needs derived from regulatory aspects (lawful interceptions). It consolidates a general framework, in line with the regulation and international standards, where technical solutions for PII protection can be plugged into.\\\hline
    
    ISO	&   ISO/IEC DIS 27555~\cite{ISOIECDIS27555} 	&   Information security, cybersecurity, and privacy protection – guidelines on personally identifiable information deletion&This document contains guidelines for developing and establishing policies and procedures for deletion of PII in organizations by specifying (1) an approach for efficiently defining deletion rules (2) a broad definition of roles, responsibilities, and processes (3) a harmonized terminology for PII deletion.\\\hline 
    
    ISO &	ISO/IEC 27007:2020 ~\cite{ISOIEC27007} 	&   Information security, cybersecurity, and privacy protection — guidelines for information security management systems auditing&This document guides the management of an Information Security Management System (ISMS) audit programme, on conducting audits, and on the competence of ISMS auditors, in addition to the guidance contained in ISO 19011.\\\hline

    ISO &   ISO/IEC 27009:2020 ~\cite{ISOIEC27009}	&   Information security, cybersecurity and privacy protection — sector-specific application of iso/iec 27001 — requirements&This document specifies the requirements for creating sector-specific standards that extend ISO/IEC 27001, and complement or amend ISO/IEC 27002 to support a specific sector (domain, application area or market). It explains how to include requirements in addition to those in ISO/IEC 27001.\\\hline

    ISO	&   ISO/TR 21332:2021 ~\cite{ISOTR21332}	&   Health informatics — Cloud computing considerations for the security and privacy of health information systems&This document provides an overview of security and privacy considerations for Electronic Health Records (EHR) in a cloud computing service that users can leverage when selecting a service provider.\\ \hline

    BSI	&   PD ISO/TR 18638:2017 ~\cite{ISOTR18638}	&   Health informatics. Guidance on health information privacy education in healthcare organizations& This standards specifies the essential educational components recommended to establish and deliver a privacy education program to support information privacy protection in healthcare organizations. The primary users of this document are those responsible for planning, establishing, and delivering healthcare information privacy education to a healthcare organization.\\\hline
\end{longtable}
\end{centering}

\section{Lessons Learned and Future Research Directions} \label{sec:lesson}

The research and work related to 6G are currently in the early stages. Therefore, it is clear that B5G/6G privacy is also in its initial state. We can hypothesize it will increasingly rise the requirements on privacy with the new technologies, a broad range of services, and a massive amount of data collected every day in beyond 5G networks. Also, since intelligence is added in future B5G/6G networks, there will be more opportunities for reducing human supervision, thus creating more space for privacy leakages. Furthermore, there are unsolved questions in privacy that can be found in the existing 5G networks. However, we can also find potential solutions for these issues from the existing research. For standardization of these efforts, there are many research and projects are in progress.

To summarise the discussions, the following subsections will briefly describe the learning of each of our discussions, the questions, and future possibilities of research.

\subsection{Privacy Taxonomy}
\subsubsection{Lessons Learned}
Considering the taxonomies of privacy, we have observed different approaches to categorizing privacy based on privacy needs, functions to fulfill those needs, and privacy based on different actions done on data subjects. This implies that privacy may be categorized according to different views, and there is no universal definition of privacy nor a taxonomy. Hence, it is based on differing opinions on what information should be excluded and what actions should be taken with data.  

\subsubsection{Open Research Questions}

The taxonomy defined in our survey suggested that the following questions may need to be addressed:
\begin{itemize}
    \item The concept of privacy is non-intuitive and subtle. This may make it difficult to understand the importance for a general user. Therefore, how is it possible to make the general user aware of privacy or make their actions not cause privacy damages?
    \item How to differentiate privacy with security aspects in B5G/6G applications and use in the design, and implementation process, as they are closely related yet distinct concepts?
    
\end{itemize}
\color{black}

\subsubsection{Possible Future Directions}
When considering the privacy taxonomy, a unified taxonomy for privacy considering all the aspects of the taxonomies mentioned above and those not included may need to be developed, which can be used as a guideline in B5G/6G for privacy requirements evaluation.

\subsection{Privacy Issues}
\subsubsection{Lessons Learned}
As we discussed, the privacy-related aspects for 6G are based mainly on three considerations. The privacy issues related to AI are the latest addition to the privacy issues. As B5G/6G is coupled with intelligence, privacy issues associated with AI will be applicable for these networks, including attacks on AI components in the network and AI as a tool for launching attacks against the network. It is also shown that privacy considerations exist in current 5G networks, which will also apply to B5G/6G. It includes the issue of non-unified privacy levels and location-dependent privacy variations. Privacy issues related to B5G/6G technologies are the other category of issues that we identified. The associated technologies rising with B5G/6G will also affect users' privacy, especially with new types of sensors, data types, and lack of investigation on privacy on them. Therefore, we can see that privacy comes with a broad range of requirements for future networks to be addressed in these categories. We can take immediate actions on specific areas, such as the issues that are already existing in 5G networks. There are different aspects of AI-based attack scenarios that should be considered before fully adopting AI to B5G/6G since it poses risks to the network with privacy leaks. The future technologies and applications for B5G/6G may need some time for complete consideration of privacy since they evolve at different speeds, where some are already mature enough privacy aspects. Still, some are at the conceptual or initial implementation stages, where privacy might not be a top priority in research and development.

\subsubsection{Open Research Questions}
Though we have considered many aspects of privacy issues applicable in B5G/6G networks, there is still space for new areas of privacy considerations that may appear over time and ones that are already available but not included. The question, therefore, exists to identify such privacy issues and their relevancy for B5G/6G.  We can identify several other important questions on privacy issues remaining as:
\begin{itemize}
    \item How to identify the impact of cross-cutting issues on each of the layers of B5G/6G networks?
    \item It is difficult to address the privacy issues based on location due to the dynamic nature of geopolitics, demographics, and other cases such as conflicts, lack of trust among neighboring states, and cultural influences. Therefore, what actions can be taken to mitigate the influence of the factors affecting the privacy of individuals utilizing B5G/6G in different regions?
\end{itemize}

\subsubsection{Possible Future Directions}
The layered model proposed for B5G/6G networks may probably get updated with new layers added in the future. Therefore, new privacy issues may need to be added by considering them. Even the existing model can be improved; for example, when considering the intelligent sensing layer, issues related to privacy in the sensing layer may require inclusion of both edge and fog computing and storage requirements. When considering the data mining and analytics layer, we can implement a quantitative measurement system for privacy during a data mining task. For this, it may be possible to use techniques such as differential privacy, as mentioned, to mitigate the issues related to this layer. The intelligent control layer should be provided with better privacy measures, including privacy-enhancing techniques and explainable interfaces for AI. It appears to be the most vulnerable to attacks due to its interfacing with external applications. Also, the adoption of new applications could be done with caution using a precise set of measurable privacy thresholds before connecting with the smart application layer. As we mentioned, since third-party technology applications are evolving at different speeds, having such a defined privacy filter before the application layer can reduce the consideration of privacy measures for each new technology.

\subsection{Privacy Solutions}
\subsubsection{Lessons Learned}
Considering the proposed solutions, it is clear that some of them cause a higher impact over the others since we can see some privacy solutions address more issues in B5G/6G as shown by Table \ref{tbl:solutionsForIssues}. We think it is important to prioritize encryption mechanisms, privacy preserved decentralized AI, XAI, and intelligent management solutions. Consequently, we also can observe the higher impact of AI-based solutions and encryption mechanisms. Considering more active solutions where the current research community is active, we can consider XAI, blockchain-based solutions, homomorphic encryption, edge AI, and decentralized AI since they have more recent related work than the other solutions proposed. Government regulatory approaches are relatively easy to adopt, but they do necessitate the involvement of the relevant legal authorities. Applying the privacy by design approaches is also feasible to adopt and implement in practice, such as at industry levels, to mitigate potential privacy threats associated with new designs.

\subsubsection{Open Research Questions}
Considering the existing issues that are outside the scope of this paper and thus not included, we list the following questions remaining in privacy solutions:
\begin{itemize}
    \item How much the influence on industries and potential impact on the economy due to possible frictions when implementing privacy solutions?
    \item We mentioned several types of costs for each privacy solution in Table~\ref{tbl:privacySolutionsSummary}. What will be the trade-offs before applying these solutions to a certain privacy issue, depending on various factors, such as use-case, costs, possible technologies, etc.?
\end{itemize} 

\subsubsection{Possible Future Directions}
Some of the solutions proposed are already implemented and well studied; meanwhile, some solutions are newly introduced that require more background check on their feasibility, considering the financial and technical capability of a certain organization implementing them. For example, blockchain is now applied practically in the real world and is continuously evolving. Also, we can see it is getting cheaper and technically much more feasible to implement since the background and infrastructure required to implement blockchain applications is reducing with the introduction of new blockchain frameworks, services, documentation, and the interest of the learning community. Similarly, we can see solutions such as differential privacy, homomorphic encryption, and edge AI applications are also increasing, based on their increasing related work in recent years. Other solutions such as new techniques of decentralized learning, personalizing of privacy, and privacy levels can be further studied to be applied for B5G/6G networks, as well as for other futuristic applications.

\subsection{Privacy Projects and Standardization}
\subsubsection{Lessons Learned}
There are numerous EU and non-EU funded projects that already started to address the privacy challenges in B5G networks. Projects listed in Section \ref{sec:projects} aim to guarantee the next generation network privacy and security using approaches beyond classical, for example XAI-based techniques to assure the privacy of future networks that play major role in most of the research projects reviewed in this paper. Projects listed above have in common, that they all consider AI-powered scenarios to approach the privacy and security concerns and it is very clear that the integrity of those solutions is going to be the greatest concern. 
Standards Developing Organizations (SDOs), such as NIST, ETSI, 3GPP, IETF, IEEE, and ISO, are expected to work on 6G security and privacy in the near future or already do so. 6G aims to merge different technologies already standardized by SDOs. AI/ML mechanisms will have to become the main elements in 6G to achieve a satisfactory level of privacy, such as automating decision-making processes and achieving a zero-touch approach.

\subsubsection{Possible Future Directions}
\begin{figure}[ht]
    \centering
    \includegraphics[width=0.7\linewidth]{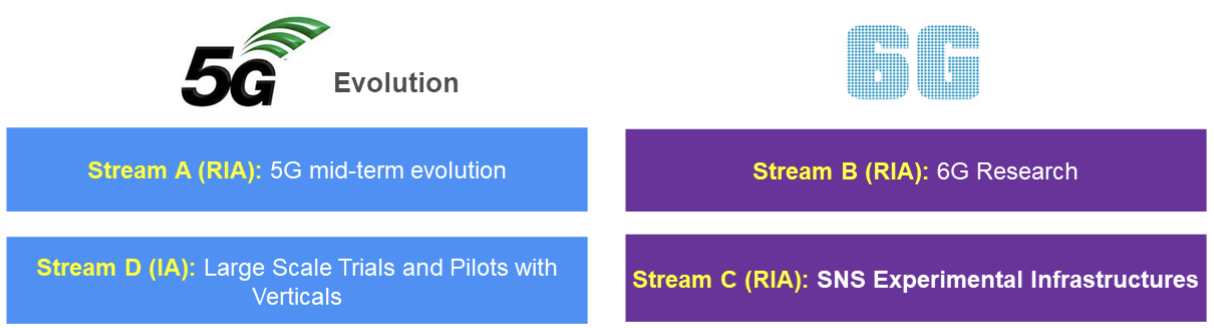} 
    \vspace{1mm}
    \caption{Four major streams of the Smart Networks and Services Joint Undertaing programme}
    \label{sns-streams}
\end{figure}

The EU established a Smart Networks and Services Joint Undertaking (SNS) programme in 2021 with the commitment to contribute €900 million from the EU over the course of the next seven years. The goal of SNS is to enable European companies and research institutions to create research and development capabilities for 6G systems and to establish themselves as leaders in the markets for 5G and 6G infrastructure, which will serve as the foundation for digital and green transformation. The SNS work program will serve as the foundation for calls for proposals that will be issued in early 2022. The SNS work programme is responsible for developing the first phase of the SNS roadmap and the expansion of the early wave of European 6G projects that were launched in January 2021 under the 5G-PPP. Figure~\ref{sns-streams} shows four main streams of the SNS programme. In terms of standards, we anticipate that projects funded under calls like ICT-52-2020 can contribute valuable insights to standardization bodies, assisting in developing advanced 6G systems and technologies. 3GPP believes that there are still features and capabilities from existing 5G systems that require full specification, which is planned to be delivered towards the end of 2023 according to the 3GPP schedule. To complete the transition from historical and existing proprietary radio protocols to next generation 3GPP protocols, it will take around 5-10 years. The development of AI/ML-assisted privacy is still in its early stages. It will take time to respond to the new security and privacy threats presented by the dynamic nature of 6G services and networks.

\subsection{Other Emerging Topics related to 6G Privacy} \label{sec:nssecurityprivacy}

6G will cohabit with emerging technologies such as metaverse, quantum computing, cloud-native technologies, telco clouds, and Open RAN. We briefly describe these currently evolving, and upcoming technologies below that could potentially impact B5G/6G networks in the future.

\subsubsection{Metaverse}
The concept of metaverse initiated from the book named Snow Crash in 1992
\cite{stephenson2003snow}, where it is regarded as a computer-generated virtual environment. There are characters in this environment, called ``Avatars'', where they can perform actions similar to or even beyond the perception of the actual physical world. We see there will be a very high likelihood of becoming the metaverse mainstream since more people are inclining towards the remote working environment. Industrial attention on metaverse has significantly increased recently. For example, rebranding the name Meta for the Facebook company~\cite{kim2021advertising} emphasizes the plans on investing in metaverse by the private sector. The current work on metaverse will be accelerated by the introduction of B5G/6G communication since metaverse will need much higher data rates for communication. Especially important when considering high-quality three-dimensional environment rendering, multi-sensory remote devices, and the real-time connectivity of billions of users in a complex virtual environment to create a seamless social experience. There will be an array of new technologies with haptics to emulate sensations for metaverse users, demanding faster communication in B5G/6G, which may be crucial for their real-time operation.

The introduction of metaverse will clearly come with privacy concerns since they will be tracking user behaviour, PIIs, emotions, preferences in a three spatial dimensional virtual environment with complex multi-dimensional data that are not currently available, with new sensors and haptic technologies. Considering the behaviour of users, the work about a metaverse game named Second Life shows the majority of players (72\% female and 68.8\% male) show the same real-world behaviour in the game~\cite{leenes2007privacy}. Therefore, it is clear that actual user behaviour can be traced by third parties which provide the metaverse. The critical factor here is that the user behaviour detected through indirect data analytics methods currently used will not be needed since the user may be virtually living and spending most of their daily life in the metaverse. In metaverse VR/XR applications will be heavily used. The work in~\cite{nguyen2021security}, for example, addresses how XR could acquire biometric data and physical motions.~\cite{9419108} also agrees that XR collects a diverse set of data. Since the sensing devices in these applications could track the user, which could be used to easily identify patterns of user behaviours, and the activity the user is doing. The possibility of privacy leakage is much higher in such an environment since other private data such as user preferences and emotions could be easily predicted through the behaviours or actions. Also, as this information will be communicated through B5G/6G networks, there could be many privacy attacks since it is one approach to get the user data in the metaverse.

\subsubsection{Quantum Computing}
Quantum computing is another emerging field related to B5G/6G privacy since the use of quantum computing operations may pose threats to the privacy of users since nearly all cryptographic schemes may likely become obsolete with the delivery of quantum computing~\cite{folger2016quantum}. For example, quantum computers can easily factor very large numbers, whereas classical computers do not. This forms the basis for asymmetric encryption, and thus it has many possibilities of decrypting the data, which could contain private data, leading to the leakage of privacy. These capabilities of quantum computers will be practical in the B5G/6G era in the upcoming decade; thus, special concern should be given to this since secure communication is a vital factor for privacy protection. It will also affect the privacy of IoT communication since these devices may also rely on lightweight cryptosystems due to their resource constraint environments~\cite{fernandez2019pre}. However, quantum computing itself can act as the solution for the issues arising through it. The techniques such as quantum key-distribution use physics principles to encode and transmit data using photons, which is an unbreakable form of cryptography~\cite{folger2016quantum}. 

The use of quantum computing not only will prevent privacy leakages but will also help enhance privacy, such as in ML operations. Quantum clustering is one such approach that can be used for privacy enhancement as discussed in~\cite{ramezani2020machine}. It shows that fewer queries to the database will be required for clustering operations with quantum clustering algorithms, which will help reduce the possibility of privacy leakages. Such approaches for ML will benefit B5G/6G since the future networks, and their applications that heavily use ML.

\subsubsection{Cloud Native Technologies}
The cloud-native applications are built in the cloud, rather than the classical application development ``on-premises''. The work in~\cite{gannon2017cloud} provides a list of properties where cloud-native applications are different from conventional applications. It includes operation globally with replicated data, serving thousands of concurrent users by scaling well, assuming that infrastructure is fluid and failure is constant, and testing without disruption. However, the authors show another characteristic that the cloud-native applications are made of aggregation of many different cloud components where they mention these components should not hold sensitive credentials. Also, in ~\cite{gannon2017cloud} it is shown the firewalls are insufficient as the access control need to be managed at multiple levels. Therefore, privacy should be a major concern for cloud-native applications. The designers of cloud-native applications could take privacy-by-design approaches, where privacy should be integrated with the application architecture. The work in~\cite{grunewald2021cloud} identifies issues in cloud-native system challenges and shows that software engineers are ill-equipped with privacy-preserving methods addressing all privacy principles and agile development of software engineering often neglect or even contradict privacy principles. There will be a further increase of cloud-native applications in the future with B5G/6G networks. Thus, novel privacy-aware development lifecycles should be investigated, and adoption techniques for privacy should be considered in software engineering practices. 

\subsubsection{Telco Clouds}
The combination of the telecommunication sector with the cloud is an ongoing evolution by combining telcos with the clouds, mainly two ways~\cite{soares2015toward}: 1) telcos supporting the cloud to provide scalable computing and storage services on demand, and 2) telcos using the cloud functions to implement virtualized service functions through hosting and cloud computing with NFV and SDN. It makes possible agile operation,  addition of services on demand, faster response to changes, and efficient management of resources is supported. Therefore, we can expect Telco cloud operations will also be extensively used for B5G/6G networks due to their dynamic fulfillment of work. 

The use of cloud technologies on network functions may also pose all privacy requirements associated with cloud infrastructure to the networks, such as a higher possibility of attacks since the traditional physical boundaries that define and protect information are transformed and disappear with the virtualization~\cite{zhiqun2013emerging}. Therefore, future research work can focus more on telco cloud privacy protection mechanisms in the absence of these physical barriers, which may also help ensure privacy in general-purpose cloud computing and storage. 

\subsubsection{Open RAN}
A Radio Access Network (RAN) is a vital component in mobile telecommunication systems that implements radio access technology. It is placed in between a remotely operated User Equipment (UE) and the core network and consists of physical infrastructure radio units such as base station antennas and processing units~\cite{singh2020evolution}. Open RAN is a concept where the RAN portion of the network is designed and built by combining hardware and software components from different vendors, using open and interoperable interfaces for multi-vendor implementation, commercial off-the-shelf hardware, and virtualization software~\cite{lee2020open,alliance2020ran}. The O-RAN Alliance is the consortium defining the specifications of the  Open RAN~\cite{lee2020open}. One of the key benefits of Open RAN is its capability of utilizing ML and AI to empower network intelligence through open and standardized interfaces~\cite{alliance2020ran}. Therefore, Open RAN will be a useful association for B5G/6G networks since these networks intend to operate in an intelligence-based control and management. It also provides flexible deployment options and service provisioning models of virtualized network elements in telco clouds~\cite{alliance2020ran}. 

 However, we see there are certain privacy-related issues here as well, since with the facilitation of AI and its possibility of having privacy attack scenarios as described in Section \ref{sec:issues}. Also, the privacy issues such as the ability to identify UE by attackers unauthorized access are some other scenarios that pose potential privacy risks for users who get services through the Open RAN~\cite{alliance2021ran}. Therefore, to mitigate the issues, these risks due to evolving threats should be identified, and periodically re-assessed~\cite{alliance2021ran}.

\section{Conclusion}\label{conclusion}
In this paper, we provide a comprehensive survey on the privacy aspects related to B5G/6G networks. Privacy is a subjective concept, which is differentiating based on numerous factors. To get an overview, we provided a set of taxonomies defined for privacy. We identified several issues with varying difficulties are available to limit ensuring privacy. The addition of intelligence, ultra-high data rates transmitting both personal and non-personal data, and adopting a range of novel technology ecosystems with new sensing capabilities make future network privacy more challenging. Emerging research work, well-established concepts, and existing technologies in different fields can be used as potential solutions to cope with these challenges. The investigation revealed that these solutions also have many gaps that need to be addressed, depending on various factors such as maturity, applicability, and costs of the solutions. Finally, we discuss lessons learned, possible future directions, and other emerging topics that appear with B5G/6G networks to lay the foundation for further research to be made in the aspects of privacy.

\section{Acknowledgments}

This work is primarily supported by European Union under the SPATIAL (Grant No: 101021808) project. In addition, this work is partly supported by European Union under projects CONCORDIA (Grant No: 830927) and  ACCORDION (Grant No: 871793);
Academy of Finland under project 6Genesis (Grant No: 318927);
Science Foundation Ireland under CONNECT phase 2 (Grant No: 13/RC/2077\_P2).

\printcredits
\bibliographystyle{unsrt}
\bibliography{privacy-survey}
\vskip3pt

\bio{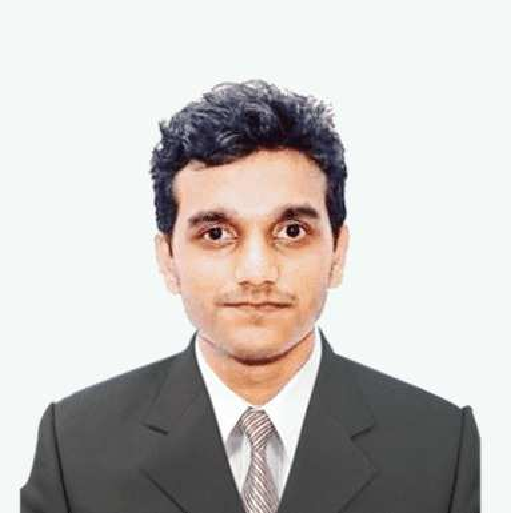}\textbf{Chamara Sandeepa} is currently a Ph.D. student in the School of Computer Science, University College Dublin, Ireland, and Doctoral Student/Research Engineer of the EU H2020 SPATIAL project. He is currently working in the field of privacy aspects of AI. He received his Bachelor's degree in Electrical and Information Engineering from the University of Ruhuna, Sri Lanka in 2020.  During his undergraduate period and later work, he actively contributed in research and published in multiple conferences and journals. He has professional experience in Software Engineering, and he also worked in the fields of IoT, and AI.
\endbio
\vspace{20pt}
\bio{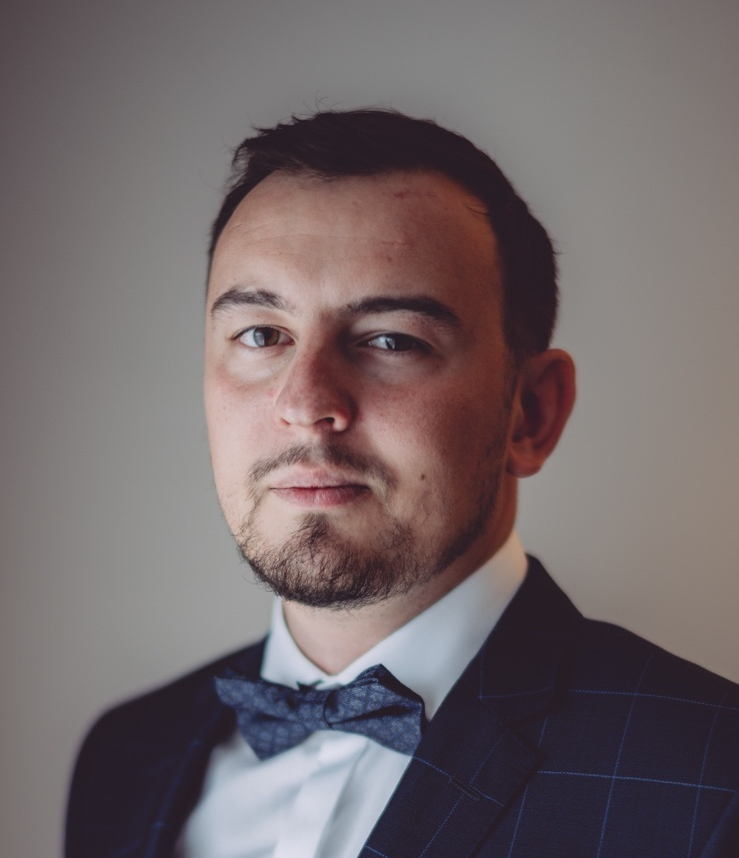}\textbf{Bartlomiej Siniarski} is currently a post doctoral researcher and a project manager for the EU H2020 SPATIAL project at University College Dublin. He completed his undergraduate studies in Computer Science at University College Dublin (Ireland) and University of New South Wales (Australia). He was awarded with a doctoral degree in 2018. He has a particular interest and experience in the design of the IoT networks and in particular collecting, storing and analysing data gathered from intelligent sensors. Furthermore, he was actively involved in MSCA-ITN-ETN, ICT-52-2020 and H2020-SU-DS-2020 projects which are focused on solving problems in the area of network security, performance and management in 5G and B5G networks.
\endbio

\vspace{20pt}

\bio{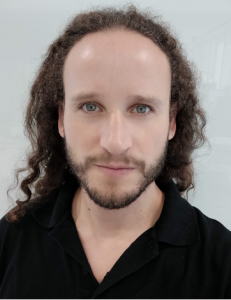}\textbf{Nicolas Kourtellis} is a Principal Research Scientist in the Telefonica R\&D team, in Barcelona.
He holds a Ph.D. in Computer Science and Engineering from the University of South Florida (2012).
He has done research at Yahoo Labs, USA and at Telefonica Research, Spain, on online user privacy and behavior modeling with machine learning, Internet measurements and distributed systems, with 80+ published peer-reviewed papers.
Lately, he focuses on CyberPrivacy (Privacy-preserving Machine Learning and Federated Learning on the edge, user online privacy and PII leaks), etc.) and CyberSafety (modeling and detecting abusive, inappropriate or fraudulent content on social media using data mining and ML methods, and how they can be applied on the edge on user-owned or network devices).
He has served in many technical program committees of top conferences and journals (e.g., WWW, KDD, CIKM, ECML-PKDD, TKDD, TKDE, TPDS, etc.), and presented his work in top academic and industrial venues such as The Web Conference (WWW), Apache BigData Europe and North America, and Mobile World Congress Barcelona.
\endbio
\bio{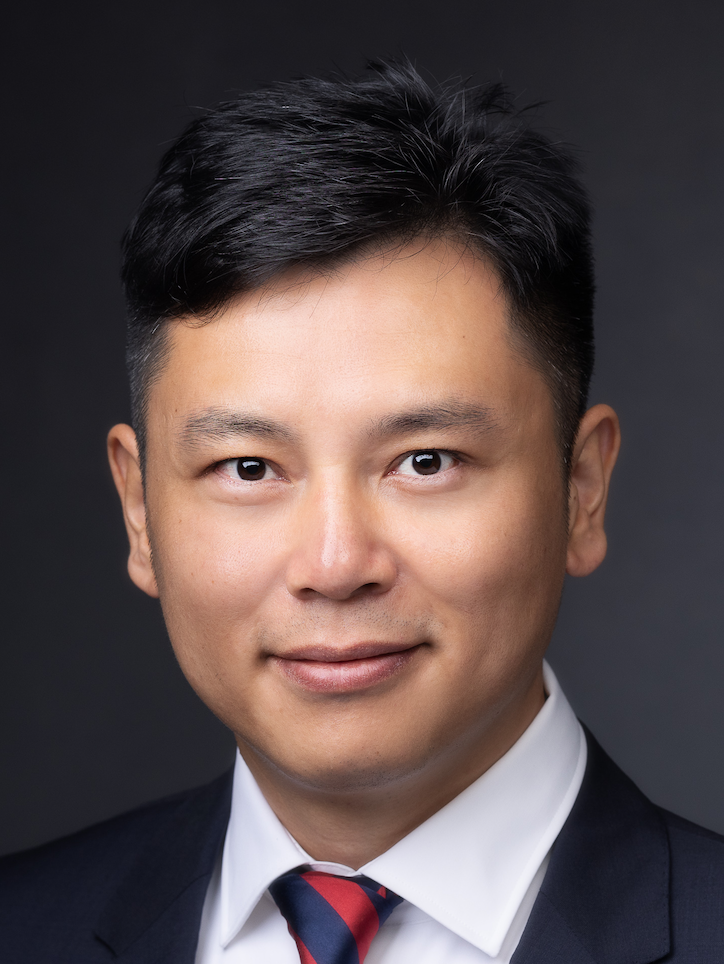}\textbf{Shen Wang} is currently an Assistant Professor with the School of Computer Science, University College Dublin, Ireland. He received the M.Eng. degree from Wuhan University, China, and the Ph.D. degree from Dublin City University, Ireland. Dr. Wang is a member of the IEEE and has been involved with several EU projects as a co-PI, WP and Task leader in big trajectory data streaming for air traffic control and trustworthy AI for intelligent cybersecurity systems. Some key industry partners of his applied research are IBM Research Brazil, Boeing Research and Technology Europe, and Huawei Ireland Research Centre. He is the recipient of the IEEE Intelligent Transportation Systems Society Young Professionals Travelling Fellowship 2022. His research interests include connected autonomous vehicles, explainable artificial intelligence, and security and privacy for mobile networks.
\endbio

\pagebreak
\vspace{35pt}
\bio{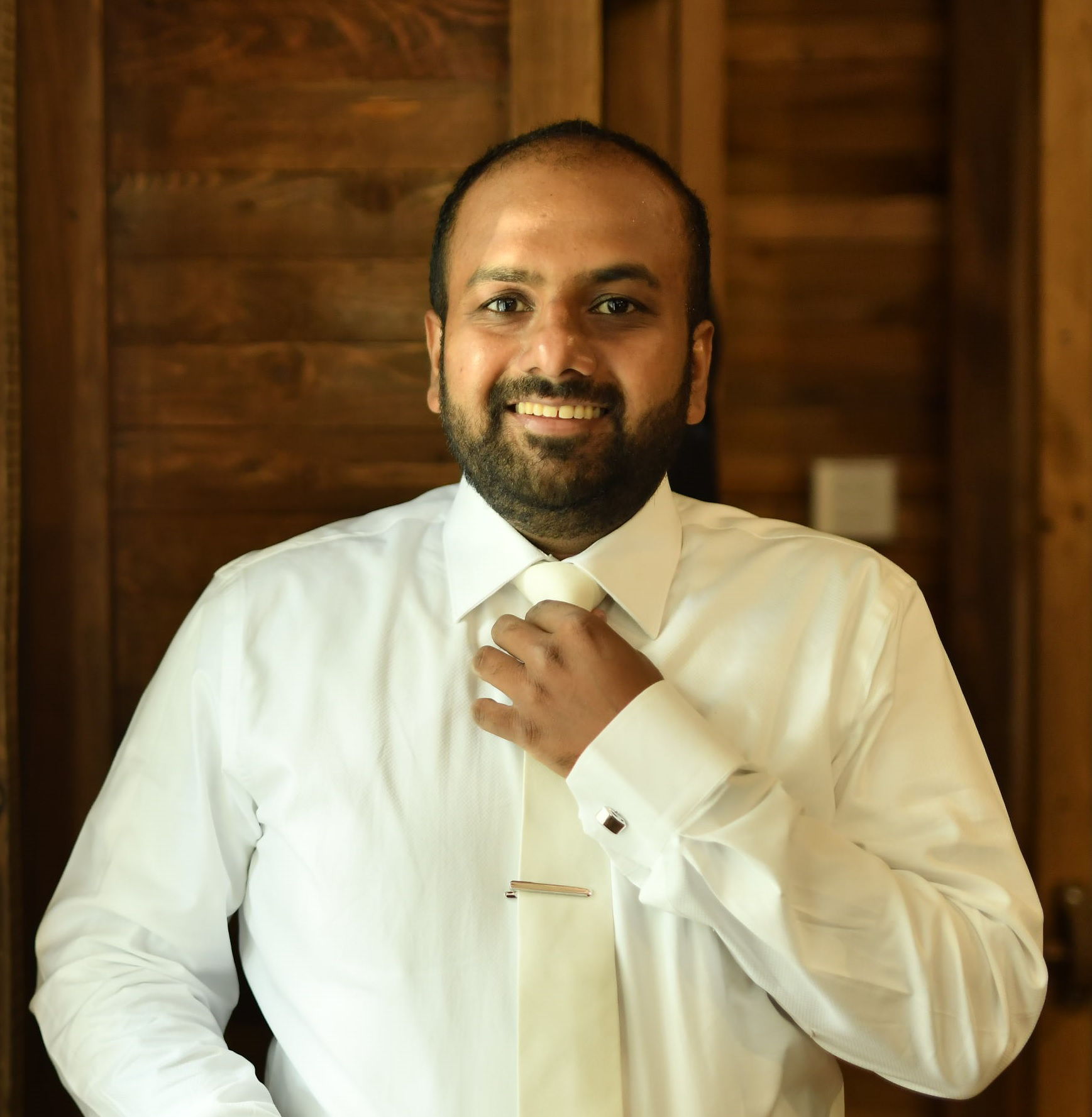}\textbf{Madhusanka Liyanage} is an Assistant Professor/Ad Astra Fellow and Director of Graduate Research at the School of Computer Science, University College Dublin, Ireland. He is also acting as a Adjunct Professor at the centre for Wireless Communications, University of Oulu, Finland, the Department of Electrical and Information Engineering, University of Ruhuna, Sri Lanka and  the Department of Electrical and Electronic Engineering, University of Sri Jayawardhenepura, Sri Lanka. He received his Doctor of Technology degree in communication engineering from the University of Oulu, Oulu, Finland, in 2016. From 2011 to 2012, he worked as a Research Scientist at the I3S Laboratory and Inria, Sophia Antipolis, France. He was also a recipient of the prestigious Marie Skłodowska-Curie Actions Individual Fellowship and Government of Ireland Postdoctoral Fellowship during 2018-2020. During 2015-2018, he has been a Visiting Research Fellow at the CSIRO, Australia, the Infolabs21, Lancaster University, U.K., Computer Science and Engineering, The University of New South Wales, Australia, School of IT, University of Sydney, Australia, LIP6, Sorbonne University, France and Computer Science and Engineering, The University of Oxford, U.K. He is also a senior member of IEEE. In 2020, he received the "2020 IEEE ComSoc Outstanding Young Researcher" award by IEEE ComSoc EMEA. In 2021, he was ranked among the World's Top 2\% Scientists (2020) in the List prepared by Elsevier BV, Stanford University, USA. Also, he was awarded an Irish Research Council (IRC) Research Ally Prize as part of the IRC Researcher of the Year 2021 awards for the positive impact he has made as a supervisor.  Dr. Liyanage's research interests are 5G/6G, Blockchain, Network security, Artificial Intelligence (AI), Explainable AI, Federated Learning (FL), Network Slicing, Internet of Things (IoT), Multi-access Edge Computing (MEC), Cyber-Physical Systems (CPS).  More info: \url{www.madhusanka.com}
\endbio
\end{document}